\documentclass[11pt,a4paper]{article}

\pdfoutput=1 

\usepackage{jheppub} 

\usepackage[T1]{fontenc} 
\usepackage{amsmath,amsthm,amssymb}
\usepackage{multicol}
\usepackage{slashed}
\usepackage[Export]{adjustbox}
\usepackage{multicol}
\usepackage{blkarray}
\usepackage{color,soul}

\newcommand{\ed}{\text{d}}

\newcommand{\Mp}{M_{\rm Pl}}

\usepackage{cancel}

\numberwithin{equation}{section}

\usepackage[utf8]{inputenc}
\usepackage[english]{babel}
\usepackage{textcomp}
\usepackage{amsmath}
\usepackage{mathtools}
\usepackage{gensymb}
\usepackage{amsbsy}
\usepackage{amssymb}
\usepackage{capt-of}
\usepackage[T1]{fontenc} 
\usepackage{physics}
\usepackage{multicol} 
\usepackage{multirow}
\usepackage[table]{xcolor}
\usepackage{lscape}
\usepackage[hang, small,labelfont=bf,up]{caption} \usepackage[title]{appendix}
\usepackage[format=plain,
            labelfont={bf}]{caption}
\usepackage[compat=1.1.0]{tikz-feynman}

\def\be{\begin{equation}}
\def\ee{\end{equation}}
\def\bea{\begin{eqnarray}}
\def\eea{\end{eqnarray}}

\newcommand{\cut}[1]{\Lambda_{\rm #1}}

\usepackage{booktabs} 
\usepackage{float} 

\usepackage{hyperref} 
\usepackage{graphicx} 
\usepackage{epstopdf} 
\usepackage{paralist} 
\usepackage{subcaption}
\usepackage[framemethod=TikZ]{mdframed}
\usepackage[shortlabels]{enumitem}

\mdfdefinestyle{Example}{%
    linecolor=black!80!white,
    outerlinewidth=0.5pt,
    roundcorner=0pt,
    innertopmargin=15pt,
    innerbottommargin=15pt,
    innerrightmargin=20pt,
    innerleftmargin=20pt,
    backgroundcolor=yellow!10!white}
\def\example{\begin{mdframed}[style=Example]}

\mdfdefinestyle{Result}{%
    linecolor=black!80!white,
    outerlinewidth=0.5pt,
    roundcorner=2pt,
    innertopmargin=15pt,
    innerbottommargin=15pt,
    innerrightmargin=20pt,
    innerleftmargin=20pt,
    backgroundcolor=blue!5!white}
\def\result{\begin{mdframed}[style=Result]}

\mdfdefinestyle{Question}{%
    linecolor=black!80!white,
    outerlinewidth=0.5pt,
    roundcorner=0pt,
    innertopmargin=15pt,
    innerbottommargin=15pt,
    innerrightmargin=20pt,
    innerleftmargin=20pt,
    backgroundcolor=red!15!white}
\def\question{\begin{mdframed}[style=Question]}

\usepackage[sorting=none,style=numeric-comp]{biblatex}
\title{\boldmath EFT \& Species Scale: Friends or foes?}

\author[a]{Bruno Valeixo Bento,}
\author[b,c]{João F. Melo}

\affiliation[a]{Instituto de F\'isica Te\'orica IFT-UAM/CSIC, \\ C/ Nicol\'as Cabrera 13-15, Campus de Cantoblanco, \\
28049 Madrid, Spain}
\affiliation[b]{Insituut voor Theoretische Fysica, \\KU Leuven, \\Celestijnenlaan 200D, B-3001 Leuven, Belgium}
\affiliation[c]{Leuven Gravity Institute, \\KU Leuven,
\\Celestijnenlaan 200D box 2415, 3001 Leuven, Belgium}

\emailAdd{bruno.bento@ift.csic.es}
\emailAdd{joao.melo@kuleuven.be}

\abstract{Recently the notion that quantum gravity effects could manifest at scales much lower than the Planck scale has seen an intense Swamplandish revival. Dozens of works have explored how the so-called species scale---at which an effective description of gravity must break down---relates to String Theory and the Swampland conjectures. In particular, the interplay between this scale and the abundant towers of states becoming lighter in asymptotic regions of moduli spaces has proved to be key in understanding the real scale of quantum gravity. Nevertheless concerns have been raised regarding the validity of using infinite towers of states when estimating this scale within Effective Field Theory and, more precisely, the consistency of cutting the tower part way through in a framework that relies on a clear separation of scales. In this work we take an EFT point-of-view and provide a detailed perturbative derivation of the species scale by computing the 1-loop graviton propagator in the presence of many fields.
Not only do we clarify the setup, assumptions and regimes of validity of the result, but more importantly apply the same methods to a particular infinite tower of states, where the different scales can be computed and contrasted. We show how each state in the tower contributes to the species scale and how the procedure of counting only ``light fields'' can be compatible with \textit{not cutting the tower}, thereby maintaining the harmony between infinite towers and EFTs even in the context of the species scale.}

\begin{document} 
\maketitle
\flushbottom

\newpage
\section{Introduction}
\label{sc:intro}

Quantum gravity is often thought of as being observationally out of reach. Unlike the other fundamental interactions that have been probed with high precision in particle accelerators, gravity appears to hide its fundamental nature at energies that are far too high for us to probe (not only now, but also in the reasonably near future). For this reason, proposals for quantum gravity such as String Theory are extremely hard to test and their validity has been so far restricted to internal self-consistency\footnote{It is however impressive how much self-consistency can restrict a physical theory, something that is clearly seen in String Theory.} and the ability to reproduce the observed low-energy physics. Regrettably, while this might be enough to rule out some clearly unacceptable possibilities, it still leaves us with an unfathomable number of new-physics scenarios to consider, each presumably just as likely to be ``true'' as any other. 

Perhaps one of the key questions regarding quantum gravity is precisely the scale at which we should expect to see its effects. At low energies, gravity is very well described by General Relativity (GR) and over more than a century we were able to use it to study many gravitational phenomena, from cosmology to black holes and gravitational waves, to name a few. We have also come to understand this intrinsically geometrical theory in the language of Effective Field Theory (EFT), which gave us a deeper understanding of the regime of validity of GR. As we will review in the next section, GR can be thought of as encoding the leading gravitational interactions in a derivative expansion that should include higher-derivative terms as corrections \cite{Donoghue_1994,Burgess_2004,donoghue2017epfllecturesgeneralrelativity,Burgess:2020tbq,Donoghue:2022eay}. We also learn from GR that the gravitational coupling is dimensionful---from the EFT point-of-view it gives a dimensionful parameter controlling an expansion and suggests that the description is only valid up to energies $E<\Mp \approx 2.1\times 10^{19}$ GeV. The Planck scale is therefore our first guess for the scale of quantum gravity. 

However it is not necessarily our last---what $\Mp$ really gives us is an upper bound on the cutoff of our gravitational EFT, not the actual cutoff itself. It is conceivable that some of those higher-derivative interactions come into play at scales much lower than $\Mp$, which would then constitute a lower cutoff for the theory. One such scale was introduced in \cite{Veneziano:2001ah,Han_2005,Dvali:2007hz,Dvali:2007wp,Arkani-Hamed:2005zuc,Distler:2005hi,Dvali:2010vm} and is induced by the gravitational interaction of a large number of fields (or species) within the EFT. The ``species scale'', $\cut{sp}$, was motivated both perturbatively and non-perturbatively (through black hole and entropy arguments), in both cases representing a scale beyond which our effective description of gravity should break down. 

In this work, we will focus our attention on the perturbative definition of $\cut{sp}$ \cite{Donoghue_1994,Han_2005,Anber:2011ut,Aydemir:2012nz,Calmet:2017omb,Caron-Huot:2022ugt}. From the perturbative point of view, the ``species scale'' arises as the scale at which the perturbative expansion of gravitational interactions breaks down. It can be determined by 
performing a 1-loop computation of the graviton propagator, with the light fields running inside the loop and renormalising the gravitational coupling. From this computation, we find a ``strong coupling'' scale rather than an EFT cutoff---we will clarify the relation between the two in the next section. 

The 1-loop graviton propagator in the context of $N$ light fields of the same mass (with ``light'' meaning much smaller than $\Mp$) was presented in \cite{Veneziano:2001ah,Han_2005,Dvali:2007hz,Dvali:2007wp,Arkani-Hamed:2005zuc,Distler:2005hi,Dvali:2010vm}, where the ``species scale'' is derived. Moreover, the framework of gravitational EFTs was discussed in detail in 
\cite{Donoghue_1994,donoghue2017epfllecturesgeneralrelativity,Donoghue:2022eay} and the quantum corrections to the gravitational coupling were also addressed in \cite{Anber:2011ut,Aydemir:2012nz}. 
Here we will compute the 1-loop graviton propagator in the presence of infinite towers of states, which to our knowledge has not yet been presented in full detail. We will also review the calculation for $N$ scalars of the same mass, adding new conclusions and clearer interpretations, to build up to the infinite tower computation. 

The upshot of the argument is that in the presence of a large number of light fields the low-energy effective (and perturbative) description of gravity breaks down at a scale much lower than $\Mp$, thus suggesting a much lower quantum gravity scale. While this gives an interesting twist to the usual discussion of quantum gravity scales, $\Lambda_{\rm sp}$ only differs significantly from $\Mp$ in the presence of an exponentially large number of light fields, which might require some motivation in the first place. Within the context of the Swampland programme \cite{Vafa:2005ui,Palti:2019pca,vanBeest:2021lhn,Brennan:2017rbf,Grana:2021zvf,Agmon:2022thq,VanRiet:2023pnx} however, light fields seem to be the rule---according to the so-called Swampland Distance Conjecture (SDC) \cite{SdC}, in a theory with a moduli space of possible scalar field expectation values, any asymptotic limit taken in this moduli space is accompanied by an infinite tower of fields whose masses are exponentially suppressed. In other words, as we move towards the asymptotic boundaries, an infinite tower of states becomes light(er). From a top-down String Theory point-of-view, this is an observed pattern; the conjecture is that the same pattern will be true for any EFT compatible with quantum gravity. 
If anything, the Distance Conjecture provided a clear motivation for the presence of light fields.  

Many works have explored and developed this connection between the ``species scale'' and the infinite towers of states that arise in String Theory over the past few years 
\cite{Castellano:2023jjt,Castellano:2023stg,Etheredge:2024tok,Rudelius:2023spc,Castellano:2024bna,Castellano:2022bvr,vandeHeisteeg:2022btw,Cribiori:2022nke,vandeHeisteeg:2023ubh,vandeHeisteeg:2023dlw,Castellano:2023aum,Cribiori:2023ffn,Basile:2023blg,Basile:2024dqq,Herraez:2024kux,Martucci:2024trp,Seo:2024zzs,Bedroya:2024uva,Fichet:2022ixi,Ashoorioon:2011aa,Andriot:2023isc,Cribiori:2023sch,Scalisi:2024jhq,Casas:2024jbw,Aoki:2024ixq,Castellano:2023jjt,Castellano:2023stg,Etheredge:2024tok,Rudelius:2023spc,Blumenhagen:2024ydy,Blumenhagen:2023xmk,Blumenhagen:2023tev,Blumenhagen:2023yws}---comparing bottom-up counting predictions with the knowledge of the UV in several concrete examples; looking at the consequences for scalar field potentials and cosmology; and in connection with the Emergence Proposal\footnote{According to the Emergence Proposal \cite{Harlow:2015lma,Grimm:2018ohb,Heidenreich:2018kpg,Palti:2019pca,Blumenhagen:2023tev}, the dynamics (kinetic terms) for all low-energy fields emerges from integrating out towers of massive states down from an ultraviolet scale, below the Planck scale, where quantum gravity takes over. The need to identify this ultraviolet scale accurately gives a theoretical motivation to understand the ``species scale''.}. We highly recommend \cite{Castellano:2024bna} for an in-depth discussion of the ``species scale'' and infinite towers in the context of String Theory. 

However, one may worry that estimates involving infinite towers of states within an EFT can never be controlled procedures. The scale separation that is required to properly define an EFT with cutoff $\cut{EFT}$ appears to be in tension with the idea of cutting an infinite tower partway so that $N$ fields remain in the theory, while all others are integrated out \cite{Burgess:2023pnk}\footnote{Note that this view is not entirely consensual. A series of works \cite{Branchina:2023rgi,Branchina:2023ogv,Branchina:2024ljd} argues that such a separation of scales is not required in the particular case of a Kaluza-Klein tower, since the Kaluza-Klein mass corresponds in reality to the components of the momentum along the compact directions.}. Indeed such a cut would require $m_N\ll\cut{EFT}<m_{N+1}$, which is not compatible with a regular tower of states (e.g. Kaluza-Klein or string-oscillator towers). Yet there appears to be a plethora of examples in String Theory where $\cut{sp}$ does in fact give the correct scale at which an effective gravitational theory breaks down \cite{Castellano:2023jjt,Castellano:2024bna}. 

Our goal here is to clarify the relation between these two observations. We shall argue below that there is no contradiction as long as one is referring to the correct EFT, dealing with the correct scales and considering convergent towers\footnote{We clarify what we mean by convergence in Section \ref{sec:towers-convergence}.}. Firstly, to run the perturbative argument for the ``species scale'' one must consider the EFT that includes the infinite tower and not one with the massive tower integrated out---integrating it out would indeed generate an EFT expansion controlled by the ratio $E/m_{\rm tower}$ that would break down at energies $E\sim m_{\rm tower}$. We will also argue that keeping the infinite tower in the computation remains consistent, even if ultimately the theory should not be valid above (at least) the scale $\cut{sp}$---since the tower states will only appear inside the loops, we are never actually considering processes with energies above $\cut{sp}$ and the result should therefore not be in tension with the effective description used to compute it. 

We intend to clarify what is meant by ``light'' in deriving the ``species scale'' and study how ``heavy'' states of the same tower contribute to this scale, thereby demystifying the \textit{apparent} cut of the infinite tower and justifying the procedure even from an EFT standpoint.  
We also compare $\cut{sp}$ with a true EFT cutoff $\cut{EFT}$, emphasizing the idea that $\cut{sp}$ should serve as an upper bound on the regime of validity of the effective description, rather than giving the EFT cutoff itself. 
In order to make the discussion concrete, we compute the 1-loop graviton propagator for a particular infinite tower of states, for which the different scales can be computed and contrasted explicitly.

Nevertheless, by computing several corrections in a controlled string-theoretic setup one finds many examples where the ``species scale'' does control higher-curvature corrections\footnote{See \cite{Calderon-Infante:2025ldq} for a more detailed characterisation of the structure of these higher-derivative corrections that distinguishes between field-theoretic and quantum gravitational contributions, each associated with their own expansions and suppression scales.} \cite{Castellano:2023aum,vandeHeisteeg:2023dlw}. This top-down \textit{observation} suggests a strong connection between the two scales, but it requires the knowledge of a specific UV completion. Identifying this scale from a bottom-up perspective instead requires assumptions about the UV theory and may provide a hint towards the fundamental principles that underlie the top-down observations \cite{Caron-Huot:2021rmr,Caron-Huot:2022ugt,Caron-Huot:2024lbf,Caron-Huot:2024tsk,Alberte:2020jsk}. While we want to emphasise that the strong coupling scale is not guaranteed to be the scale suppressing higher-curvature corrections, this is mostly of no consequence for these top-down studies. 
In contrast, it is rather crucial if we wish to use bottom-up arguments to estimate the EFT scale.

In fact, recent work has discussed how three different manifestations of the ``species scale'' in string theory settings relate to each other \cite{Calderon-Infante:2023uhz}: through the presence of towers of exponentially light states that lower the strong-coupling scale (the case of interest in this work); through the scale that suppresses higher-derivative corrections that arise from integrating such towers; and as the size of the smallest possible black hole described by the EFT due to higher derivative terms that modify the solution. 
Note that only the first of these proposals in relevant for us. Our results will have no impact on the alternative definitions of this scale.

The paper is structured as follows. In Section \ref{sc:EFT-background} we review some of the necessary background on EFTs, using well-known examples such as QED and the Fermi theory to work our way towards a gravitational EFT in Section \ref{sc:EFT-gravity}. In Section \ref{sec:1-loop-computation} we compute in detail the 1-loop correction to the graviton propagator due to a massive scalar field coupled to gravity, leading in Section \ref{sec:renormalisation} to the determination of the strong-coupling scale in the presence of the scalar. In Section \ref{sec:original-species-scale}, we apply these results to a theory with $N$ identical scalar fields and relate them to the original proposal for the ``species scale''. We analyse different limits for the mass of the scalars that will be useful when considering infinite towers of states, recovering the well-known behaviour of $\cut{sp}$ in the small mass limit. In Section \ref{sec:towers} we finally consider infinite towers of scalar fields. After briefly discussing the convergence of the results for different types of towers in Section \ref{sec:towers-convergence}, in Section \ref{sec:towers-summing} we explicitly analyse the contribution of different states in the tower.
Finally, in Section \ref{sec:conclusion} we briefly summarise the key points of our analysis and contextualise it within the existing literature. We also include in Appendix \ref{ap:details-1loop-computation} some details of the 1-loop computation. 
\section{Effective Field Theories: from QED to gravity}
\label{sc:EFT-background}

Many phenomena of gravitational EFTs are commonly attributed to the idiosyncrasies of gravity. Although some are due to the properties of gravity, we can find many in more familiar effective theories. In this section we will introduce several relevant concepts related to Effective Field Theories and perturbation theory, making use of theories where these ideas are particularly clear. After giving a brief overview of the EFT framework itself, we consider QED in turn as a UV theory and as a low-energy EFT, introducing and contrasting concepts such as the strong coupling scale and the EFT cutoff. We also use them to introduce the idea of an EFT with many ``light'' fields that will be key for the notion of species scale. It will also be useful to consider a theory of photons and neutrinos, where the particular interactions of neutrinos result in naively unexpected Wilson coefficients in the low-energy EFT. This is an illustration of how effective interactions depend heavily on the UV theory, in ways that might seem to go against EFT intuition. Finally, having all these concepts and examples at hand, we end the section by giving a brief overview of gravitational EFTs, as well as the setup that will be the basis of the remaining sections.

\subsection{Effective Field Theories}

Nowadays the average theoretical physicist's toolbox includes the powerful machinery of Effective Field Theory (EFT) \cite{Burgess:2020tbq}, which is understood to various levels of depth to mean the following: the low-energy physics of a system may often be described in terms of a theory (the low-energy EFT) that is somewhat agnostic as to what lies at much higher energy scales. This notion of decoupling between IR and UV physics turns out to be extremely useful, and successful theories like the Standard Model (SM) of particle physics and General Relativity (GR) have been progressively reframed in the context of EFTs. 

From the point of view of an EFT, the low-energy physics is described by a finite set of renormalisable interactions that dominate in the IR together with an infinite set of non-renormalisable interactions suppressed by the ratio $E/\cut{UV}$, for some high-energy scale $\cut{UV}$,\footnote{We consider the case $d=4$, for concreteness, but this discussion could be generalised to higher dimensions.}
\begin{equation}
    \mathcal{L}_{\rm EFT} = \mathcal{L}_{\rm ren} + \sum_{n=5}^\infty \frac{\mathcal{O}^{(n)}}{\cut{UV}^{n-4}} \,.
    \label{eq:EFT}
\end{equation}
At energies $E\sim\cut{UV}$ contributions from this infinite series of terms can no longer be treated perturbatively in powers of $E/\cut{UV}$ and the EFT no longer gives a good description of the physics at these scales.
One should include in the EFT all terms (renormalisable and non-renormalisable) that can be built out of the field content and derivatives of these fields, while respecting the symmetries one wants to preserve \cite{Burgess:2020tbq}.

It is clear that the scale $\cut{UV}$ is a key feature of an EFT, telling us the range of validity of this low-energy description---but what exactly is $\cut{UV}$? The answer will depend crucially on the amount of information one has either from the UV theory or the IR physics, in the absence of which this scale remains unknown. In other words, one could fix it using either a bottom-up or a top-down approach, which in practice correspond to an experimental and theoretical determination, respectively.  

From a bottom-up perspective, we can use the EFT to compute some observable (e.g. the amplitude of a specific scattering process) that depends on $\cut{UV}$ and then measure it in a laboratory. Comparing the measurement with the EFT prediction, one can \textit{deduce} the value of $\cut{UV}$---this is, for example, how the scale related to the Fermi theory of weak interactions \cite{Fermi:1934hr} was determined to be $\Lambda_{\rm Fermi}\sim 360$ GeV. Alternatively, from a top-down perspective, we could \textit{derive} the form of these non-renormalisable interactions from a known UV theory and express the cutoff $\cut{UV}$ in terms of parameters of this parent theory---doing this we learn that $\cut{Fermi}$ is related to the electroweak coupling $g_w$ and the $W$-boson mass $M_W$ as
\begin{equation}
    \Lambda_{\rm Fermi} \sim \frac{2\sqrt{2}\cdot M_W}{g_w} \,.
\end{equation}
Of course, an actual number would still require some measurement, now of the values of $g_w\sim 0.63$ and $M_W\sim 80$ GeV in the SM \cite{navas2024review}. However, it gives new information because it tells us that this interaction is due to the exchange of a massive state\footnote{More precisely, one should also include $Z$-boson exchanges, whose mass is similar to the $W$-boson.} (the $W$-boson) and that new physics appears \textit{not exactly} at the inferred scale of $360$ GeV (where the EFT description would certainly break down), but at a somewhat lower scale of $M_W\sim 80$ GeV. We understand the discrepancy between the two scales in terms of the weak coupling $g_w$ with which the $W$-boson will couple to fermions. 

But the appearance of the coupling $g_w$ is not merely a precision factor that corrects the UV scale and it is not a peculiarity of the Fermi theory or the weak interactions. It hints at an even more important point of EFTs: the low-energy theory and, in particular, the suppression of the non-renormalisable interactions is dictated by the way different particles interact in the UV theory \cite{Burgess:2020tbq}.  

\vskip 1em
\subsection{QED as the UV theory} 
\label{sec:QED}

Let us illustrate some key points using the simple case of QED and a low-energy EFT describing photons at energies $E\ll m_e$. In this example, the UV theory is given by the QED Lagrangian,
\begin{equation}
    \mathcal{L}_{\rm QED} = -\frac{1}{4g^2_e} F_{\mu\nu}F^{\mu\nu} - \Bar{\psi}(\slashed{\partial}+m_e)\psi - \Bar{\psi} \slashed{A}\psi \,, 
    \label{eq:LQED}
\end{equation}
where $F_{\mu\nu} = \partial_\mu A_\nu - \partial_\nu A_\mu$ and $\slashed{X} = \gamma^\mu X_\mu$, while the photon EFT is \cite{Burgess:2020tbq}
\begin{equation}
    \mathcal{L}_{\rm EFT}^{\rm (photon)} = -\frac{1}{4g_{\rm eff}^2} F_{\mu\nu}F^{\mu\nu} + \frac{1}{(\sqrt{12\pi}\cdot m_e)^4}\left[\frac{1}{10}(F_{\mu\nu}F^{\mu\nu})^2 + \frac{7}{40}(F_{\mu\nu}\tilde{F}^{\mu\nu})^2 \right] \,.
    \label{eq:Lphotons}
\end{equation}
Since the photon kinetic term belongs to the renormalisable part of the EFT, it already appears in the UV theory---the effect of integrating out electrons will be equivalent to the usual quantum corrections due to electron loops in QED, apart from the fact that we are now mainly interested in energies $E\ll m_e$. Using dimensional regularisation, the one-loop vacuum polarisation diagram contributing to this term gives\footnote{Note that due to the non canonical normalisation we chose, the vertices do not carry any powers of the coupling. Nevertheless the tree-level propagator will be enhanced by a factor fo $g_e^{-2}$ and therefore this 1-loop result is relatively suppressed.}
\begin{equation}
    \Pi(p^2)_{\rm 1-loop} = \frac{1}{2\pi^2}\int_0^1 dx ~x(1-x) \left\{
        \frac{2}{\varepsilon} + \gamma + \log\qty(\frac{m_e^2 + p^2 x(1-x)}{4\pi\mu^2}) + O(\varepsilon)
    \right\} \,,
    \label{eq:QED-1loop-result}
\end{equation}
where $\varepsilon = d - 4$, $\mu$ is an arbitrary scale introduced for dimensional consistency and $\gamma$ the Euler-Mascheroni constant. As we should expect, there is a divergence as $\varepsilon\to 0$, which tells us that we must renormalise the theory by introducing the appropriate counterterm with parameter $\delta Z$, so that the renormalised (finite) result becomes
\begin{equation}
    \Pi(p^2)_{\rm ren} = \Pi(p^2)_{\rm 1-loop} - \delta Z \,.
    \label{eq:QED-renormalised-general}
\end{equation}
Because we want it to contribute at tree-level, the counterterm in question should be of the form $\delta Z \cdot F_{\mu\nu}F^{\mu\nu}$, so that in practice adding it corresponds to shifting the coefficient of the photon kinetic term
\begin{equation}
    -\frac{1}{4g_e^2}F_{\mu\nu}F^{\mu\nu} \rightarrow -\qty(\frac{1}{4g_e^2}+\delta Z)F_{\mu\nu}F^{\mu\nu} \,.
\end{equation}
But how do we fix $\delta Z$? Note that the total contribution to the 1PI 2-point function is
\begin{equation}
    p^2\cdot \left[-\frac{1}{g_e^2} + \Pi(p^2)_{\rm ren}\right] \,.
    \label{eq:1PI-QED-p}
\end{equation}
Let's say we \textit{know} what the result should be at some scale $p^2 = p_\star^2$, 
\begin{equation}
    p_\star^2\cdot \left[-\frac{1}{g_e^2} + \Pi(p_\star^2)_{\rm ren}\right] = \text{measured value}\,,
\end{equation}
where 
\begin{equation}
    \Pi(p_\star^2)_{\rm ren} =  \frac{1}{2\pi^2}\int_0^1 dx ~x(1-x) \left\{
        \frac{2}{\varepsilon} + \gamma + \log\qty(\frac{m_e^2 + p_\star^2 ~ x(1-x)}{4\pi\mu^2})
    \right\} - \delta Z \,.
\end{equation}
We can thus express $\delta Z$ in terms of this quantity
\begin{equation}
    \delta Z = - \Pi(p_\star^2)_{\rm ren} + \frac{1}{2\pi^2}\int_0^1 dx ~x(1-x) \left\{
        \frac{2}{\varepsilon} + \gamma + \log\qty(\frac{m_e^2 + p_\star^2 ~ x(1-x)}{4\pi\mu^2})
    \right\} \,.
\end{equation}
Plugging it back into the renormalised result \eqref{eq:QED-renormalised-general}, we find
\begin{equation}
    \Pi(p^2)_{\rm ren} = \Pi(p_\star^2)_{\rm ren} +
    \frac{1}{2\pi^2}\int_0^1 dx ~x(1-x) \log\qty(\frac{m_e^2 + p^2 ~ x(1-x)}{m_e^2 + p_\star^2 ~ x(1-x)}) \,.
    \label{eq:QED-renormalised-physical}
\end{equation}
This result is now finite and expressed in terms of some physical scale $p_\star^2$. However, writing it in terms of $\Pi(p_\star^2)_{\rm ren}$ still hides a crucial point---we did \textit{not} really measure $\Pi(p_\star^2)_{\rm ren}$, but instead the combination $[-\frac{1}{g_e^2} + \Pi(p_\star^2)_{\rm ren}]$ and so there is a redundancy between $g_e^2$ and $\Pi(p_\star^2)_{\rm ren}$. Although we have been implicitly assuming that $g_e^2$ is fixed and independent of $p_\star^2$, if $g_e^2 = g_e^2(p_\star^2)$ we can choose $\Pi(p_\star^2)_{\rm ren} = 0$ and let the coupling carry the information about our measurement, in which case \eqref{eq:QED-renormalised-physical} becomes
\begin{equation}
    \Pi(p^2)_{\rm ren} =
    \frac{1}{2\pi^2}\int_0^1 dx ~x(1-x) \log\qty(\frac{m_e^2 + p^2 ~ x(1-x)}{m_e^2 + p_\star^2 ~ x(1-x)}) \,,
    \label{eq:QED-renormalised-physical-fixed}
\end{equation}
in terms of which \eqref{eq:1PI-QED-p} is
\begin{equation}
    p^2\cdot \left[-\frac{1}{g_e^2(p_\star^2)} + \Pi(p^2)_{\rm ren}\right] \,.
    \label{eq:1PI-QED-p-running-g}
\end{equation}
Note that we could have chosen any scale $p_\star^2$ (as long as we had a way to determine what $\Pi(p_\star^2)_{\rm ren}$ should be) and the outcome could not have depended on our choice of reference. Therefore, $p_\star^2$ is just as arbitrary as the parameter $\mu$, although it now has the clear physical meaning of providing the scale at which we will \textit{match} the result with something we \textit{know}. 
It also becomes clear that the quality of our perturbative expansion will not depend only on how small the coupling $g_e$ is, but also on how small the $\log$ remains; and it is clear from \eqref{eq:QED-renormalised-physical-fixed} that it will only remain small as long as $p^2$ is close enough to $p_\star^2$. 
Moreover, the fact that the physics should be independent of our choice of $p_\star^2$ while $g_e(p_\star^2)$ depends on this scale, leads to the following equation for this coupling's energy dependence,
\begin{equation}
    \frac{\ed}{\ed p_\star^2}\qty(p^2\cdot \left[-\frac{1}{g_e^2(p_\star^2)} + \Pi(p^2)_{\rm ren}\right]) = 0 \,,
    \label{eq:QED-running-condition}
\end{equation}
which yields
\begin{equation}
     p_\star^2~ \frac{\ed g_e^2}{\ed p_\star^2} = + \frac{g_e^4}{2\pi^2}\int_0^1 dx ~ \frac{p_\star^2 ~ x^2(1-x)^2}{m_e^2 + p_\star^2 ~x(1-x)} \,.
     \label{eq:QED-general-running}
\end{equation}
This equation allows us to track continuously the evolution of $g_e(p_\star^2)$ and 
quantifies exactly \textit{how} the coupling depends on the reference scale $p_\star^2$. Recalling that we are interested in the low-energy EFT obtained from integrating out the electron, we take the limit $p_\star^2\ll m_e^2$, 
\begin{equation}
    p_\star^2 ~ \frac{\ed g_{e}^2}{\ed p_\star^2} \approx + \frac{g_e^4}{60\pi^2}\cdot \frac{p_\star^2}{m_e^2} \,,
\end{equation}
where we can see that already the leading term is suppressed in powers of $p_\star^2/m_e^2 \ll 1$. This is to be expected, since it signals the decoupling of the electron at energies $p_\star^2 \ll m_e^2$. Therefore, at these energies the coupling is approximately constant, which fixes the effective coupling in the low-energy EFT to be 
\begin{equation}
    g_{\rm eff} = g_e(p_\star^2=m_e^2) + O\qty(\frac{p_\star^2}{m_e^2}) \,,
\end{equation}
essentially the QED coupling evaluated at $m_e^2$, the mass of the lightest particle to be integrated out---it gives the familiar value $\frac{g_{\rm eff}^2}{4\pi}\approx \frac{1}{137}$.

This covers the renormalisable part of the EFT action, which simply gives the photon kinetic term. Interactions between photons within the EFT are captured by the non-renormalisable set of terms generated by integrating out the electron and for those the effective coupling can be made explicit by canonically normalising the photon,
\begin{equation}
    \mathcal{L}_{\rm EFT}^{\rm (photon)} = -\frac{1}{4} F_{\mu\nu}F^{\mu\nu} + \frac{1}{\cut{QED}^4}\left[\frac{1}{10}(F_{\mu\nu}F^{\mu\nu})^2 + \frac{7}{40}(F_{\mu\nu}\tilde{F}^{\mu\nu})^2 \right] \,,
    \label{eq:Lphotons-canonical}
\end{equation}
in terms of the scale
\begin{equation}
    \cut{QED} = \frac{\sqrt{12\pi}\cdot m_e}{g_{\rm eff}} \,.
\end{equation}
The first thing to point out is that, just like for the Fermi theory, the scale that suppresses these terms is not directly the mass of the integrated-out particle. Rather than simply corresponding to the electron mass, $m_e$, this scale is enhanced by $1/g_{\rm eff}$, and so will be somewhat bigger. Again this is reflecting the weak coupling of the UV theory, i.e. the fact that electrons and photons interact weakly at the electron mass scale. On the other hand, the scale $\cut{QED}\sim 10$ MeV is what tells us about the validity of the EFT---the actual coupling for these interactions will be $p^4/\cut{QED}^4$ and so treating them perturbatively will no longer be allowed once we reach an energy scale $E\sim\cut{QED}$. Because the EFT intrinsically encodes the effect of integrated-out electrons as a perturbative expansion of induced interactions, this description simply breaks down at this scale. Fixing it requires us to go back to the UV theory that contains the electrons mediating these photon interactions. 

It is important to distinguish between this break-down scale $\cut{QED}$ and a ``strong coupling'' scale corresponding to the physical scale at which photons (and electrons) interact strongly. 
At $E\sim\cut{QED}$ the QED coupling is still small. We can deduce at which scale the interactions become strong by taking the opposite limit $p_\star^2\gg m_e^2$ in \eqref{eq:QED-general-running}, 
\begin{equation}
     p_\star^2~\frac{\ed g_{e}^2}{\ed p_\star^2} \approx + \frac{g_e^4}{12\pi^2} \,.
\end{equation}
Choosing $g_{\rm eff}$ as a reference scale (or initial condition), we can solve the equation and track the evolution of $g_e$ as
\begin{equation}
    g_e^2(p_\star^2) = \frac{g_{\rm eff}^2}{1 - \frac{g_{\rm eff}^2}{12\pi^2}\log\frac{p_\star^2}{m_e^2}} \,.
    \label{eq:QED-running-coupling}
\end{equation}
Since the energy scale $p_\star^2$ appears logarithmically, only at very high energies will the QED coupling actually become strong---in fact it would become of $O(1)$ at an energy scale\footnote{Here we focus our attention on the scale at which the coupling becomes of $O(1)$, where QED interactions can no longer be treated perturbatively. It differs from the famous Landau pole, at which an extrapolation of the relation \eqref{eq:QED-general-running} to arbitrarily high-energies would suggest the theory becomes \textit{infinitely} strongly-coupled (i.e. $g_e(p_\star^2)\to \infty$ as $p_\star^2\to \cut{Landau}^2$). This scale will be even higher than $\cut{strong}^{\rm QED}$, at $\cut{Landau}\sim 10^{280}\cdot m_e$. Since all we did was within perturbation theory, the conclusions should be trusted only while $g_e < 1$, and one would be hardly justified in trusting it all the way up to the Landau pole.}
\begin{equation}
    \sqrt{p_\star^2} = \cut{strong}^{\rm QED} \equiv m_e \cdot e^{6\pi^2\cdot\qty(\frac{1 - g_{\rm eff}^2}{g_{\rm eff}^2})}  \sim 10^{254}\cdot m_e   \,.
    \label{eq:QED-strong-coupling-scale}
\end{equation}
This is indeed an absurdly high energy, much higher than our EFT cutoff $\cut{QED}\sim 10\cdot m_e$. 
If we are handed an EFT such as \eqref{eq:Lphotons-canonical}, with no reference to electrons or electron loops, the scale $\cut{strong}^{\rm QED}$ is out of reach. The scale at which loop expansions will break down will be $\cut{QED}$ and what this signals is that the EFT expansion is no longer valid. It is only by bringing back the electron and using QED as the UV completion of our EFT that one can compute the running of the QED coupling and find that the theory becomes strongly coupled at $\cut{strong}^{\rm QED}$.  

As a final comment, note that we can actually solve \eqref{eq:QED-running-condition} exactly. To do this, rather than considering it as a differential equation (which does have its benefits) we can just match two choices of $p_\star$---one generic that we want to solve for, and a known one (e.g. $p_\star=m_e$ or $p_\star=0$). More specifically \eqref{eq:QED-running-condition} implies (emphasising the dependence of $\Pi(p^2)_{\rm ren}$ on $p_\star$)
\begin{align}
    p^2\cdot \left[-\frac{1}{g_e^2(p_\star^2)} + \Pi(p^2,p_\star^2)_{\rm ren}\right]=p^2\cdot \left[-\frac{1}{g_e^2(p_{\star\star}^2)} + \Pi(p^2,p_{\star\star}^2)_{\rm ren}\right]\Longleftrightarrow&\\
    \Longleftrightarrow-\frac{1}{g_e^2(p_\star^2)} - \frac{1}{2\pi^2}\int_0^1 dx ~x(1-x) \log\qty(1 + \frac{p_\star^2}{m_e^2} ~ x(1-x))&=\nonumber\\
    =-\frac{1}{g_e^2(p_{\star\star}^2)}  - \frac{1}{2\pi^2}\int_0^1 dx ~x(1-x) \log\qty(1 + \frac{p_{\star\star}^2}{m_e^2} ~ x(1-x))&
\end{align}
where in going to the second line we cancelled the common $p^2$ dependence on both sides and added $\frac{1}{2\pi^2}\log m_e^2$ on both sides to make the argument of the logarithms dimensionless. 
Choosing $p_{\star\star}^2=0$, which makes the logarithm on the RHS vanish, and using the fact that $g_{\rm eff}^2\equiv g_e^2(p_\star^2=0) = \frac{4\pi}{137}$ we find
\begin{equation}
    -\frac{1}{g_e^2(p_\star^2)} - \frac{1}{2\pi^2}\int_0^1 dx ~x(1-x) \log\qty(1 + \frac{p_\star^2}{m_e^2} ~ x(1-x))=-\frac{1}{g_\text{eff}^2}
    \label{eq:QED-1-loop-correction-exact}
\end{equation}
which we can solve to obtain
\begin{equation}
    g_e^2(p_\star^2) = \frac{g_{\rm eff}^2}{1 - \frac{g_{\rm eff}^2}{2\pi^2}\int_0^1 dx~x(1-x)\log\qty(1 + \frac{p_\star^2}{m_e^2}x(1-x))} \,.
    \label{eq:QED-running-coupling-exact}
\end{equation}
This is the exact solution. This expression gives the same limits we found before, from which we recover the freezing of the coupling at energies below $m_e$ and the running at high-energies \eqref{eq:QED-running-coupling} when $p^2\gg m_e^2$. From this we can then extract both the ``strong coupling'' scale $\cut{strong}^{\rm QED}$ and the Landau pole $\cut{Landau}$.

\subsubsection*{With an eye on the species scales}

Let us make a few comments that will be useful to keep in mind when we discuss our main topic with gravity. First, one may wonder what would happen if we tried to extract the scale at which perturbation theory breaks down by simply comparing the 1-loop result \eqref{eq:QED-renormalised-physical-fixed} with a tree-level contribution. Choosing again the reference scale to be $p_\star^2 = 0$ and focusing our attention on the high-energy limit $p^2\gg m_e^2$,
\begin{align}
    1 &\overset{!}{=} \frac{g_{\rm eff}^2}{2\pi^2}\int_0^1~dx ~x(1-x) \log\qty(1 + \frac{p^2}{m_e^2}~x(1-x)) \nonumber \\
    &\approx \frac{g_{\rm eff}^2}{12\pi^2}\log\qty(\frac{p^2}{m_e^2}) + O\qty(\frac{m_e^2}{p^2}) \,,
\end{align}
we would conclude that perturbation theory breaks down when 
\begin{equation}
    \sqrt{p^2} \approx m_e \cdot e^{\frac{12\pi^2}{g_{\rm eff}^2}} = \cut{Landau} \,,
\end{equation}
i.e. at the Landau pole that appears when we follow the running coupling all the way up to $g_e\to\infty$. Since we know from the running that $\cut{strong}^{\rm QED} \ll \cut{Landau}$, estimating the ``strong coupling'' scale by immediately comparing the tree-level and the 1-loop terms provides a(n exponentially) misleading result. In reality, the theory becomes strongly coupled much earlier than our estimate predicts. In order to accurately estimate this scale, one should find the solution to $g_e^2(p^2)=1$ using \eqref{eq:QED-running-coupling-exact}, which will lead to \eqref{eq:QED-strong-coupling-scale} and the correct ``strong-coupling'' scale. 

A second question we might pose is what would happen if there were $N$ copies of the electron, rather than one. Assuming these are actual copies, with the same mass and couplings to the photon, each of them would contribute the same amount to the photon self-energy and the result \eqref{eq:QED-renormalised-physical-fixed} would simply come multiplied by $N$, 
\begin{equation}
    \Pi(p^2)_{\rm ren} =
    N\cdot \frac{1}{2\pi^2}\int_0^1 dx ~x(1-x) \log\qty(\frac{m_e^2 + p^2 ~ x(1-x)}{m_e^2 + p_\star^2 ~ x(1-x)}) \,.
    \label{eq:QED-renormalised-physical-fixed-N-electrons}
\end{equation}
Estimating the ``strong coupling'' scale for this setup, we now find that the perturbative expansion breaks down when 
\begin{equation}
    \sqrt{p^2} = m_e \cdot e^{\frac{6\pi^2}{N}\cdot\qty(\frac{1 - g_{\rm eff}^2}{g_{\rm eff}^2})}  \,. 
\end{equation}
This tells us that having $N$ copies of the electron would bring down the ``strong coupling'' scale exponentially. Note that the effect of $N$ electrons with coupling constant $g_e$ is equivalent to that of 1 electron with a rescaled coupling
$g_e\to \sqrt{N}\cdot g_e$, which can be seen from \eqref{eq:QED-1-loop-correction-exact}\footnote{One should be careful to perform this rescalling in both couplings appearing in \eqref{eq:QED-1-loop-correction-exact}, $g_e$ and $g_{\rm eff}$.}. This leads to
\begin{equation}
    \frac{\cut{strong}^{\rm QED}}{m_e}\to \qty(\frac{\cut{strong}^{\rm QED}}{m_e})^{\frac1N} \,.
\end{equation}
For example, if there were $N=40$ electrons, $\cut{strong}^{\rm QED}\sim m_e\cdot 10^{6} \sim 1$ TeV, much lower than the absurdly high scale \eqref{eq:QED-strong-coupling-scale}. 

What about the theory at low-energies? Were we to integrate out these $N$ electrons, the coefficients in front of the non-renormalisable interactions generated for the photon at energies $E\ll m_e$ would also come multiplied by $N$. Our photon EFT would then read 
\begin{equation}
    \mathcal{L}_{\rm EFT}^{\rm (photon)} = -\frac{1}{4} F_{\mu\nu}F^{\mu\nu} + \frac{N\cdot g_{\rm eff}^4}{(\sqrt{12\pi}\cdot m_e)^4}\left[\frac{1}{10}(F_{\mu\nu}F^{\mu\nu})^2 + \frac{7}{40}(F_{\mu\nu}\tilde{F}^{\mu\nu})^2 \right] \,,
\end{equation}
for canonically normalised photons, resulting in a cutoff scale 
\begin{equation}
    \cut{QED} = \frac{\sqrt{12\pi}\cdot m_e}{N^{1/4}g_{\rm eff}} \,,
    \label{eq:QED_N1/4}
\end{equation}
suggesting that we should expect our EFT to break down at energies $N^{1/4}$ lower that in the case where 1 electron had been integrated out. We do not find the same $\sqrt{N}\cdot g_{\rm eff}$ dependence as before, since the scaling suggested by the photon self-energy result does not scale all terms in the QED Lagrangian equally. Instead, factors in front of the non-renormalisable interactions depend on the details of the UV physics that generates them.  

\subsection{``QED'' as a low-energy EFT}
\label{sec:QED-EFT}

It is instructive to compare this with the scenario where this $N$-electron QED is itself an EFT at energies $E\ll m_\mu$ that arises from integrating out the muon,
\begin{align}
    \mathcal{L}_{\rm QED} &= -\frac{1}{4g^2_e} F_{\mu\nu}F^{\mu\nu} 
    - N\cdot \Bar{\psi}(\slashed{\partial}+m_e)\psi - N\cdot\Bar{\psi} \slashed{A}\psi  \nonumber \\
    &~~~ + \frac{1}{(\sqrt{12\pi}\cdot m_\mu)^4}\left[\frac{1}{10}(F_{\mu\nu}F^{\mu\nu})^2 + \frac{7}{40}(F_{\mu\nu}\tilde{F}^{\mu\nu})^2 \right] \nonumber \\ 
    &
    + \frac{1}{(\sqrt{240\pi} \cdot m_\mu)^2} F_{\mu\nu}\Box F^{\mu\nu} 
     \,.
\end{align} 
There are two key differences in this setup. One is that the coupling $g_e$ is not effectively frozen anymore, since we still have electrons that are allowed to run in the loop, and the coupling will run at the (logarithmic) rate we have computed above. Hence we expect that our EFT should break down at a scale that depends on this running coupling after we canonically normalise the photon,
\begin{equation}
    \frac{p^4}{(\sqrt{12\pi}\cdot m_\mu)^4}\cdot g_e^4(p^2) \overset{!}{=} 1 \,.
\end{equation}
Since this should happen at $p^2\gg m_e$, we have 
\begin{align}
    \cut{muon} \approx \frac{\sqrt{12\pi}\cdot m_\mu}{g_{\rm eff}}\qty(1 - N\cdot \frac{g_{\rm eff}^2}{12\pi^2}\log\qty(\frac{\cut{muon}^2}{m_e^2})) \approx \frac{\sqrt{12\pi}\cdot m_\mu}{g_{\rm eff}} \,.
    \label{eq:muon_cut}
\end{align}
While it is true that the running coupling will \textit{affect} the scale at which the EFT breaks down, it will not by itself \textit{determine} this scale. In fact, when this effect is strong enough to be important, we reach scales at which the full theory becomes strongly coupled rather than one at which the low-energy description simply breaks down (i.e. the UV completion that includes the muon is itself becoming strongly coupled). The extremely slow logarithmic running allows the corrections to remain small even though the ratio of scales is large. 

This example is instructive because there is a clear distinction between $\cut{strong}$ and $\cut{EFT}$, which can even be hierarchically well separated. Furthermore, we see explicitly that computing $\cut{strong}$ within the $N$-electron EFT does not determine $\cut{EFT}$, although it provides a scale beyond which the EFT would not be perturbatively well-defined \textit{regardless} of $\cut{EFT}$.

The second key difference is that there is a new non-renormalisable term that did not exist in our theory of photons. This is a redundant operator (quadratic in $A_\mu$) which is better expressed in terms of the current sourcing the photons. When there is no source left (as in the photon EFT), this term vanishes. If the electrons remain in the EFT, they contribute to the current and therefore this operator is effectively a contact 4-electron interaction \cite{Burgess:2020tbq},
\begin{equation}
    \frac{g_e^4}{(\sqrt{120\pi^2}\cdot m_\mu)^2}(\Bar{\psi}\gamma^\mu\psi)(\Bar{\psi}\gamma_\mu\psi) \,.
\end{equation}
This term is not only less suppressed than the other two, but also appears suppressed by a formally different scale,
\begin{align}
    \cut{muon}' \approx \frac{\sqrt{120\pi}\cdot m_\mu}{g_e^2} > \cut{muon} \,.
\end{align}
The key point is that the precise dependence of $\cut{EFT}$ on the coupling $g_e$ depends on the details of the UV physics generating these interactions. Although it is still true that the exact scale that  suppresses each non-renormalisable operator will be bigger than the mass of the particle whose integrating out generates the interaction, the explicit dependence on the coupling depends on the UV process that generates it. 

This effective theory is the closest analogue to the case we study in the main body of the paper, where the graviton plays the role of the photon, the scalars play the role of the electrons and the higher-curvature operators correspond to the photon effective couplings suppressed by $\cut{\rm muon}$. The species scale we will introduce for gravity will correspond to $\cut{strong}$ in this theory of photons and $N$ electrons, while the scale suppressing higher-curvature corrections to GR, which constitutes $\cut{EFT}$ in the gravitational EFT, will correspond to $\cut{muon}$. One may refer back to this example later on, when the distinction between these two scales becomes less obvious. 

\vskip 1em
\subsection{A theory of photons and neutrinos}

When the UV theory is as simple as QED, with only one heavy state to be integrated out and one light state to remain in the EFT, tracking scales is relatively simple. Most often, however, the UV theory is richer, with more states to integrate out and more states in the low-energy EFT. 

For example, consider the EFT of photons and neutrinos that arises after integrating out all other massive states in the SM. What will be the scale suppressing non-renormalisable neutrino interactions? One may naively say that $\cut{UV}$ should be close to the electron mass, $m_e$, since the electron is the lightest particle to be integrated out and does indeed interact with neutrinos. Unfortunately, one would be wrong. All neutrino interactions would in fact be suppressed by at least one power of $1/M_W^2$, which gives a much bigger suppression. This is because in the Standard Model, neutrinos only have renormalisable couplings involving $W$ (and $Z$) bosons and thus all effective neutrino interactions will be suppressed by $1/M_W^2$ once the massive bosons are integrated out. Of course, one will then integrate out lighter particles (like quarks and electrons) and the electron will indeed provide the lowest suppression scale out of these states---what we will find then are leading neutrino interactions at low-energies that are at least suppressed by some power of $M_W^2$ \cite{Burgess:2020tbq}. 

The simplest way to understand this is by considering what happens when we integrate out each of the relevant heavy states. Let us write schematically only the couplings of interest for the argument, involving the light photons and neutrinos $(A^\mu,\nu)$, and the heavy electrons and W-bosons $(\psi,W)$,
\begin{align}
    \mathcal{L}_{\rm SM} \supset - g_e\cdot\Bar{\psi}\slashed{A}\psi - g_e^2 (A_\mu W^\mu)^2 + i g_w \cdot \Bar{\nu}~\slashed{W}~\psi \,.    
\end{align}
We will take this as a simple toy model. For the photons, we already know what happens. All particles that couple to it like the electron will generate interactions analogous to the ones in \eqref{eq:Lphotons}, which will be suppressed by a scale proportional to their mass. Since the electron is the lightest out of these, it will provide the leading order terms as expected. For neutrinos the situation is different because the neutrino never interacts with the electron alone, without also interacting with a $W$-boson. One could make this explicit by taking an intermediate step and integrating out the $W$-boson,
\begin{align}
    \mathcal{L}_{\rm Fermi} \supset - \Bar{\psi}\slashed{A}\psi + \frac{c_w}{M_W^4}(F_{\mu\nu}F^{\mu\nu})^2 + \frac{1}{\cut{Fermi}^2} ~  (\Bar{\nu}\gamma^\mu \psi)(\Bar{\psi}\gamma_\mu \nu) \,.    
\end{align}
The last term describing interactions between electrons and neutrinos is nothing but the kind of interactions described by the Fermi theory and will be suppressed by $\cut{Fermi}\sim\frac{M_W}{g_w}$. 
The photon self-coupling term will be suppressed by $M_W^4$, but by now we already know \textit{not to expect} the coefficient $c_w$ to be $O(1)$. In fact, since the photon couples to the $W$-boson through a coupling $g_e$, we should expect $c_w \sim g_e^4$. Although the two terms arise from integrating out the same massive particle, this gives a scale slightly higher that $\cut{Fermi}$ since $g_e < g_w$. 

Now we take the next step and integrate out the electron to find the EFT of photons and neutrinos,
\begin{align}
    \mathcal{L}_{\rm EFT}^{\rm (photon,\nu)} \supset - \Bar{\psi}\slashed{A}\psi - \qty(\frac{g_e^4}{m_e^4} + \frac{g_e^4}{M_W^4})(F_{\mu\nu}F^{\mu\nu})^2 + \frac{c_\nu}{m_e^6} ~  (\Bar{\nu}~\nu)^2 \,.    
    \label{eq:EFT-photons-neutrinos}
\end{align}
The photon interaction terms show clearly that the leading term is suppressed by $m_e$, so these are indeed dominated by the lightest particle to be integrated out. But once again we should ask what the coefficient $c_\nu$ should be---from the lessons above, rather than expecting it to be $O(1)$, it should depend on the coupling in the UV theory, which in this case is a dimensionful coupling. What one finds is that $c_\nu \sim (g_w\cdot m_e/M_W)^4 \sim 10^{-22}$, which is reflecting the fact that the suppression by $\cut{Fermi}$ trickles down from EFT to EFT as we integrate out more particles.  

The key lesson here is that, if our effective coupling arises from interactions that are themselves non-renormalisable, then the naive expectation that the integrated out particle should give the suppression of the generated coupling might fail. In fact, in such cases we should instead expect the original coupling suppression to appear in the generated coupling, in the way that $M_W^4$ appears in the low-energy neutrino coupling \cite{Burgess:2020tbq}.

\subsection{Effective Field Theories of Gravity}
\label{sc:EFT-gravity}

We have so far kept gravity out of the picture, even though our main focus will be an EFT with gravity. While all the phenomena described above are also present in these theories, they tend to appear in more complicated and intermingled ways. Keeping in mind everything we introduced previously in this section in the context of more familiar theories, we end it by giving a brief overview of gravitational EFTs. 

What should one expect of an EFT describing gravity? Classically, gravity is well described by Einstein's General Relativity,
\begin{equation}
    \mathcal{L}^{\rm (GR)} = \sqrt{-g}\cdot \frac{\Mp^2}{2}~R \,.
    \label{eq:GR-action}
\end{equation}
The Ricci scalar $R$ is quadratic in the metric $g_{\mu\nu}$ and contains 2 derivatives, which looks a lot like a field kinetic term. In order to use perturbation theory, we must rewrite this action in terms of fluctuations around a given background, $g_{\mu\nu} = \Bar{g}_{\mu\nu} + \frac{2}{\Mp}h_{\mu\nu}$. Choosing the background to be Minkowski for simplicity, the $R$ term becomes
\begin{equation}
    \sqrt{-g}\cdot \frac{\Mp^2}{2}~ R = -\frac{1}{2}h^{\mu\nu}\mathcal{E}_{\mu\nu}^{\alpha\beta}h_{\alpha\beta} 
    + \sum_{n=3}^{\infty} c_n\cdot(\partial h)^{2}\frac{h^{n-2}}{\Mp^{n-2}} \,,
    \label{eq:GR-action-expanded}
\end{equation}
where $\mathcal{E}_{\mu\nu}^{\alpha\beta}$ is the Lichnerowicz tensor defined as
\begin{equation}
    \mathcal{E}_{\mu\nu}^{\alpha\beta}h_{\alpha\beta} = \Box h_{\mu\nu} - 2\partial_\alpha\partial_{(\mu}h^\alpha_{\phantom{\alpha}\nu)} + \partial_\mu\partial_\nu h 
        - \eta_{\mu\nu}\qty(\Box h - \partial_\alpha\partial_\beta h^{\alpha\beta}) \,.
\end{equation}
The first term corresponds to the graviton kinetic term. When it comes to pure gravity, this is the only term that is renormalisable. All graviton self-interactions correspond to operators of order bigger than 4 and will therefore be non-renormalisable. 
This means that higher-derivative interactions are not negotiable, they are generated even if one does not include them in the original action. 
Therefore even in its simplest form, at least perturbatively (see however \cite{Weinberg:1980gg,Niedermaier:2006wt,Gomis:1995jp} for a non-perturbative approach), gravity \textit{must} be treated as an EFT that includes all possible non-renormalisable interactions consistent with diffeomorphism invariance \cite{Donoghue_1994,Burgess_2004,donoghue2017epfllecturesgeneralrelativity,Burgess:2020tbq,Donoghue:2022eay}, 
\begin{equation}
    \mathcal{L}_{\rm EFT}^{\rm (gravity)} = -\frac{1}{2}h^{\mu\nu}\mathcal{E}_{\mu\nu}^{\alpha\beta}h_{\alpha\beta} 
    + \sum_{n} \frac{\mathcal{O}^{(n)}}{\cut{UV}^n} \,.
    \label{eq:EFT-gravity}
\end{equation}

Higher-derivative interactions in \eqref{eq:EFT-gravity} correspond to higher-curvature terms that can be added to \eqref{eq:GR-action}, but the scale $\cut{UV}$ need not be $\Mp$---as we saw in our quick review of EFTs, the suppression of these non-renormalisable terms will depend on the details of the UV theory. Since gravity couples to \textit{everything}, we might even expect the opposite of what we found for neutrinos: an EFT scale which is lower than expected rather than higher. Just like integrating out the electron generated photon interactions that dominated over the ones generated by the muon, the lightest states to be integrated out should provide the scale of the dominant interaction terms, unless some symmetry prevents it (e.g. supersymmetry). 

It is worth emphasising that from a bottom-up perspective, not knowing what states (if any) were integrated out to generate higher-derivative interactions, we cannot determine $\cut{UV}$ in \eqref{eq:EFT-gravity}. However, at least in QED, the coupling appears in the interactions, and we even argued that reaching a scale $\cut{strong}$ below $\cut{UV}$ would render the question ill-defined (since we would reach strong coupling before the EFT expansion breaks down and require a new description anyway). Thus one may try to use $\cut{strong}$ instead to bound the validity of our effective description.

This is where the notion of \textit{species scale} may help us extract some information from the EFT itself. ``Species scale'' is just another name for gravitational strong coupling scale, $\cut{strong}^{\rm grav}$, which emphasises the role of the number of different species coupling to gravity in determining strong coupling \cite{Han_2005,Dvali:2007hz,Dvali:2007wp,Arkani-Hamed:2005zuc,Distler:2005hi,Dvali:2010vm}. %
It corresponds to the scale at which any quantum field theory of gravitons (as massless spin-2 fields) becomes strongly coupled---it might still be that a weakly coupled theory exists in terms of different degrees of freedom (e.g. strings in string theory). This might be reminiscent of what happens in QCD below $\cut{QCD}$, in which a theory of quarks will be strongly coupled, but one can describe the physics using a theory of weakly coupled hadrons (quark bound states) (cf. chapter 8 of \cite{Burgess:2020tbq}). In contrast with string theory, this is still a quantum field theory.

In the following we will carefully study the simplest possible system---a scalar field minimally coupled to gravity in 4d,
\begin{align}
    \mathcal{L}_{\rm EFT}^{\rm (gravity)} &= 
    -\frac{1}{2}h^{\mu\nu}\mathcal{E}_{\mu\nu}^{\alpha\beta}h_{\alpha\beta}
    -\frac{1}{2}(\partial_\mu\phi)(\partial^\mu\phi) - \frac{1}{2}m^2\phi^2 \nonumber \\
    &~ +\frac{1}{2\Mp} h_{\mu\nu}T^{\mu\nu} + \frac{1}{4\Mp^2}h_{\mu\nu}h_{\rho\sigma}\mathbb{T}^{\mu\nu\rho\sigma}
    + \sum_{n} \frac{\mathcal{O}^{(n)}}{\cut{UV}^n} \,.
    \label{eq:EFT-GR-scalars}
\end{align}
with $T^{\mu\nu}$ and $\mathbb{T}^{\mu\nu\rho\sigma}$ the scalar energy-momentum tensor and its metric variation. We will discuss in detail the computation of the graviton vacuum polarisation at leading (loop) order, i.e. the 1-loop correction to the graviton propagator due to the presence of the scalar field. This correction is of particular interest precisely because it is universal, in the sense that every graviton exchange receives a correction from the vacuum polarization. As soon as this specific correction becomes too large to be treated as a perturbation, no gravitational interaction can be studied perturbatively. At the scale at which this happens, any perturbative expansion of gravitational interactions necessarily breaks down. 

Carefully going through this computation will allow us to illustrate and clarify several of the concepts and statements made here, and analyse different regimes and the validity of such computations.  
The results will then be generalised in two ways. First, we will consider the effect of $N$ identical scalar fields, which will lead us to recover the original species scale \cite{Dvali:2007hz,Dvali:2007wp}. Finally we will consider a tower of scalar fields as has been done in the context of the Swampland programme \cite{Castellano:2023jjt,Castellano:2023stg,Etheredge:2024tok,Rudelius:2023spc,Castellano:2024bna,Castellano:2022bvr,vandeHeisteeg:2022btw,Cribiori:2022nke,vandeHeisteeg:2023ubh,vandeHeisteeg:2023dlw,Castellano:2023aum,Martucci:2024trp,Seo:2024zzs,Bedroya:2024uva,Fichet:2022ixi,Andriot:2023isc,Cribiori:2023sch,Scalisi:2024jhq,Casas:2024jbw,Aoki:2024ixq,Castellano:2023jjt,Castellano:2023stg,Etheredge:2024tok,Rudelius:2023spc,Blumenhagen:2024ydy,Blumenhagen:2023xmk,Blumenhagen:2023tev,Blumenhagen:2023yws} and discuss in detail the interplay between the infinite tower and the validity of the EFT.

\section{The 1-loop computation}
\label{sec:1-loop-computation}

We begin by examining how a single scalar field contributes to the graviton propagator at 1-loop order. This will serve as the building block for the cases of interest---the contribution to the species scale of $N$ scalars of equal mass and a massive tower of states.

\subsection{Setup and conventions}
\label{sec:setup-and-conventions}

The action describing GR and a scalar field $\phi$ of mass $m$ minimally coupled to gravity is
\begin{equation}
    S = 
    \int \dd[4]{x}\sqrt{-g} \left\{\frac{\Mp^2}{2}R
    -\frac{1}{2}\nabla_\mu\phi\nabla^\mu\phi-\frac{1}{2}m^2\phi^2 \right\} \,.
\end{equation}
There are two equivalent ways to interpret our calculations, from a path integral point-of-view \cite{Burgess:2020tbq}. Although they yield the same answer in the end, it is useful to keep both in mind in intermediate calculations.

\begin{enumerate}
    \item \textbf{The metric is non-dynamical:} from this perspective we only integrate over $\phi$ in the path integral, the metric is an external source and we are merely computing the term in the partition function/effective action which is quadratic in the metric.
    \item \textbf{The metric is dynamical:} from this perspective we are computing a two-point function for the metric and zeroing in on the scalar contribution to that correlator.
\end{enumerate}
In either case we must start by changing variables in the metric, defining
\begin{align}
    g_{\mu\nu}&=\eta_{\mu\nu}+\frac{2}{\Mp} ~h_{\mu\nu} \,,
    && \eta_{\mu\nu} = \text{diag}(-1,+1,+1,+1) \,,
\end{align}
so that our calculations will be valid for a Minkowski background.
We then expand the scalar field action to quadratic order in $h_{\mu\nu}$ (or equivalently, to quadratic order in $1/\Mp$)\footnote{One might worry about expanding in a parameter with non-zero mass dimension. To make it more palatable one can think of this as an expansion in $\frac{q^2}{M_P^2}$ and $\frac{m}{M_P}$.}
\begin{align}
    S_\phi=\int\dd[4]{x}\Bigg\{&-\frac{1}{2}\partial_\mu\phi\partial^\mu\phi-\frac{1}{2}m^2\phi^2+\frac{1}{\Mp}h_{\mu\nu}\qty[\partial^\mu\phi\partial^\nu\phi-\frac{1}{2}\eta^{\mu\nu}\qty(\partial_\rho\phi\partial^\rho\phi+m^2\phi^2)]+\nonumber\\
    &+\frac{1}{\Mp^2}h_{\mu\nu}h_{\rho\sigma}\qty[\frac{1}{2}P^{\mu\nu\rho\sigma}\qty(\partial_\lambda\phi\partial^\lambda\phi+m^2\phi^2)+\eta^{\mu\nu}\partial^\rho\phi\partial^\sigma\phi-2\eta^{\mu\rho}\partial^\nu\phi\partial^\sigma\phi]\Bigg\}+\nonumber\\
    &+O\qty(\frac{1}{\Mp^3}) \,, 
\end{align}
where we have raised and lowered the indices using the background Minkowski metric $\eta_{\mu\nu}$ and define

\begin{equation}
    P^{\mu\nu\rho\sigma}=\frac{1}{2}\qty(\eta^{\mu\rho}\eta^{\nu\sigma}+\eta^{\mu\sigma}\eta^{\nu\rho}-\eta^{\mu\nu}\eta^{\rho\sigma}) \,.
\end{equation}

From this action one can derive the Feynman rules needed to compute the 1-loop correction to the graviton propagator due to the scalar field,

\begin{equation}
    \includegraphics[valign=c,scale=0.3]{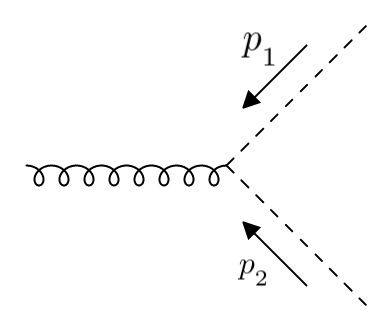}=-i\frac{1}{\Mp}\qty(p_1^\mu p_2^\nu+p_1^\nu p_2^\mu+\eta^{\mu\nu}\qty(-p_1\vdot p_2+m^2))
\end{equation}

\begin{align}
    \includegraphics[valign=c,scale=0.25]{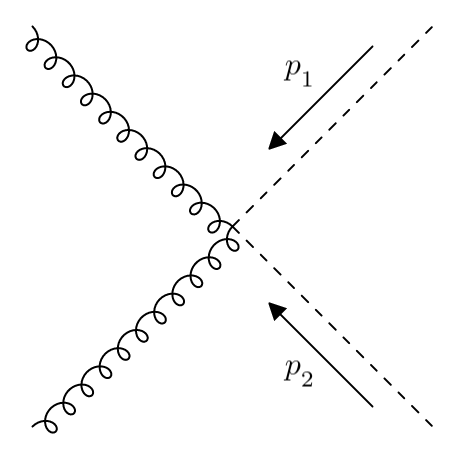}
    =\frac{i}{\Mp^2}\Big[&2P^{\mu\nu\rho\sigma}\qty(-p_1\cdot p_2+m_\phi^2)-\nonumber\\
    &-\eta^{\mu\nu}\qty(p_1^\rho p_2^\sigma+p_1^\sigma p_2^\rho)
    -\eta^{\rho\sigma}\qty(p_1^\mu p_2^\nu+p_1^\nu p_2^\mu)-\nonumber\\
    &+\eta^{\mu\rho}\qty(p_1^\nu p_2^\sigma+p_1^\sigma p_2^\nu)
    +\eta^{\mu\sigma}\qty(p_1^\nu p_2^\rho+p_1^\rho p_2^\nu)+\nonumber\\
    &+\eta^{\nu\rho}\qty(p_1^\mu p_2^\sigma+p_1^\sigma p_2^\mu)
    +\eta^{\nu\sigma}\qty(p_1^\mu p_2^\rho+p_1^\rho p_2^\mu)\Big] \,.
\end{align}

With these ingredients we can now compute the relevant Feynman diagrams and evaluate the necessary integrals.

\subsection{Integrals to compute}
\label{sec:integrals}

\begin{figure}[]
    \centering
    \begin{subfigure}[b]{0.3\textwidth}
         \centering
         \includegraphics[width=\textwidth]{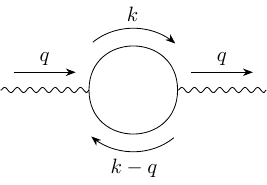}
         \label{fig:d33}
     \end{subfigure}
     \hspace{2em}
     \begin{subfigure}[b]{0.3\textwidth}
         \centering
         \includegraphics[width=\textwidth]{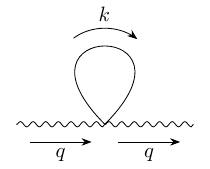}
         \label{fig:d4}
     \end{subfigure}
    \caption{Feynman diagrams contributing to the graviton propagator involving only scalar field loops.}
    \label{fig:feynman-diagrams}
\end{figure}

There are two diagrams contributing at 1-loop (Fig.\ref{fig:feynman-diagrams}). The one with 2 cubic couplings contributes as 
\begin{equation}
    (d_{33})_{\mu\nu\rho\sigma} = \frac{1}{\Mp^2}\int\frac{\dd[4]{k}}{(2\pi)^4}\frac{A_{\mu\nu\rho\sigma}}{(k^2+m^2-i\varepsilon)((k-q)^2+m^2-i\varepsilon)} \,,
\end{equation}
where
\begin{align}
    A_{\mu\nu\rho\sigma} &= \qty(-k_\mu(k_\nu-q_\nu)-(k_\mu-q_\mu)k_\nu+\eta_{\mu\nu}(k\vdot(k-q)+m^2)) \nonumber \\
    &~~~~ \times\qty(-k_\rho(k_\sigma-q_\sigma)-(k_\rho-q_\rho)k_\sigma+\eta_{\rho\sigma}(k\vdot(k-q)+m^2)) \\
    &= \qty(2k_\mu k_\nu-k_\mu q_\nu-q_\mu k_\nu-\eta_{\mu\nu}(k^2-k\vdot q+m^2)) \nonumber\\
    &~~~~ \times\qty(2k_\rho k_\sigma-k_\rho q_\sigma-q_\rho k_\sigma-\eta_{\rho\sigma}(k^2-k\vdot q+m^2)) \,.
\end{align}
The diagram with 1 quartic coupling contributes as 

\begin{equation}
    (d_{4})_{\mu\nu\rho\sigma} = \frac{1}{\Mp^2}\int\frac{\dd[4]{k}}{(2\pi)^4}\frac{B_{\mu\nu\rho\sigma}}{k^2+m^2-i\varepsilon} \,,
\end{equation}
where
\begin{align}
    B_{\mu\nu\rho\sigma} &= P_{\mu\nu\rho\sigma}(k^2+m^2)+\eta_{\mu\nu}k_\rho k_\sigma+\eta_{\rho\sigma}k_\mu k_\nu-2\eta_{\mu(\rho}k_{\sigma)}k_\nu-2\eta_{\nu(\rho}k_{\sigma)}k_\mu \,.
\end{align}
Here we can use rotational invariance to replace $k^\mu k^\nu\to\frac{1}{d}k^2 \eta^{\mu\nu}$ under the integral and obtain (after some algebra)

\begin{align}
    B_{\mu\nu\rho\sigma}(q) &= \left[\qty(1-\frac{4}{d})k^2 + m^2\right] \cdot P_{\mu\nu\rho\sigma} \,.
\end{align}

From here onwards it is a matter of tedious and technical computations to perform the above integrals. We have used dimensional regularisation to maintain gauge invariance and have checked our work by verifying the Ward identity was preserved throughout.

\subsection{The result}
\label{sec:1-loop-result}

It is convenient to write down the result in terms of the following index structures,
\begin{align}
    \Mp^2\cdot\Pi^\text{1-loop}_{\mu\nu\rho\sigma}(q) 
        =~
        & c_1\cdot (\eta_{\mu\sigma}\eta_{\nu\rho} + \eta_{\mu\rho}\eta_{\nu\sigma} ) 
        + c_2\cdot \eta_{\mu\nu}\eta_{\rho\sigma} \nonumber \\
        & + c_3\cdot \qty(\eta_{\mu\rho}\frac{q_\nu q_\sigma}{q^2} + \eta_{\mu\sigma}\frac{q_\nu q_\rho}{q^2} + \eta_{\nu\rho}\frac{q_\mu q_\sigma}{q^2} + \eta_{\nu\sigma}\frac{q_\mu q_\rho}{q^2})  \nonumber \\
        & + c_4\cdot \qty(\eta_{\mu\nu}\frac{q_\rho q_\sigma}{q^2} + \eta_{\rho\sigma}\frac{q_\mu q_\nu}{q^2}) \nonumber \\
        & + c_5\cdot \frac{q_\mu q_\nu q_\rho q_\sigma}{q^4} \,.
    \label{eq:index-structure}
\end{align}
for which the Ward identity can be written as
\begin{subequations}
    \begin{align}
        c_3 &= -c_1 \,, \\
        c_4 &= -c_2 \,, \\
        c_5 &= 2c_1 + c_2 \,.
    \end{align}
    \label{eq:ward-identity-relations}
\end{subequations}
As shown in Appendix \ref{ap:details-1loop-computation}, our 1-loop computation does obey the above relations and we can therefore rewrite the result as 
\begin{align}
    \Mp^2\cdot\Pi^\text{1-loop}_{\mu\nu\rho\sigma}(q) 
        =~ c_1\cdot\qty(L_{\mu\rho}L_{\nu\sigma} + L_{\mu\sigma}L_{\nu\rho} - 2L_{\mu\nu}L_{\rho\sigma}) 
        + c_5 \cdot L_{\mu\nu}L_{\rho\sigma} \,,
        \label{eq:index-structure-L}
\end{align}
where we define
\begin{equation}
    L_{\mu\nu} = g_{\mu\nu} - \frac{q_\mu q_\nu}{q^2} \,.
\end{equation}
Now it is evident that the first term has the exact index structure of the tree-level (gauge invariant) graviton propagator. 

Using dimensional regularisation, to leading order in $\varepsilon=4-d$ we find\footnote{Note that only writing $c_1$ and $c_5$ is sufficient as they uniquely determine the full result through \eqref{eq:ward-identity-relations}.}

\begin{subequations}
\begin{align}
     c_1 =&~ \frac{m^4}{30\pi^2} \qty(1 - \text{F}(\alpha)) + \frac{m^2q^2}{15\pi^2}\qty(\frac{5}{16\varepsilon} + \frac{43}{96} - \frac14 \text{F}(\alpha) - \frac{5}{32}\gamma - \frac{5}{32}\log\qty(\frac{m^2}{4\pi\mu^2})) \nonumber \\
    &+ \frac{q^4}{15\pi^2}\qty(\frac{1}{32\varepsilon} + \frac{23}{480} - \frac{1}{32}\text{F}(\alpha) - \frac{1}{64}\gamma - \frac{1}{64}\log\qty(\frac{m^2}{4\pi\mu^2})) \,, \\
    c_5 =&~ \frac{m^4}{10\pi^2}\qty(1 - \text{F}(\alpha)) + \frac{m^2q^2}{15\pi^2}\qty(-\frac38 + \frac12~\text{F}(\alpha)) \nonumber \\
    & +\frac{q^4}{15\pi^2}\qty(\frac{1}{4\varepsilon} + \frac{47}{240} - \frac14 \text{F}(\alpha) - \frac18\gamma - \frac18\log\qty(\frac{m^2}{4\pi\mu^2})) \,.
\end{align}
\label{eq:1-loop-propagator-c1-c4-result}
\end{subequations}
where we define the variable $\alpha=\frac{4m^2}{q^2}$ and the function
\begin{align}
    \text{F}(x) = \sqrt{1 + x}\cdot\text{arctanh}\qty(\frac{1}{\sqrt{1+x}}) \,,
    \label{eq:F-function}
\end{align}
satisfying the following properties
\begin{align}
    F(x) &= F\qty(\frac{4m^2}{q^2}) = 1 + \frac{q^2}{12m^2} - \frac{q^4}{120m^4} + \mathcal{O}\qty(\frac{q^2}{4m^2})^3 \,, \\
    F(x) &\sim - \frac12\log x \,, \quad \text{as} \quad x\to 0 \,.
\end{align}

\begin{figure}
    \centering
    \includegraphics[width=0.75\linewidth]{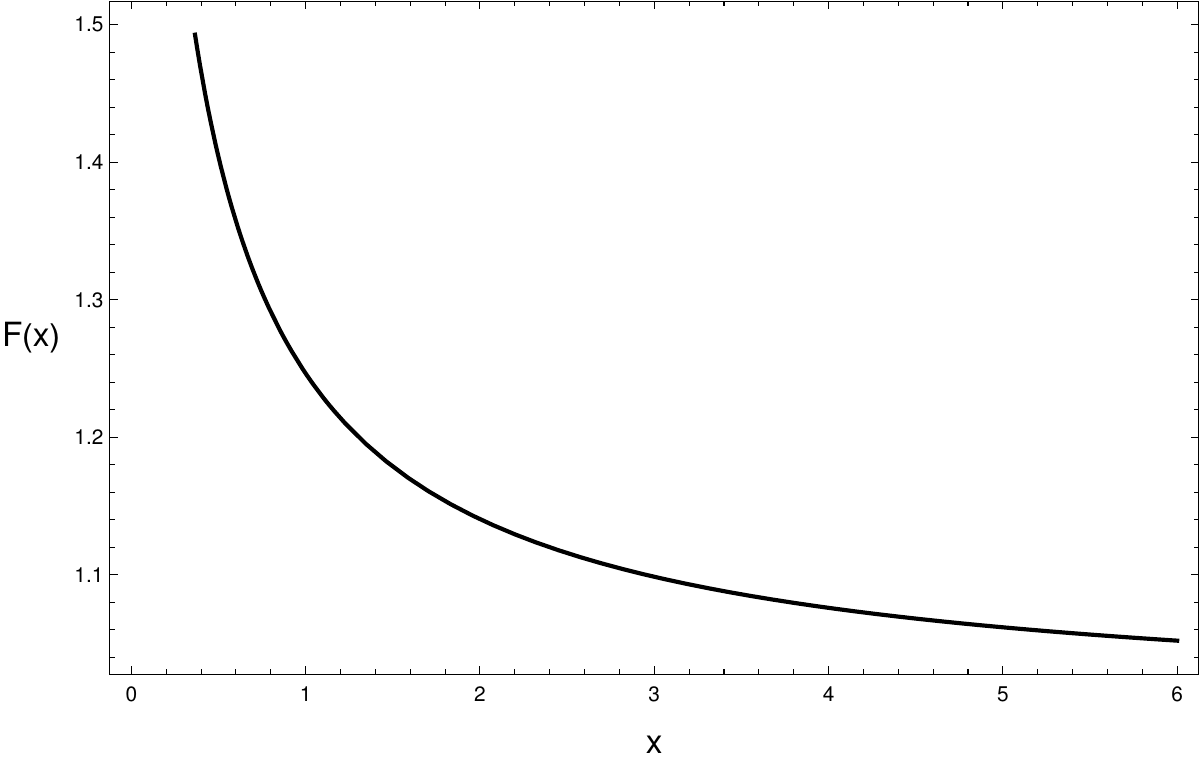}
    \caption{Plot of the function $F(x)$ defined in \eqref{eq:F-function}. The argument of $F(x)$ is always $\alpha = \frac{4m^2}{q^2}$, so that the limit $q^2\to 0$ corresponds to $x\to\infty$ and $q^2=m^2$ corresponds to $x = 4$.}
    \label{fig:F-function}
\end{figure}

We immediately identify in \eqref{eq:1-loop-propagator-c1-c4-result} the divergences as $\varepsilon\to 0$ that need to be cancelled by appropriate counterterms in order to render the result finite and physical. 

\subsection{Counterterms}
\label{sec:counterterms}

\newcommand{\dm}{\delta m_{\phi}^2}
\newcommand{\Zg}{\delta Z_\text{g}}
\newcommand{\Zphi}{\delta Z_{\phi}}
\newcommand{\dL}{\delta\Lambda}
\newcommand{\F}[1]{\text{F}\qty(#1)}

In order to cancel these divergences, one must introduce counterterms in the original action and choose their coefficients appropriately. 
The counterterm action is
\begin{align}
    S_\text{c.t.} = \int d^4x \sqrt{-g}\Big\{& \dL + \frac{\Mp^2}{2}~\Zg\cdot R 
    - \frac12\Zphi\cdot \nabla_\mu\phi\nabla^\mu\phi - \frac12 \dm ~\phi^2 \nonumber \\
        &
    + a_1 R^2 + a_2 R_{\mu\nu}R^{\mu\nu} 
    \Big\} \,.
\end{align}
The first line corresponds to a renormalisation of all the parameters that appear in our original action, while the second line introduces higher-derivative terms that are required\footnote{We have used the Gauss-Bonnet topological invariant $\mathcal{G}=R_{\mu\nu\rho\sigma}R^{\mu\nu\rho\sigma} - 4R_{\mu\nu}R^{\mu\nu} + R^2$ to trade $R_{\mu\nu\rho\sigma}R^{\mu\nu\rho\sigma}$ with a linear combination of $R^2$ and $R_{\mu\nu}R^{\mu\nu}$.} to cancel the divergences at order $q^4$ in graviton momentum. This generation of additional higher-order divergences is a consequence of the fact that GR contains irrelevant interactions, i.e. that it is a non-renormalisable theory \cite{Burgess:2020tbq}.
Since we are only interested in terms quadratic in $h_{\mu\nu}$ with no other fields, the scalar counterterms $(\Zphi,\dm)$ will not be relevant to our discussion and will be discarded. They would of course be crucial were we discussing loop corrections to the scalar field propagator instead.

Expanding the counterterm action to second order in $h^{\mu\nu}$,
\begin{align}
    \label{eq:counter-terms}
    S_\text{c.t.} = \int  &d^4x \sqrt{-g} ~\frac{1}{\Mp^2}~h^{\mu\nu}h^{\rho\sigma} 
    \bigg\{
        -\dL \qty(g_{\mu\rho}g_{\nu\sigma} + g_{\mu\sigma}g_{\nu\rho} - g_{\mu\nu}g_{\rho\sigma})   \nonumber \\
        &+ \Mp^2\cdot\frac{\Zg}{2} \cdot q^2\Big\{
             - \qty(g_{\mu\rho}g_{\nu\sigma} + g_{\mu\sigma}g_{\nu\rho}) + 2 g_{\mu\nu}g_{\rho\sigma} \nonumber \\
            &\hspace{4em} + \qty(g_{\mu\rho}\frac{q_\nu q_\sigma}{q^2} + g_{\mu\sigma}\frac{q_\nu q_\rho}{q^2} + g_{\nu\rho}\frac{q_\mu q_\sigma}{q^2} + g_{\nu\sigma}\frac{q_\mu q_\rho}{q^2}) 
            - 2\qty( g_{\mu\nu}\frac{q_\rho q_\sigma}{q^2} + g_{\rho\sigma}\frac{q_\mu q_\nu}{q^2})
        \Big\} \nonumber \\
        &+ q^4\Big\{
        {\delta a_2} (g_{\mu\rho}g_{\nu\sigma} + g_{\mu\sigma}g_{\nu\rho}) + (8{\cdot\delta a_1} + 2{\cdot\delta a_2})g_{\mu\nu}g_{\rho\sigma} \nonumber \\
        &\hspace{3.5em} - {\delta a_2}\qty(g_{\mu\rho}\frac{q_\nu q_\sigma}{q^2} + g_{\mu\sigma}\frac{q_\nu q_\rho}{q^2} + g_{\nu\rho}\frac{q_\mu q_\sigma}{q^2} + g_{\nu\sigma}\frac{q_\mu q_\rho}{q^2}) \nonumber \\
        &\hspace{3.5em}- (8{\cdot\delta a_1} + 2{\cdot\delta a_2})\qty( g_{\mu\nu}\frac{q_\rho q_\sigma}{q^2} + g_{\rho\sigma}\frac{q_\mu q_\nu}{q^2}) \nonumber \\
        &\hspace{3.5em}+ (8{\cdot\delta a_1} + 4{\cdot\delta a_2}) \frac{q_\mu q_\nu q_\rho q_\sigma}{q^4} 
    \Big\}\bigg\} +O\qty(h_{\mu\nu}^3)\,,
\end{align}
we find their contribution to the coefficients $c_i$,
\begin{subequations}
\begin{align}
    \delta c_1 &= -\dL - \Mp^2\cdot\frac{\Zg}{2}\cdot q^2 + {\cdot\delta a_2}\cdot q^4 \,, \\
    \delta c_2 &= \dL + 2\cdot \Mp^2\cdot\frac{\Zg}{2}\cdot q^2  + (8{\cdot\delta a_1} + 2{\cdot\delta a_2})\cdot q^4 \,, \\
    \delta c_3 &= \Mp^2\cdot\frac{\Zg}{2}\cdot q^2 - {\cdot\delta a_2}\cdot q^4\,, \\
    \delta c_4 &= -2\cdot \Mp^2\cdot\frac{\Zg}{2}\cdot q^2 - (8{\cdot\delta a_1} + 2{\cdot\delta a_2})\cdot q^4 \,, \\
    \delta c_5 &= (8{\cdot\delta a_1} + 4{\cdot\delta a_2})\cdot q^4 \,.
    \label{eq:delta-c-coeff}
\end{align}    
\end{subequations}

We can check that the $\delta c_i$ satisfy the same identities as the 1-loop result \eqref{eq:ward-identity-relations} \textit{only when} $\dL = 0$---this tells us that $\delta\Lambda$ is not required for this 1-loop computation. Additionally, were we to take this contribution seriously we would obtain a tadpole term that gives a non-zero vev to $h_{\mu\nu}$. Consequently we should have expanded around a different background, \textit{i.e.} around a $\bar{g}_{\mu\nu}$ such that $R\big|_{\bar{g}_{\mu\nu}} = \Lambda$, which is of course inconsistent with expanding around Minkowski. All in all, we must disregard the presence of a cosmological constant and therefore set $\dL=0$.

\subsection{Renormalisation and the strong coupling scale}
\label{sec:renormalisation}

Adding the counterterm contribution to our previous 1-loop results we find
\begin{subequations}
\begin{align}
    \bar{c}_1=c_1+\delta c_1=&- \Mp^2\cdot\frac{\Zg}{2}\cdot q^2 + {\delta a_2}\cdot q^4+\frac{m^4}{30\pi^2} \qty(1 - \text{F}(\alpha)) \nonumber\\
    &+\frac{ m^2q^2}{15\pi^2}\qty(\frac{5}{16\varepsilon} + \frac{43}{96} - \frac14 \text{F}(\alpha) - \frac{5}{32}\gamma - \frac{5}{32}\log\qty(\frac{m^2}{4\pi\mu^2})) \nonumber \\
    &+\frac{q^4}{15\pi^2} \qty(\frac{1}{32\varepsilon} + \frac{23}{480} - \frac{1}{32}\text{F}(\alpha) - \frac{1}{64}\gamma - \frac{1}{64}\log\qty(\frac{m^2}{4\pi\mu^2})) \,, \\
    \bar{c}_5=c_5+\delta c_5=&~(8{\cdot\delta a_1} + 4{\cdot\delta a_2})\cdot q^4+\frac{m^4}{10\pi^2}\qty(1 - \text{F}(\alpha)) +\frac{m^2q^2}{15\pi^2} \qty(-\frac38 + \frac12~\text{F}(\alpha)) \nonumber \\
    & +\frac{q^4}{15\pi^2}\qty(\frac{1}{4\varepsilon} + \frac{47}{240} - \frac14 \text{F}(\alpha) - \frac18\gamma - \frac18\log\qty(\frac{m^2}{4\pi\mu^2})) \,.
\end{align}
\end{subequations}
The counterterms must be chosen such that all divergences are cancelled. There is however the usual freedom in how to fix the finite pieces---we must choose a renormalisation scheme. Let us impose the following 3 conditions in order to fix the 3 counterterms. 
\begin{align}
    \frac{\ed\bar{c}_1}{\ed q^2}\Bigg|_{q^2 = q_\star^2} = 0 \,,
    \quad\quad \frac12\frac{\ed^2\bar{c}_1}{\ed (q^2)^2}\Bigg|_{q^2 = q_\star^2} = a_2^{\rm phys} \,,
    \quad\quad \frac12\frac{\ed\bar{c}_5}{\ed (q^2)^2}\Bigg|_{q^2 = q_\star^2} = 8a_1^{\rm phys} + 4 a_2^{\rm phys} \,. 
\end{align}

The first condition ensures that at some reference scale $q_\star^2$ the 1-loop contribution to the $q^2$ term vanishes, which leaves the normalisation of the graviton fixed by the tree-level contribution when $q^2=q_\star^2$. In other words, it defines the scale at which we are choosing our graviton kinetic term normalisation.  

Contrary to the $q^2$ term, there is no $q^4$ contribution at tree-level since we did not include higher-derivative terms in the original action. Had we done so, their tree-level contribution would take exactly the same form as the counter-terms $\delta a_i$. The EFT expansion (necessary to make sense of a non-renormalisable theory) puts the tree-level 4-derivative terms at the same order as the 1-loop 2-derivative contribution \cite{Burgess:2020tbq}. Therefore, our remaining two conditions are actually analogous to the first one---we are imposing that, at the scale $q_\star^2$, the 4-derivative contribution is purely tree-level. Nevertheless, having not directly measured these contributions prevents us from assigning a value to the constants $a_i^\text{phys}$ at low-energies, the way we do for the Planck scale . At most one can require that they remain negligible/subleading, compatible with the fact that their effect has so far not been detected. The EFT suppression then implies that the $a_i^{\rm phys}$ coefficients themselves should not be much bigger than $O(1)$.

Solving these 3 equations for the 3 counter-terms $(\Zg,\delta a_1,\delta a_2)$ and plugging them back into the 1-loop result we obtain 
\begin{subequations}
\begin{align}
    \bar{c}_1 =&~ q^4\cdot a_2^{\rm phys} 
    + \frac{m^4}{30\pi^2} \qty(1 - \F{\alpha})
    \nonumber\\
    & + \frac{m^2q^2}{480\pi^2}\left\{1 - 8\F{\alpha} + 10\F{\alpha_\star}\qty(1 - \frac{2m^2}{q_\star^2}) + \frac{20m^2}{q_\star^2} - \frac{q_\star^2}{m^2}\right\} \nonumber \\ 
    &+ \frac{q^4}{480\pi^2} \left\{\frac{3}{4} - \F{\alpha} + \F{\alpha_\star}\qty(1 - \frac{2m^2}{q_\star^2} + \frac{6m^4}{q_\star^4}) + \frac{3m^2}{2q_\star^2} - \frac{6m^4}{q_\star^4}\right\}, 
    \label{eq:renormalised-graviton-result-a}
    \\
    \bar{c}_5 =&~ q^4\cdot (8a_1^{\rm phys} + 4a_2^{\rm phys})
    +\frac{m^4}{10\pi^2}\qty(1 - \F{\alpha}) 
    -\frac{m^2q^2}{30\pi^2} \qty(\frac34 - \F{\alpha}) \nonumber \\
    &+\frac{q^4}{60\pi^2}\qty(\frac{\alpha_\star}{1+\alpha_\star})^2
    \Bigg\{\frac{11}{32} - \frac{9m^4}{4q_\star^4} - \frac{19m^2}{16q_\star^2} + \frac{5q_\star^2}{16m^2} + \frac{3q_\star^4}{64m^4} \nonumber \\
    &~~-\F{\alpha}\qty(1 + \frac{q_\star^2}{2m^2} + \frac{q_\star^4}{16m^4})
    + \frac{3\F{\alpha_\star}}{8}\qty(1 + \frac{6m^4}{q_\star^4} + \frac{8m^2}{3q_\star^2} + \frac{q_\star^2}{m^2} + \frac{q_\star^4}{6m^4})  
     \Bigg\} \,.
     \label{eq:renormalised-graviton-result-b}
\end{align}
\label{eq:renormalised-graviton-result}
\end{subequations}
This is clearly finite when $\varepsilon\to 0$, as it should, and it depends very explicitly on the reference scale $q_\star^2$. A natural choice for this reference scale would be $q_\star^2 = 0$, which would not only correspond to the scale where the graviton propagator has a pole (therefore corresponding to an on-shell condition) and one at which we can certainly perform an experiment, but it would also simplify the result significantly, 
\begin{subequations}
\begin{align}
    \bar{c}_1=&~q^4\cdot a_2^\text{phys}+\frac{m^4}{30\pi^2} \qty(1 - \F{\alpha}) +\frac{ m^2q^2}{60\pi^2}\qty( \frac{7}{6} - \F{\alpha} )+\frac{q^4}{480\pi^2} \qty(\frac{23}{15} - \F{\alpha}) \,, \\
    \bar{c}_5=&~q^4\cdot (8a_1^\text{phys}+4a_2^\text{phys})+\frac{m^4}{10\pi^2}\qty(1 - \F{\alpha}) -\frac{m^2q^2}{30\pi^2} \qty(\frac34 - \F{\alpha})\nonumber\\
    &+\frac{q^4}{60\pi^2}\qty(\frac{47}{60} -  \F{\alpha}) \,.
\end{align}    
\end{subequations}
The downside of making this choice is that we would lose track of the arbitrary reference scale $q_\star^2$. Keeping this scale and 
following the reasoning we developed for QED (cf. Section \ref{sec:QED}), we can consider the 1PI 2-point function
\begin{equation}
    \frac{q^2}{2} \cdot \left[-\Mp^2 + 2\cdot \frac{\bar{c}_1(q^2)}{q^2} \right]_{q^2 = 0} \,,
    \label{eq:1PI-graviton}
\end{equation}
and demand that it does not depend on our choice of $q_\star^2$. Note that this means the parameter $\Mp$ that appears in \eqref{eq:1PI-graviton} is not the usual $\Mp\sim 10^{18}$ GeV, but instead a function of $q_\star^2$---let us therefore rename it $M_{\rm grav}(q_\star^2)$ and call it the gravitational mass scale. We then deduce the following equation
\begin{align}
    -q_\star^2\frac{\ed M_{\rm grav}^2}{\ed q_\star^2} + 2\cdot q_\star^2 \frac{\ed}{\ed q_\star^2}\bigg[\frac{\bar{c}_1}{q^2}\Big|_{q^2=0}\bigg] &= 0 \,.
\end{align}
which gives         
\begin{align}
    q_\star^2\frac{\ed M_{\rm grav}^2}{\ed q_\star^2} 
    &= \frac{q_\star^2}{960\pi^2}\cdot\qty(\frac{\alpha_\star}{1+\alpha_\star})\left\{
        1
        -\frac{4}{\alpha_\star}
        -\frac{5\alpha_\star}{2}
        -\frac{15\alpha_\star^2}{2}\qty(1-\F{\alpha_\star})  
    \right\} \,.
    \label{eq:running-gravitational-coupling}
\end{align}

\begin{figure}
    \centering
    \includegraphics[width=0.7\linewidth]{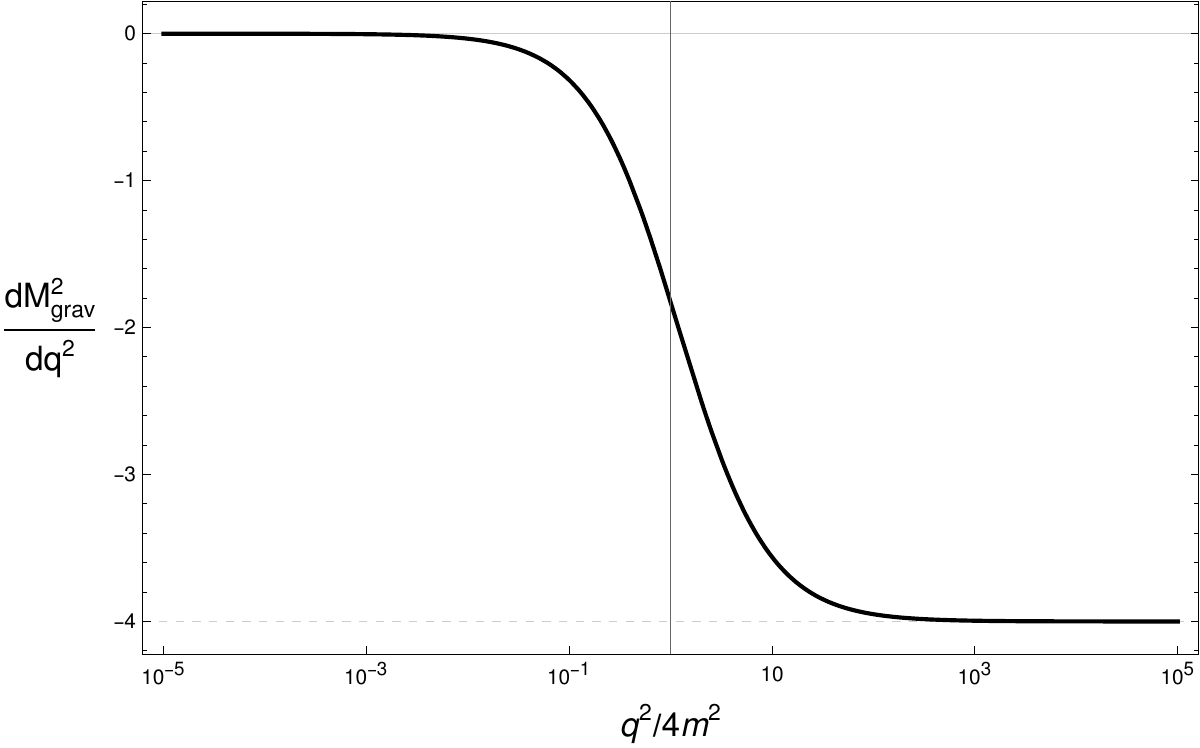}
    \caption{Plot of \eqref{eq:running-gravitational-coupling}, rescalled by the factor $960\pi^2$. We can see that in the limit $q^2\gg 4m^2$ (i.e. $\alpha\ll 1$), $M_{\rm grav}$ decreases at a constant rate, while in the limit $q^2\ll 4m^2$ (i.e. $\alpha\gg 1$), $M_{\rm grav}$ is approximately constant. Physically, we can interpret this as the scale $M_{\rm grav}$ running (to smaller values) at energies much higher than $m^2$ and not running at energies much smaller than $m^2$, at which the scalar could be integrated out and no longer run in the loops.}
    \label{fig:Mgrav-running}
\end{figure}

It is instructive to look at different limits for equation \eqref{eq:running-gravitational-coupling}. When $\alpha_\star\gg 1$, which corresponds to scales below the mass of the scalar (i.e. $q_\star^2 \ll 4m^2$), we find
\begin{equation}
    \frac{\ed M_{\rm grav}^2}{\ed q_\star^2} 
    = -\frac{1}{280\pi^2}\cdot\frac{1}{\alpha_\star} + O\qty(\frac{1}{\alpha_\star^2}) \,,
\end{equation}
which is suppressed by $\alpha_\star\gg 1$. Therefore, $M_{\rm grav}$ is constant at zeroth order in $1/\alpha_\star$ and determined by the measured Planck scale at low-energies $M_{\rm grav}(q_\star^2=0) =\Mp$. This defines the Planck scale and gives us an initial condition for $M_{\rm grav}(q^2)$. 

In the high-energy limit $\alpha_\star\ll 1$ (i.e. $q_\star^2 \gg 4m^2$), we find
\begin{equation}
    M_{\rm grav}^2(q_\star^2) = \Mp^2 - \frac{q_\star^2}{240\pi^2}\qty(1+O(\alpha_\star)) \,.
    \label{eq:Mgrav-running-large-q}
\end{equation}
This tells us that at high enough energies $q_\star^2$, the gravitational mass scale $M_{\rm grav}$ vanishes. The equivalent of this phenomenon in QED was the infinite coupling scale, where $g_e\to\infty$, which we identified as the Landau pole. For our present case, we can make this explicit by canonically normalising the graviton $h_{\mu\nu}\to \frac{h_{\mu\nu}}{M_{\rm grav}(q^2)}$ and looking at the gravitational \textit{coupling}\footnote{An attentive reader might be confused with the passage from $q_\star$ to $q$. The philosophy of RG-improvement dictates that one should choose a $q_\star$ as close as possible to our experimental scale $q$, which is why in this analysis we make the choice $q_\star=q$. Note however that we can only do this at the very end, since otherwise we could not have derived the equation for $M_\text{grav}(q^2)$.}
\begin{equation}
    \frac{q^2}{M_{\rm grav}^2(q^2)}  = \frac{q^2}{\Mp^2\cdot\qty(1 - \frac{1}{240\pi^2}\cdot\frac{q^2}{\Mp^2})} \,,
\end{equation}
for $q^2\gg 4m^2$. The divergence of the coupling at $q^2 = 240\pi^2\cdot\Mp^2$ now becomes explicit. Just like for QED, our interest is the ``strong coupling'' scale,
\begin{equation}
    \frac{q^2}{\Mp^2\cdot\qty(1 - \frac{1}{240\pi^2}\cdot\frac{q^2}{\Mp^2})} \overset{!}{=} 1
    \implies 
    \cut{grav}^{\rm strong} = \sqrt{\frac{240\pi^2}{1 + 240\pi^2}}\cdot\Mp \approx \Mp \,.
    \label{eq:cut-1-scalar}
\end{equation}
Note that the ``strong coupling'' scale is significantly smaller than the ``divergent coupling'' scale. For gravity with a scalar field of mass $m$, we find that $\cut{grav}^{\rm strong}\approx\Mp$ as long as $m\ll\Mp$ and the approximation $q^2\gg 4m^2$ is valid up to this scale. 

In the opposite limit, $4m^2\gg q^2$ ($\alpha_\star\gg 1$), the ``strong coupling'' scale is still approximately given by the Planck scale, which is consistent if $m\gg\Mp$. 
All in all, for any value of the scalar mass, as long as there is only a single scalar the strong coupling scale is approximately given by the Planck scale.

As was the case for QED we can note that $\qty[\frac{\bar{c}_1(q^2)}{q^2}]_{q^2=0}$ is zero for $q_\star^2=0$ to solve \eqref{eq:running-gravitational-coupling} more expediently,
\begin{align}
    M_\text{grav}^2(q_\star^2) &= \Mp^2 + 2\qty[\frac{\bar{c}_1(q^2)}{q^2}]_{q^2=0} \\
    &=\Mp^2 - \frac{q_\star^2}{240\pi^2}\qty(1+\frac{5}{2}\alpha_\star\qty(\frac{5}{6}-\F{\alpha_\star})-\frac{5}{4}\alpha_\star^2(1-\F{\alpha_\star}))
\end{align}
where we have used the boundary condition $M_{\rm grav}(q_\star^2=0) =\Mp$. If we define the function
\begin{equation}
    {\rm H}(\alpha) = 1+\frac{5}{2}\alpha\qty(\frac{5}{6}-\F{\alpha})-\frac{5}{4}\alpha^2(1-\F{\alpha}) \,,
\end{equation}
we can write the result as 
\begin{align}
    M_\text{grav}^2(q_\star^2) =\Mp^2 - \frac{q_\star^2}{240\pi^2}{\rm H}(\alpha) \,,
\end{align}
and recover the limits found above.

\begin{figure}
    \centering
    \includegraphics[width=0.7\linewidth]{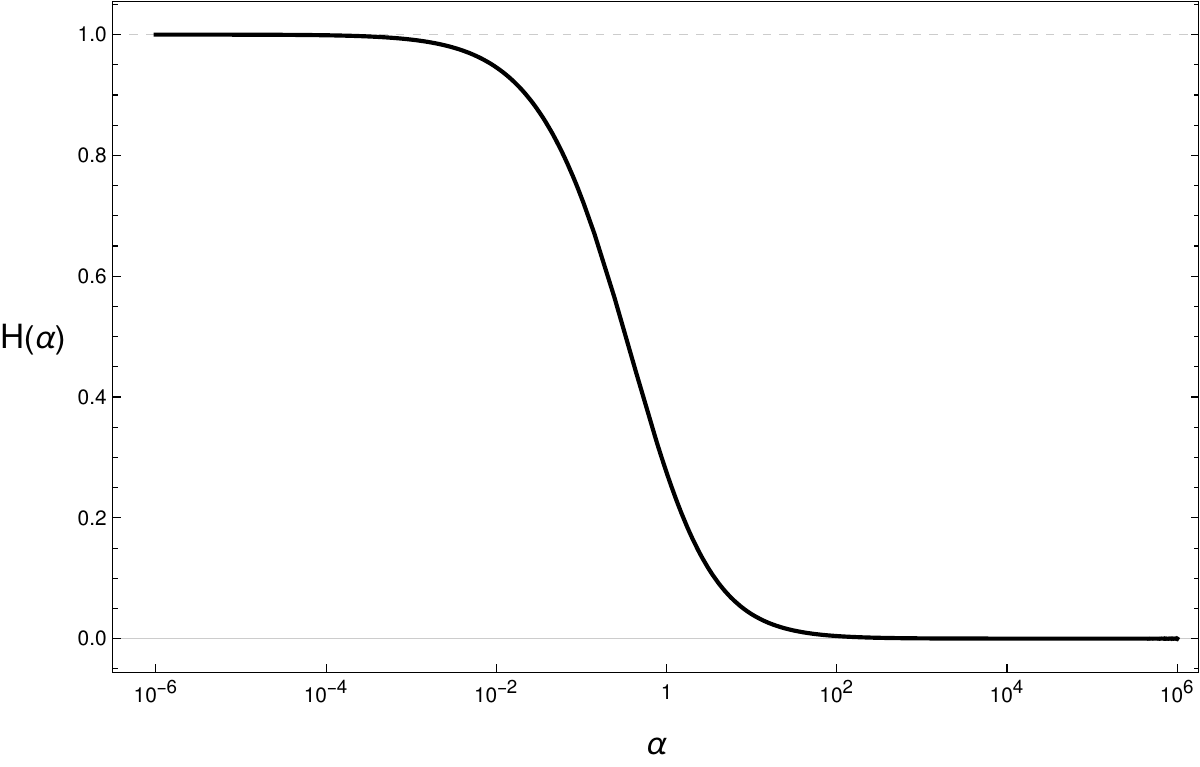}
    \caption{In terms of the function ${\rm H}(q^2)$ we can clearly still see the behaviour from Figure \ref{fig:Mgrav-running}: at low momenta this function approaches zero, which means $M_\text{grav}$ will approach $\Mp$; for $\alpha_\star\sim 1$ there is a sharp derivative; for high momenta it approaches a constant, which means $M_\text{grav}^2-\Mp^2\propto q_\star^2$.}
    \label{fig:exact-solution}
\end{figure}

Since ${\rm H}(\alpha)$ is bounded between 0 and 1 (Fig.\ref{fig:exact-solution}), the correction to $M_\text{grav}^2$ due to a single scalar field is always smaller than $\frac{q^2}{240\pi^2}\approx 4\times10^{-4}~q^2 < 4\times10^{-4}~\Mp$ at energies below $\Mp$. Therefore the correction to the tree-level value of $\Mp$ is negligible for any scalar field mass, even if it is slightly bigger for smaller masses. 

As we anticipated in Section \ref{sc:EFT-background}, there is a key difference with respect to QED. While at tree level QED appears to be valid at all energy scales, only revealing its breakdown at sufficiently high energies once we include loop contributions, gravity breaks down immediately at tree-level. This is because gravity only has non-renormalisable interactions---it should itself be treated as an EFT with an associated cutoff and it cannot be defined at arbitrarily high-energies, not even at tree-level. 

Although the above strong coupling scale is no longer exponentially high, it is still conceptually different from the EFT scale one usually refers too in the context of gravitational EFTs---the scale at which the $a_i^\text{phys}$ terms become as important as the tree-level 2-derivative terms. 
The details will depend on the precise measured value of these coefficients but we can combine \eqref{eq:renormalised-graviton-result-a} with \eqref{eq:1PI-graviton} to find
\begin{equation}
    \cut{EFT}^\text{grav}\sim\frac{M_{\rm grav}}{\sqrt{2\cdot a_2^\text{phys}}}
    \label{eq:gravity-eft-scale}
\end{equation}

If $a_2^\text{phys}\sim O(1)$ this scale will coincide with the strong coupling scale---this is not an accident, it follows directly from our definition of $a_2$. In fact, $a_1$ and $a_2$ are defined to control the ratio between $\cut{strong}^{\rm grav}$ and $\cut{EFT}^{\rm grav}$, and therefore computing $\cut{strong}^{\rm grav}$ does not tell us anything about $\cut{EFT}^{\rm grav}$ unless we also know the values of $a_1$ and $a_2$ from the UV theory\footnote{The terms associated to $a_i$ are somewhat special in the sense that the dimension of the operators $R^2$ and  $R_{\mu\nu}R^{\mu\nu}$ does not require a dimensionful parameter $\cut{EFT}$ for consistency. It might therefore seem that these operator do not have an associated scale either. Thinking of $a_i$ as the ratio between $\cut{EFT}$ and $\Mp$ makes it more transparent that a scale is still involved.}. Note that the EFT we consider includes both the graviton and the scalar field, which has not been integrated out. 

Taking $a_1,a_2\sim O(1)$ would be the \textit{technically natural} choice, but as we illustrated with the neutrino example this is not necessarily the case. In the neutrino case this coupling was ``unnaturally'' small due to the fact that it only couples to other particles via the weak force; given that gravity couples to everything, we might not be surprised if these couplings are ``unnaturally'' large.

As a final point you might be slightly concerned regarding the running of $a_2^\text{phys}$. There are two points that can be made regarding this: firstly, its value can only be fixed from knowledge of the UV theory or experimental measurements so at most we are adding a known value on an unknown constant and the total result would still be unknown; secondly, the running would appear at higher loop order and would be a subleading effect if the EFT expansion is to be trusted.

\section{The original species scale}
\label{sec:original-species-scale}

The original species scale calculation concerned $N$ identical scalars for very large $N$ \cite{Dvali:2007hz,Dvali:2007wp}. In this case the strong coupling scale will change rather dramatically, potentially shifting it away from a naive expectation near $\Mp$.

\subsection{The strong coupling scale for $N$ identical scalars}
\label{sec:sp-scale-N-scalars}

Since the $N$ scalars are identical, each of them will contribute as a scalar $\phi$ of mass $m$ as the one in Section \ref{sec:1-loop-computation}. Their effect is therefore captured by multiplying $\bar{c}_1$ in \eqref{eq:renormalised-graviton-result-a} by $N$ so that \eqref{eq:running-gravitational-coupling} becomes  
\begin{align}
    q_\star^2\frac{\ed M_{\rm grav}^2}{\ed q_\star^2} 
    &= N\cdot\frac{q_\star^2}{960\pi^2}\cdot\qty(\frac{\alpha_\star}{1+\alpha_\star})\left\{
        1
        -\frac{4}{\alpha_\star}
        -\frac{5\alpha_\star}{2}
        -\frac{15\alpha_\star^2}{2}\qty(1-\F{\alpha_\star})  
    \right\} \,.
\end{align}
With the simple rewriting 
\begin{align}
    q_\star^2\dv{q_\star^2}(\frac{M_\text{grav}^2}{N})
    &= \frac{q_\star^2}{960\pi^2}\cdot\qty(\frac{\alpha_\star}{1+\alpha_\star})\left\{
        1
        -\frac{4}{\alpha_\star}
        -\frac{5\alpha_\star}{2}
        -\frac{15\alpha_\star^2}{2}\qty(1-\F{\alpha_\star})  
    \right\} \,,
\end{align}
it seems that our previous analysis will hold for a rescaled mass parameter $M_\text{grav}\to M_\text{grav}/\sqrt{N}$. This was the crux of the original reasoning behind the species scale \cite{Han_2005,Dvali:2007hz,Dvali:2007wp,Arkani-Hamed:2005zuc,Distler:2005hi,Dvali:2010vm}. However, since we must still impose $M_\text{grav}(q_\star^2=0)=\Mp$ as a boundary condition, the solution for $N$ scalars is given by
\begin{equation}
    M_\text{grav}^2(q_\star^2)=\Mp - N\cdot\frac{q_\star^2}{240\pi^2}\qty(1+\frac{5}{2}\alpha_\star\qty(\frac{5}{6}-\F{\alpha_\star})-\frac{5}{4}\alpha_\star^2(1-\F{\alpha_\star})) \,,
\end{equation}
which means that the well-known behaviour $M_\text{grav}\to M_\text{grav}/\sqrt{N}$ is not an exact statement.

To find the scale at which the coupling becomes $O(1)$, i.e.
\begin{equation}
    \frac{q^2}{M_\text{grav}^2(q^2)} \overset{!}{=} 1 \,,
\end{equation}
we must solve the equation
\begin{align}
    q^2 = \Mp^2 - N\cdot\frac{q^2}{240\pi^2}\cdot {\rm H}(\alpha) \,,
    \label{eq:Nscalars_cut_eq}
\end{align}
with
\begin{equation}
    {\rm H}(\alpha) = 1+\frac{5}{2}\alpha\qty(\frac{5}{6}-\F{\alpha})-\frac{5}{4}\alpha^2(1-\F{\alpha}) \,.
\end{equation}
\begin{figure}
    \centering
    \includegraphics[width=0.75\linewidth]{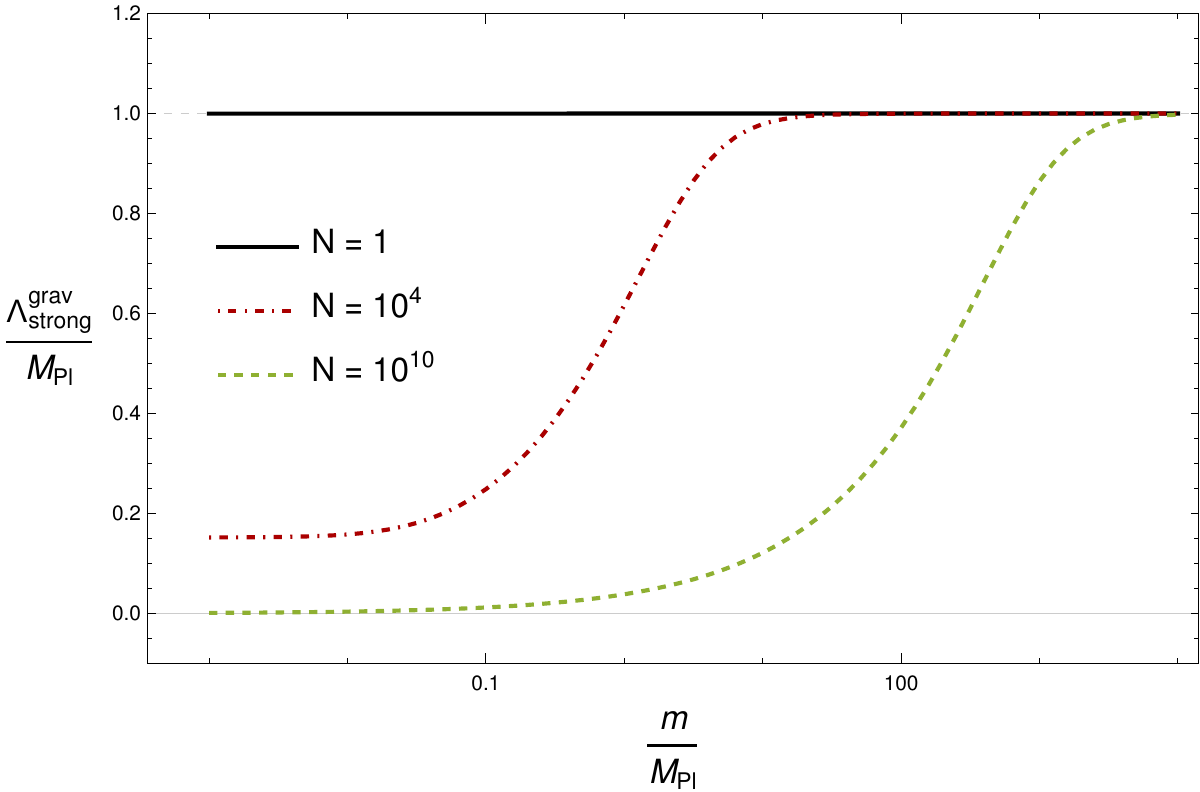}
    \caption{Strong coupling scale as a function of $m/\Mp$ for different values of $N$.}
    \label{fig:cutoff-vs-mass-different-N}
\end{figure}

\begin{figure}
    \centering
    \includegraphics[width=0.75\linewidth]{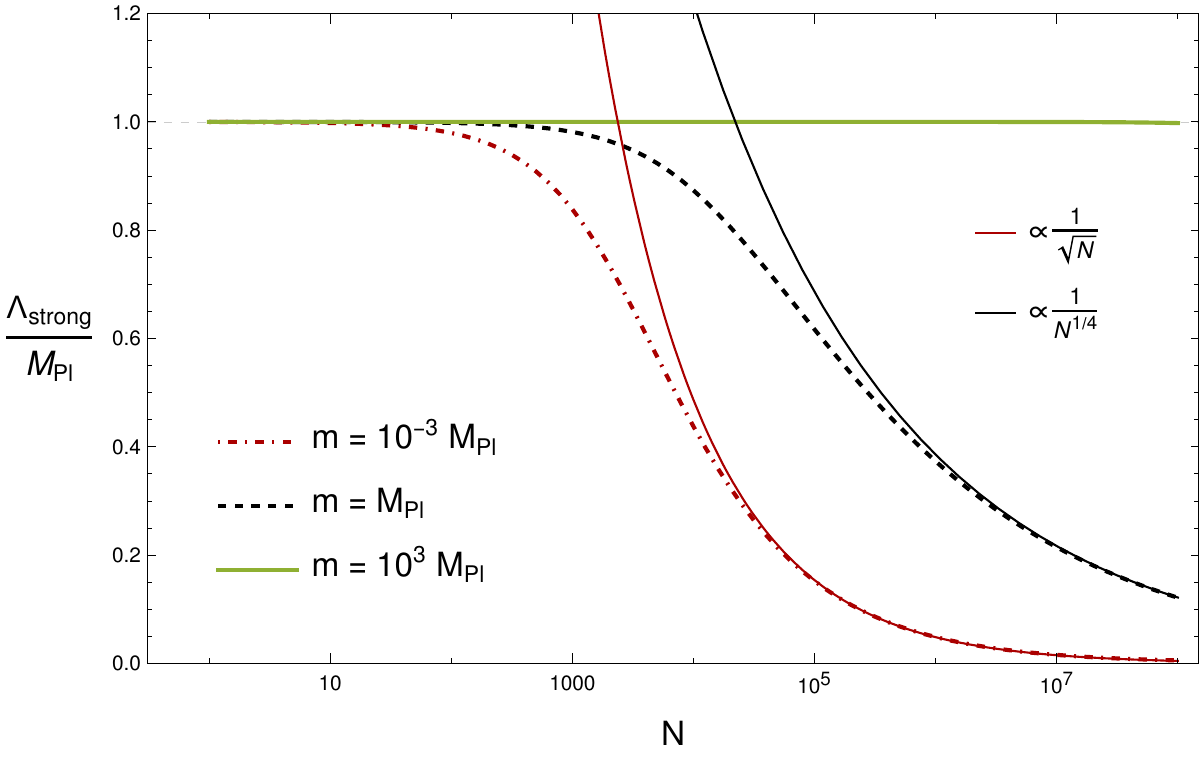}
    \caption{Strong coupling scale as a function of $N$ for different values of $m/\Mp$.}
    \label{fig:cutoff-vs-N-different-masses}
\end{figure}

Writing it fully in terms of $\alpha$, it becomes clear that the solution depends only on the parameters $m/\Mp$ and $N$,
\begin{equation}
    \frac{\alpha}{4\qty(1+ \frac{N}{240\pi^2}\cdot {\rm H}(\alpha))} = \frac{m^2}{\Mp^2} \,. 
    \label{eq:strong_alpha}
\end{equation}

Solving the equation numerically for different values of $(N,\frac{m}{\Mp})$, we can see how the strong coupling scale depends on these parameters (Fig. \ref{fig:cutoff-vs-mass-different-N} and \ref{fig:cutoff-vs-N-different-masses}). When $m\gg\Mp$, the scalars do not affect the strong coupling scale, regardless of their number $N$. In contrast, when $m\ll\Mp$ the scalars bring down the strong coupling scale, more significantly for a larger number $N$ of scalars. This means that if we have scalars with different masses, we should expect the lightest ones to have the strongest effect on $\cut{strong}^{\rm grav}$. Moreover, exactly how light these scalars need to be to behave as if they were massless depends on the their number $N$.

In the remainder of this section we will strive to make these statements more quantitative.

\subsection{Massless limit}
\label{sec:N-scalars-massless-limit}

In the limit of zero mass we have $\rm{H}(\alpha)\to1$ and therefore
\begin{equation}
    M_{\rm grav}^2(q_\star^2) = \Mp^2 -N\cdot \frac{q_\star^2}{240\pi^2}\qty(1+O\qty(\alpha_\star\log \alpha_\star)) \,,
    \label{eq:Mgrav-running-large-q-N}
\end{equation}
from which we find the strong coupling scale
\begin{equation}
    \cut{strong}^{\rm grav} = \sqrt{\frac{240\pi^2}{N + 240\pi^2}}\cdot\Mp \,.
    \label{eq:strong-large-q-N}
\end{equation}
Taking the limit when $N\gg240\pi^2\approx2369$ we obtain
\begin{equation}
    \cut{strong}^{\rm grav} \sim \sqrt{\frac{240\pi^2}{N}}\cdot\Mp+O\qty(\frac{1}{N^{3/2}})
    \label{eq:strong-large-q-large-N}
\end{equation}
which reproduces the scaling found in the literature. It is worth noting nevertheless that this is only recovered for $N$ quite large (cf. $N\sim 10^{32}$ in \cite{Dvali:2007hz}), and that the prefactor is not exactly $O(1)$ as $\sqrt{240\pi^2}\approx49$.

For us to trust the solution given by \eqref{eq:strong-large-q-N} not only do we need $m\ll\Mp$ but also $m\ll\cut{strong}^{\rm grav}$. Even if $m\ll\Mp$, for $N$ sufficiently large and non-zero masses, the strong coupling scale may not obey \eqref{eq:strong-large-q-N}.

\subsection{Large mass limit}
\label{sec:N-scalars-large-mass-limit}

For larger masses, \textit{i.e.} in the $\alpha_\star\gg1$ limit we obtain
\begin{equation}
    M_\text{grav}^2(q_\star^2)=\Mp^2 - N\cdot\frac{q_\star^2}{240\pi^2}\cdot\qty(\frac{3}{7\alpha_\star}+O\qty(\frac{1}{\alpha_\star^2}))
    \label{eq:large-alpha-limit}
\end{equation}

The factor of $N$ is common to all terms in the expansion in large $\alpha_\star$, so the validity of this expansion is unaffected. However, in this limit neglecting the leading order term in $1/\alpha_\star$ in comparison with $\Mp$ will be less accurate for sufficiently large $N$.

At leading order in $\frac{1}{\alpha_\star}$, the strong coupling scale becomes

\begin{equation}
    \cut{strong}^{\rm grav}=\Mp\sqrt{\frac{1120\pi^2}{N}\cdot\frac{m^2}{\Mp^2}\qty(-1+\sqrt{1+\frac{N}{560\pi^2}\frac{\Mp^2}{m^2}})} \,.
    \label{eq:largeN-strong-scale}
\end{equation}
Note that $\cut{strong}^{\rm grav}$ is really a function of the combination
\begin{equation}
    \xi=\frac{560\pi^2}{N}\frac{m^2}{\Mp^2} \,,
    \label{eq:parameter-xi}
\end{equation}
in terms of which we can write
\begin{equation}
    \cut{strong}^{\rm grav}=\Mp\sqrt{2\xi\qty(-1+\sqrt{1+\frac{1}{\xi}})} \,.
    \label{eq:strong-coupling-scale-vs-xi}
\end{equation}
Since this solution was derived under the assumption that $\alpha_\star\gg1\implies 4m^2\gg q_\star^2$, we should only trust it if $\cut{strong}^{\rm grav}\ll 2 m$. 
Thus it is worth checking under which circumstances this hierarchy holds. Being a function of $\xi$, which depends on the ratio $m^2/N$, $\cut{strong}^{\rm grav}$ may satisfy this condition in different limits in $N$ and $m/\Mp$. 
Let us then examine these different limits one by one.

\begin{figure}
    \centering
    \includegraphics[width=0.7\linewidth]{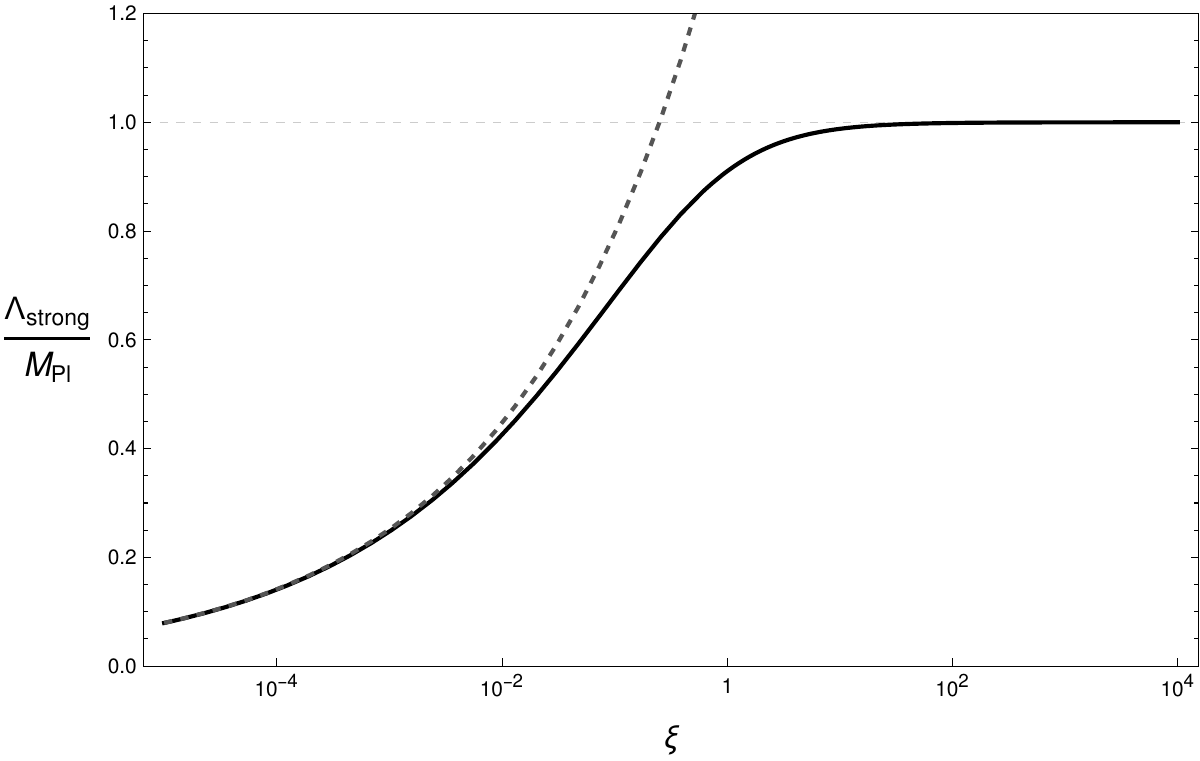}
    \label{fig:cutoff-N-scalars}
    \caption{Strong coupling scale as a function of the parameter $\xi$ \eqref{eq:parameter-xi}, which depends on both the number of scalars $N$ and their mass $m/\Mp$. The black line corresponds to the exact solution from \eqref{eq:strong-coupling-scale-vs-xi} and the dashed gray line to the first order approximation for small $\xi$ \eqref{eq:strong-coupling-scale-vs-xi-small-xi}.}
\end{figure}

\begin{description}
    \item[$N$ fixed, $m^2$ large:] 

In this limit $\xi$ is large and we have asymptotically
\begin{equation}
    \cut{strong}^{\rm grav} = \Mp\qty(1-\frac{1}{8\xi}+O\qty(\frac{1}{\xi^2})) \,.
    \label{eq:strong-coupling-scale-vs-xi-large-xi}
\end{equation}
As is to be expected, in the infinite mass limit the strong coupling scale freezes at $\Mp$, a constant. If $m\gg\Mp$ then \eqref{eq:largeN-strong-scale} is valid.
In fact, $m\to\infty$ also corresponds to $\alpha\to\infty$ and therefore $\rm{H}(\alpha)\to0$. Since it is always true that $\rm{H}(\alpha)>0$, it follows directly from \eqref{eq:Nscalars_cut_eq} that 

\begin{equation}
    \text{H}(\alpha)>0\implies \frac{240\pi^2}{N}\qty(\frac{\Mp^2}{q^2}-1)>0\implies q^2<\Mp^2 \,.
\end{equation}
The strong coupling scale is always bounded above by the Planck scale and approaches it as $m\to\infty$ for any finite $N$.

\item[$m^2$ fixed, $N$ large:]
In this case $\xi\to0$ and therefore

\begin{equation}
    \cut{strong}^{\rm grav}\approx\Mp\qty(\sqrt{2}\xi^{1/4}+O\qty(\xi^{3/4})) \,,
    \label{eq:strong-coupling-scale-vs-xi-small-xi}
\end{equation}
which is valid for
\begin{equation}
    \xi\ll1\implies N\gg560\pi^2\frac{m^2}{\Mp^2} \,.
\end{equation}
In this case, the strong coupling scale approaches zero, which is not entirely surprising.
The hierarchy $\cut{strong}^{\rm grav}\ll 2 m$ is respected in this regime, not because $m$ is arbitrarily large, but because $\cut{strong}^{\rm grav}$ is very small at large $N$. 

This is the situation alluded to in the paragraph below \eqref{eq:strong-large-q-large-N}. Even if the mass is parametrically smaller than the Planck mass, as long as it is non-zero, there will be an $N$ sufficiently large to push us outside of the regime of validity of \eqref{eq:strong-large-q-N}.

\item[$N$ and $m^2$ large, $\xi$ fixed:]
In this case, the hierarchy is satisfied because $\frac{\cut{strong}^{\rm grav}}{\Mp}$ will be some fixed number between 0 and 1, whereas $m$ will be large.

\end{description}
We conclude that \eqref{eq:largeN-strong-scale} is valid for any value of $\xi$ as long as either $m^2$ or $N$ is large.

One may worry about having $\cut{strong}^{\rm grav}\ll 2 m$, since this surely implies that the mass of the scalars is above the EFT scale. Intuitively, one should not consider these states. This intuition might not be justified for two reasons.

Firstly, as emphasised in the previous section we do not have access to the EFT scale without first measuring or calculating the $a_i$. It is therefore possible that despite the scalars lying above the strong coupling scale they would still lie below the EFT scale. 

Secondly, note that there is no specific mathematical inconsistency with including states with masses above $\cut{EFT}$, as long as they do not appear in external legs to avoid considering processes with energies at these scales. The only problem we might run into is that when we calculate the effects of those states they would contribute at the same order as unknown UV physics, and adding a known quantity to an unknown one does not give us more knowledge. At its most basic level, an EFT is agnostic with respect to the physical states whose masses lie above $\cut{EFT}$. Therefore, when we add a number of scalars with mass above $\cut{EFT}$ we have no way of knowing what other states could appear at the same scales. In fact, this is precisely what happens in the $\xi\to\infty$ limit, where the $q^2$ contribution vanishes. The only contribution is a $q^4$ term, where the effect of these scalars appears intertwined with the unknown $a_2$ coefficient. There is therefore no physical significance in the scalar contribution in this case.

However, we find a different behaviour in the limit $\xi\to 0$. When $N$ is sufficiently large, even for masses above the strong coupling scale, we cannot neglect the contribution from the scalars. Despite a single scalar with mass above $\cut{strong}^{\rm grav}$ having a completely negligible contribution, if there are a sufficient number of them, their combined behaviour will no longer be subleading in the EFT expansion and we cannot neglect these states. 
This behaviour is somewhat reminiscent of dangerously irrelevant contributions, when we believe we can fully neglect a term in the IR, which ends up contributing towards certain observables. 

\subsection{Intermediate case}
\label{sec:N-scalars-intermediate-case}

It may happen that $N$ and $m$ are delicately balanced to produce $\cut{strong}^{\rm grav}\sim m$ and therefore, neither of the above approximations holds. In this case the interesting question is not ``what is the cutoff'' because we know it: it is close to $m$. What is interesting is the relationship between $m$ and $N$ required to make this happen. Setting $\alpha=1$ in \eqref{eq:strong_alpha}, we have

\begin{equation}
    \frac{1}{4\qty(1+\frac{N}{240\pi^2}\rm{H}(1))}=\frac{m^2}{\Mp^2}\implies N=\frac{240\pi^2}{\rm{H(1)}}\qty(\frac{\Mp^2}{4m^2}-1) \,,
    \label{eq:min-N-alpha-1}
\end{equation}
which tells us how many particles of mass $m$ are needed to push the strong coupling scale all the way down to $m$. 
Note that the RHS is only positive for
\begin{equation}
    m<\frac{\Mp}{2} \,,
\end{equation}
so that for larger masses $\cut{strong}^{\rm grav}$ will always be below $m$.

\subsection{Summary of results}
\label{sec:N-scalars-summary}

We summarise our results in the table below, including all the different regimes studied in this section.
{
    \def\arraystretch{2.75}
    \setlength\tabcolsep{1em}
    \newcolumntype{a}{>{\columncolor{gray!25}}c}
    \begin{table}[H]
    \centering
    \begin{tabular}{|c|c|c|}
        \hline
        \multirow{2}{*}[-1em]{$\cut{strong}^{\rm grav}\gg 2m$} & \multicolumn{2}{c|}{$\cut{strong}^{\rm grav} \approx \sqrt{\frac{240\pi^2}{N + 240\pi^2}}\cdot\Mp$ \quad\eqref{eq:strong-large-q-N}}  \\[6pt]
        \cline{2 - 3}
        & $N\gg240\pi^2$ & $\cut{strong}^{\rm grav} \approx \sqrt{\frac{240\pi^2}{N}}\cdot\Mp$ \quad\eqref{eq:strong-large-q-large-N} \\[6pt]
         \hline
        \multirow{3}{*}[-1em]{$\cut{strong}^{\rm grav}\ll 2m$} & \multicolumn{2}{c|}{$\cut{strong}^{\rm grav}\approx\Mp\sqrt{\frac{1120\pi^2}{N}\cdot\frac{m^2}{\Mp^2}\qty(-1+\sqrt{1+\frac{N}{560\pi^2}\frac{\Mp^2}{m^2}})}$ \quad\eqref{eq:largeN-strong-scale}} \\[6pt]
        \cline{2 - 3}
         & $m\gg\frac{\Mp}{8}\sqrt{\frac{N}{70\pi^2}}$  & $\cut{strong}^{\rm grav} \approx \Mp$ \quad\eqref{eq:strong-coupling-scale-vs-xi-large-xi}\\[6pt]
         \cline{2 - 3}
         & $N\gg\frac{560\pi^2}{\sqrt{2}}\frac{m^2}{\Mp^2}$ & $\cut{strong}^{\rm grav}\approx\Mp\sqrt{2}\cdot\qty(\frac{560\pi^2}{N}\frac{m^2}{\Mp^2})^{1/4}$ \quad\eqref{eq:strong-coupling-scale-vs-xi-small-xi} \\[6pt]
         \hline
         $\cut{strong}^{\rm grav}\approx 2m$ & \multicolumn{2}{c|}{$N\approx\frac{240\pi^2}{\rm{H(1)}}\qty(\frac{\Mp^2}{4m^2}-1)$  \quad\eqref{eq:min-N-alpha-1}}
          \\[6pt] \hline 
    \end{tabular}
    \caption{Summary of results for $N$ identical scalars in different mass regimes.}
    \label{tab:summary-N-identical}
    \end{table}
}

\section{A Tower of States}
\label{sec:towers}

Having gone through the detailed analysis of the previous section, our main goal is to consider an infinite tower of scalars with mass $m_n=f(n)\cdot m$, rather than $N$ scalars of equal mass.
Their contribution to $M_{\rm grav}(q^2)$ is such that the coupling becomes strong when 
\begin{align}
    q^2 = \Mp^2 - \sum_{n=1}^\infty\frac{q^2}{240\pi^2}\cdot d_n\cdot {\rm H}(\alpha_n) \,,
    \label{eq:cutoff-tower}
\end{align}
where $\alpha_n = \frac{4m_n^2}{q^2} = f(n)^2\cdot\alpha$ and we allow for a degeneracy of states at each level $d_n$.

\subsection{Convergence of the sum}
\label{sec:towers-convergence}

Since we are now dealing with an infinite number of states, the first question one must ask is whether this is a sensible mathematical procedure. There is in fact some hope that it might be sensible because states with masses greater than $\Mp$ barely contribute to the loop. Let us make this more precise.

If the solution to \eqref{eq:cutoff-tower} is finite, there will be an $n_H$ such that all states with masses $m_{n} \geq f(n_H)\cdot m$ obey $\alpha_n\gg1$. In this case, we an approximate \eqref{eq:cutoff-tower} by (cf. \eqref{eq:large-alpha-limit})

\begin{align}
    q^2 &\approx \Mp^2 - \frac{q^2}{240\pi^2}\cdot\left\{\sum_{n=1}^{n_H-1} d_n\cdot H(\alpha_n) + \frac{3q^2}{28m^2}\sum_{n=n_H}^{\infty} \frac{d_n}{f(n)^2} \right\} \,.
\end{align}
The first sum is finite because $n_H$ is a finite number. Whether the latter sum converges will of course depend on the specific $f(n)$ we choose to analyse. For the famous example of a simple Kaluza-Klein tower with $f(n)=n$ and $d_n = 2$, the sum does indeed converge
\begin{equation}
    \sum_{n=1}^\infty \frac{2}{n^2} = \frac{\pi^2}{3} \,.
\end{equation}

This tower arises from the compactification of a single extra dimension on a circle. When more dimensions are compact, the structure of the tower can become more complicated. If instead one compactifies $p$ extra dimensions on an isotropic torus, a generic tower corresponds in fact to a lattice of charges $(n_1,...,n_p)$ such that a state has mass $m_{n_1,...,n_p} = \sqrt{n_1^2 + ... + n_p^2}\cdot m$. Such lattice can be encoded in a single tower with $f(n)\sim n$ and $d_n\sim n^{p-1}$ asymptotically (i.e. at large $n$). This is also the tower behaviour one would find by compactifying $p$ extra dimensions on the $p$-dimensional sphere ($S^p$). Note that, more generally, Weyl's law implies that the degeneracy of eigenstates of the Laplace operator with a given eigenvalue (i.e. for a given level $n$) is asymptotically $d_n\sim f(n)^p - f(n-1)^p$ \cite{DeLuca:2024fbc}, which for $f(n)=n$ does indeed reproduce $d_n \sim n^{p-1}$ asymptotically. For this type of tower, one finds
\begin{align}
    \sum_{n=1}^{\infty} \frac{n^{p-1}}{n^2} 
    = \sum_{n=1}^{\infty} \frac{1}{n^{3-p}} 
    \,,
\end{align}
which is divergent for $p>1$.

In the literature  one often finds towers parameterised as $f'(n) = n^{\frac{1}{p}}$ and $d_n'\sim {\rm const}$ \cite{Castellano:2022bvr}, for which the sum becomes
\begin{equation}
    \sum_{n=1}^{\infty} \frac{1}{n^{\frac{2}{p}}} \,.
\end{equation}
From the point of view of state counting, this tower is equivalent to a KK tower associated with $p$ extra dimensions. To see this, let us count the number of states with masses up to $\Lambda_s$, i.e. $f'(n_{\rm max}') = \Lambda_s/m =  f(n_{\rm max}) = n_{\rm max}$, in both cases,
\begin{align*}
    N &= \sum_{n=1}^{n_{\rm max}} d_n = \sum_{n=1}^{n_{\rm max}} n^{p-1} \sim n_{\rm max}^{p}  \\
    &\overset{!}{=} \sum_{n=1}^{n_{\rm max}'} d_n' 
    = \sum_{n=1}^{n_{\rm max}'} 1 = n_{\rm max}' \,.
\end{align*}
It follows that $n_{\rm max}' = n_{\rm max}^p$, so that $f'(n_{\rm max}') = f'(n_{\rm max}^p) \overset{!}{=} n_{\rm max} \implies f'(n) = n^{1/p}$, since this must be true for any $n_{\rm max}$. While this nicely encodes the total number of states up to a certain level, it does carry all the information we need for our sum. 

For a string-like tower $f(n) = \sqrt{n}$ and $d_n \sim e^{\sqrt{n}}$ \footnote{More precisely, to leading order at large $n$ the degeneracy behaves as $d_n\sim n^{-11/2} e^{4\pi\sqrt{2n}}$ for Type II strings and $d_n\sim n^{-11/2}e^{2\pi(2+\sqrt{2})\sqrt{n}}$ for Heterotic strings \cite{Green:2012oqa}.}, which makes the series exponentially divergent,
\begin{equation}
    \sum_{n=1}^\infty \frac{e^{\sqrt{n}}}{n} \to \infty \,.
\end{equation}
Likewise, for a gonion-like tower \cite{Aldazabal:2000cn,Casas:2024ttx}, which behaves as $f(n)=\sqrt{n}$ and $d_n\sim O(1)$, the series becomes
\begin{equation}
    \sum_{n=1}^\infty \frac{d_n}{n} \geq \sum_{n=1}^\infty \frac{1}{n} \to \infty \,, 
\end{equation}
since the harmonic series is divergent. 

Let us clarify the way in which we use these towers as examples. When we refer to a string-like tower, we only mean a tower whose masses and degeneracies behave as an actual string tower would, but we \textit{do not consider} the actual physics of a tower of \textit{string states} in a full string theory (e.g. not all of those would be scalars). Likewise, when we refer to gonion-like towers, we do not mean to consider a tower of massive states that arise from open strings stretched between intersecting D-branes in a chiral D-brane model \cite{Aldazabal:2000cn,Casas:2024ttx}, but only an EFT tower of scalar fields with masses and degeneracies like those of such towers. It is important to keep this in mind, since considering the string theory and the D-brane models will introduce subtleties that we are ignoring in our computations and results\footnote{For example, if one naively considers a string-like tower in QFT and computes its contribution to the cosmological constant, one gets a divergent result. Nevertheless, computing this contribution in string theory renders the result finite due to properties that are inherently stringy and not captured by the field theory description.}. Here we want to assume that a tower of scalars with these properties appears and contributes to the graviton self-energy, and that we know \textit{nothing} more about this tower. This way we can analyse how much can be inferred from a ``tower state counting'' procedure akin to the original proposal of \cite{Dvali:2007wp} compared to what can be deduced only by knowing the UV origin of such towers.

It is interesting to note that the only convergent tower in this series of examples is the simplest KK tower associated with a circle compactification, while all stringy towers were divergent. Even if an infinite tower of states can be compatible with an EFT computation, it is probably pushing it too far to ask such a computation to be compatible with towers whose origin we know to be stringy in nature. While for a KK tower, both the tower and its UV origin can be described with appropriate EFTs, string oscillator towers originate from a vibrating string whose full description cannot be captured by a field theory. Our conclusion is that we cannot use a field theoretic description to study stringy towers. 

These observations are consistent with results found in the literature. From a top-down perspective, specific string theory examples identify the string scale itself, $m_s$, as the gravitational EFT cutoff\footnote{This is still often referred to as ``species scale''. However, in our nomenclature it corresponds to the EFT cutoff rather than the strong coupling scale.} in limits associated with string towers becoming light (e.g. see \cite{Castellano:2024bna} and references therein). This contradicts state counting arguments, which would predict a species scale above $m_s$. 
Conversely, if the string scale was actually the species scale, there would be no states in the string tower below that scale to use in the counting argument to begin with. 

Wanting to remain as agnostic and general as possible, it suffices for our purposes to identify one concrete example for which the sum converges and the procedure is well-defined. This is enough to argue that infinite towers of states are not inherently inconsistent with an EFT description, with the strong coupling scale lying partway along the tower. It is however tempting to expect more general KK towers to also be amenable to a field theory computation such as the one we are using here---while we cannot exclude this possibility, our current approach does not apply to those cases.
From this point onwards we will assume that $f(n)$ and $d_n$ is such that the sum in \eqref{eq:cutoff-tower} is convergent, so that the procedure is meaningful. 

\subsection{Computing the sum}
\label{sec:towers-summing}

Computing the relevant sum is more complicated for a tower of states than it was for $N$ identical scalars, in particular because different parts of the tower belong to different mass regimes. Nevertheless some progress can be made by splitting the sum into different sectors, in each of which we can use the approximations from the previous section.

\subsubsection{Light states}
\label{sec:towers-summing-light}

Given that $\rm{H}(\alpha)$ is a strictly decreasing function (cf. Fig.\ref{fig:exact-solution}), the leading contribution should come from the lightest states. To figure out how they contribute, imagine adding each state one by one, starting with the lightest, rather than including the full tower at once. 

If $m_1 = m\ll\Mp$, then the first state we add can be approximated as massless. Given $\alpha\ll 1\implies {\rm H}(\alpha)\approx 1$, we can use \eqref{eq:strong-large-q-N} with $N=1$ to determine $\cut{strong}^{\rm grav}$. If $m_2\ll\cut{strong}^{\rm grav}$ we can repeat the procedure, now using  \eqref{eq:strong-large-q-N} with $N=2$, and so on and so forth for each state in the tower. This procedure will be valid for the level $n$ as long as $\alpha_n\ll 1$ and therefore
\begin{equation}
    2\cdot f(n)\cdot m\ll\Mp\sqrt{\frac{240\pi^2}{N+240\pi^2}},
    \label{eq:ineq-light-states}
\end{equation}
for $N=\sum_{k=1}^nd_k$. We \textit{define} $n_L$ to be the highest level for which the above approximation is still valid, so that 
$N_L=\sum_{n=1}^{n_L}d_n$ can be interpreted as the number of light states in our theory.

Our crudest approximation for the strong coupling scale will then be given by
\begin{equation}
    \cut{strong}^\text{tower}\approx\Mp\sqrt{\frac{240\pi^2}{N_L+240\pi^2}} \,,
    \label{eq:cut-tower-light}
\end{equation}
which for large enough $N_L$ recovers the well-known behaviour $\cut{sp} \approx \frac{\Mp}{\sqrt{N_L}}$. Fig. \ref{fig:cutoff-KK} exemplifies this approximation for the KK tower case when $d_n=1$ and $f(n)=n$.

Notice that there are two approximations in the above expression. Firstly, we assume that the states above $n_L$ do not contribute at all. Secondly, we take \textit{all} the modes below $n_L$ to have masses $m_n\ll \cut{strong}^{\rm tower}$. In some sense the choice of $n_L$ should strike a balance between these two concerns---for $n_L$ closer to saturating \eqref{eq:ineq-light-states}, the assumption that all modes above $n_L$ are heavy (and thus negligible) improves, but it means that there might be modes below it that are not approximately massless; conversely if $n_L$ is smaller, then the modes below $n_L$ could be reasonably approximated as massless, but we would be less justified in neglecting all modes above $n_L$, since some might not be sufficiently heavy. The next two sections will improve on this approximation by including the modes above $n_L$ in the discussion as well.

It is worth remarking that the definition of $n_L$ is somewhat circular---we need to know $n_L$ to find $\cut{strong}^{\rm tower}$ but, simultaneously, we need to know the solution to verify \eqref{eq:ineq-light-states}. This circularity is present in most estimates of the species scale for a tower of states \cite{Castellano:2022bvr}. To some extent, we can justify it by including the states one by one starting from the lightest state until \eqref{eq:ineq-light-states} stops being valid, thereby defining $n_L$ and computing $\cut{strong}^{\rm tower}$ simultaneously. The key assumptions are then that any heavier states will have negligible contributions and all states up to level $n_L$ can be accurately approximated as massless. We shall revisit these two assumptions in turn in the following sections.

\subsubsection{Heavy states}
\label{sec:towers-summing-heavy}

In the previous section we argued that the modes above $n_L$ could be neglected. While it is true that $\rm{H}(\alpha)\to0$ as $\alpha\to\infty$ there is still an infinite number of them and even for a convergent tower their contribution might be significant. In this subsection we include them as very massive states by approximating $\alpha_n\gg1$, \textit{i.e.} we take $n_H=n_L+1$ using the notation introduced at the beginning of this section. The equation that determines $\cut{strong}^{\rm grav}$ is then
\begin{equation}
    q^2\approx\Mp^2-\frac{q^2}{240\pi^2}\qty(N_L+\frac{3 q^2}{28 m^2}\sum_{n=n_H}^\infty\frac{d_n}{f(n)^2}) \,.
\end{equation}
For ease of notation let us define
\begin{equation}
    \sigma_H=\frac{3}{7}\sum_{n=n_H}^\infty\frac{d_n}{f(n)^2} \,,
    \label{eq:def-sigmaH}
\end{equation}
in terms of which the strong coupling scale is given by
\begin{align}
    \cut{strong}^\text{tower}
    &\approx\Mp\sqrt{\frac{240\pi^2}{N_L+240\pi^2}}\cdot\sqrt{2\xi_H\qty(-1+\sqrt{1+\frac{1}{\xi_H}})}
    \label{eq:cut-tower-light-heavy}
\end{align}
with
\begin{equation}
    \xi_H = \qty(\frac{N_L+240\pi^2}{240\pi^2})^2\cdot\frac{240\pi^2}{\sigma_H}\cdot\frac{m^2}{\Mp^2}
    \label{eq:def-xiH}
\end{equation}

This solution is very similar to what we found for $N$ identical scalars with masses above the cutoff \eqref{eq:strong-coupling-scale-vs-xi}, except that the prefactor is $\cut{strong}^{\rm tower}$ in the light-state-only approximation \eqref{eq:cut-tower-light} rather than $\Mp$ and $\xi\to\xi_H$ as defined in \eqref{eq:def-xiH} through a combination of the number of light states and the sum over heavy states.

In this solution, since $n_H = n_L + 1$, both $N_L$ and $\sigma_H$ are functions of $n_L$. The number of light states $N_L$ obviously increases as $n_L$ increases and $\sigma_H$ decreases as fewer modes are included in the series \eqref{eq:def-sigmaH}. Therefore the parameter $\xi_H$ increases for larger $n_L$%
---if we choose to include more modes as light-states, the $\xi_H$-dependent factor in \eqref{eq:cut-tower-light-heavy} will approach $1$, so that $\cut{strong}^{\rm tower}$ is approximately given by \eqref{eq:cut-tower-light}. Conversely, if we include fewer states as massless then $n_L$ will be smaller, which implies that $\xi_H$ will be smaller. In this case $\cut{strong}^\text{tower}$ will approach $0$ and the solution will significantly deviate from \eqref{eq:cut-tower-light}.

\begin{figure}
    \centering
    \includegraphics[width=0.7\linewidth]{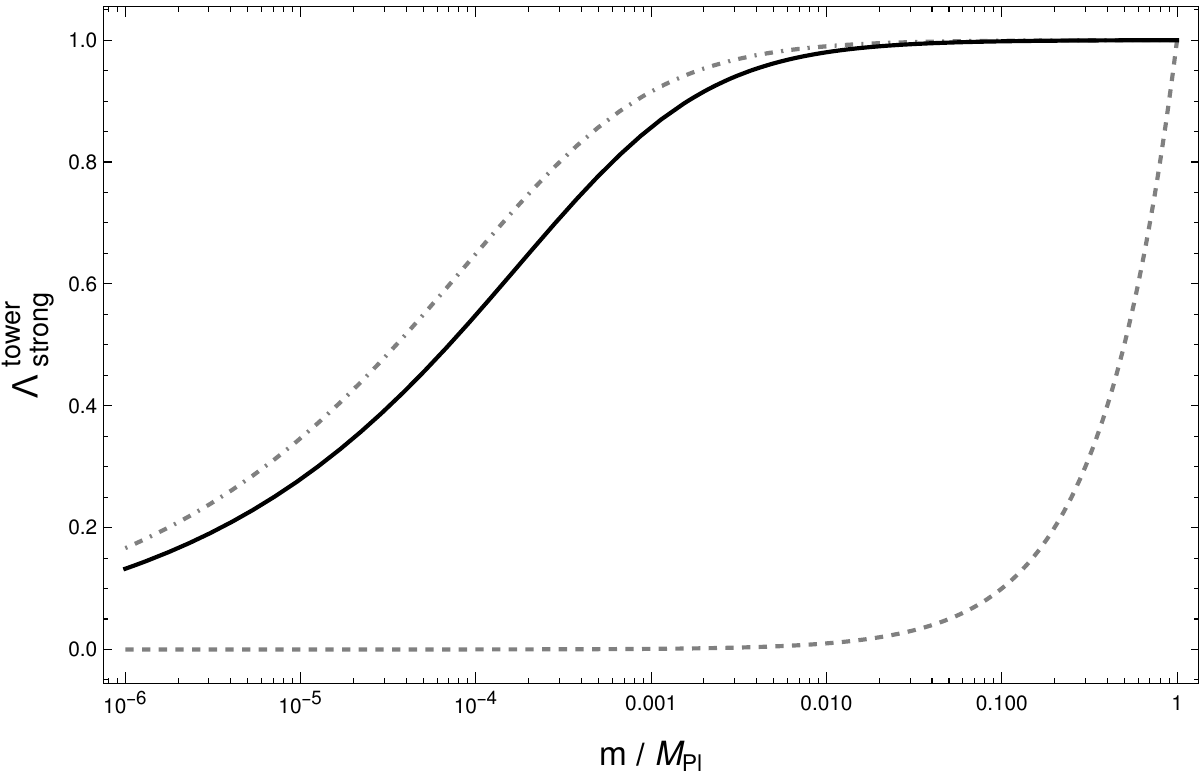}
    \caption{Strong coupling scale as a function of the mass of the lightest scalar for a KK tower. The solid curve includes both light and heavy states, corresponding to the solution \eqref{eq:cut-tower-light-heavy}. The dot-dashed line only includes the light states and corresponds to the solution \eqref{eq:cut-tower-light}.
    Note how the effect of the heavy states is to always lower the cutoff and that this effect becomes negligible for heavier towers. The dashed line represents the curve $\cut{strong}^\text{tower}=m$ to illustrate that the cutoff always lies part way along the tower.}
    \label{fig:cutoff-KK}
\end{figure}

To minimize the error in our approximations we should choose $n_L$ such that all states below it obey $\alpha_n\ll1$, \textit{i.e.} lie far below $\cut{strong}^\text{tower}$, and such that all states above it obey $\alpha_n\gg1$, \textit{i.e.} lie far above the $\cut{strong}^\text{tower}$. This is of course not possible in the way we have set up our problem. By choosing $n_H=n_L+1$ there will definitely be modes for which $\alpha_n\approx 1$ and therefore neither approximation holds.
The best we can do is to demand
\begin{equation}
    f(n_L)\cdot m 
    < \Mp\sqrt{\frac{240\pi^2}{N_L+240\pi^2}} \cdot\sqrt{2\xi_H\qty(-1+\sqrt{1+\frac{1}{\xi_H}})}
    < f(n_L+1)\cdot m \,,
\end{equation}
to balance the error in the light and heavy states. Fig. \ref{fig:cutoff-KK} exemplifies this method for the KK tower with $d_n=1$ and $f(n)=n$, in comparison with the ligh-states-only approximation. While the rough shape of the solution is similar, the contribution of the heavy states lowers $\cut{strong}^{\rm tower}$ with respect to the light-states-only result.
\begin{figure}
    \centering
    \includegraphics[width=0.7\linewidth]{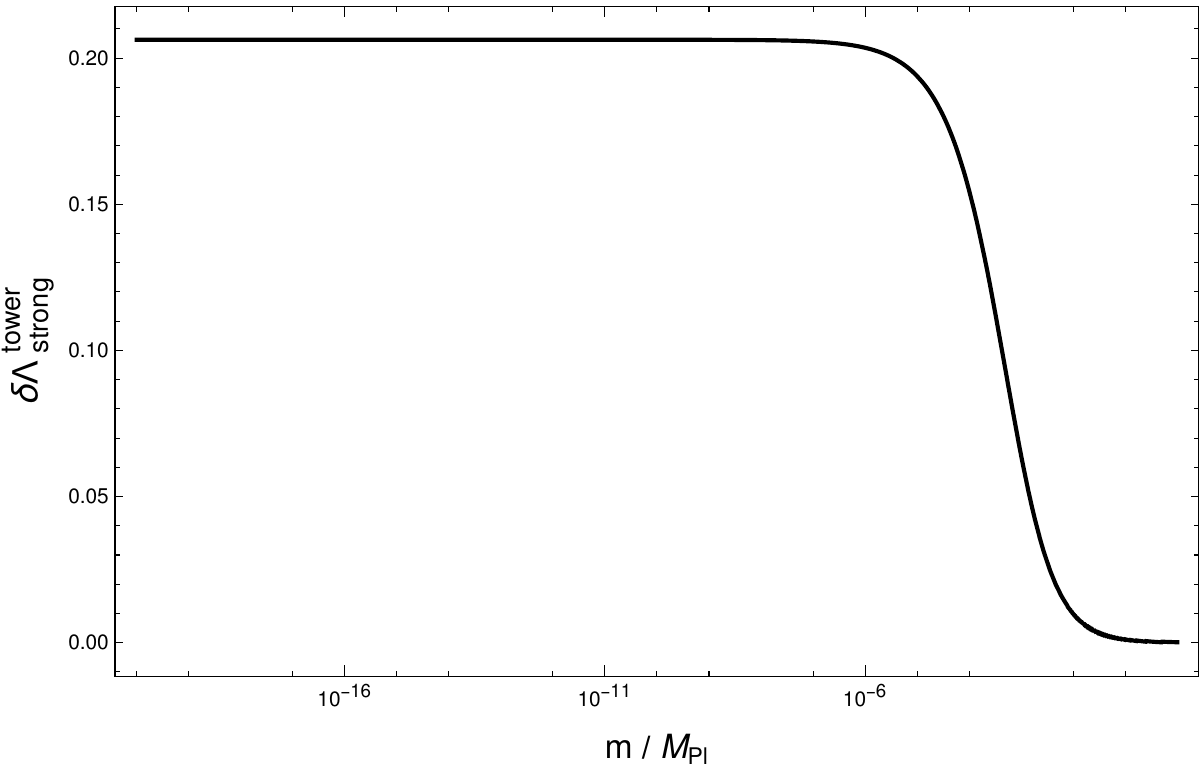}
    \caption{Plot of the difference between $\cut{strong}^{\rm grav}$ obtained from \eqref{eq:cut-tower-light-heavy} including the correction from the heavy states and the one obtained from the light-states-only estimate \eqref{eq:cut-tower-light}. Note that for $m\sim\Mp$, the two result agree---this is simply reflecting the fact that $\cut{strong}^{\rm grav}\to\Mp$ in this limit. In the opposite limit, for $m\ll\Mp$, the difference saturates at $\sim 20\%$.}
    \label{fig:tower-ratio-heavy-vs-light}
\end{figure}
In the following section we will attempt to improve on this approximation by treating the modes for which $\alpha_n\approx 1$ separately.

Before we do so, let us just note a curious result. Setting $N_L=0$ and $n_H=1$ corresponds to a $\cut{strong}^\text{tower}$ lying below the mass of the first state, $m$. The consistency of the approximation $\alpha_{n_H}=\alpha_1\gg1$ then requires 
\begin{equation}
    \frac{m^2}{\Mp^2}\gg\frac{60\pi^2}{\sigma_H+240\pi^2} \,.
\end{equation}
Unless $\sigma_H$ is comparable to $240\pi^2$ the RHS is approximately $1/4$ and thus the strong coupling scale can only lie below the whole tower if the first state is already close to $\Mp$, otherwise $\cut{strong}^{\rm grav}$ will always lie part way along the tower (\textit{e.g.} for a KK tower with $m \ll\Mp$ we have $\sigma_H=\frac{\pi^2}{7}$ and therefore $\cut{strong}^{\rm grav}$ \textit{should} be higher than the KK scale).

This might seem quite disturbing. Cutting the tower part way seems to leave very little separation between the last mode included and the first not included in the computation of $\cut{strong}^{\rm grav}$. The key point is that we are \textit{not} excluding the modes above $\cut{strong}^\text{tower}$---every mode in the tower is included in the theory. The scale separation we require is between the external graviton momenta and the EFT scale\footnote{And this could also differ from the strong coupling scale! Without knowledge of the $a_i$ we have no access to the true EFT scale. In any case, the validity of the loop computation (i.e. the perturbative expansion) requires the graviton momentum to at least be below this strong coupling scale.}. As long as we do not include tower states in external legs, there is no need for scale separation between the masses of the scalars above and below $\cut{strong}^\text{tower}$. Furthermore, let us reiterate that there is no mathematical inconsistency in including modes above $\cut{strong}^\text{tower}$ in internal lines. As we discussed at the end of Section \ref{sec:N-scalars-large-mass-limit}, one simply has to keep in mind the possibility that these only contribute to UV sensitive observables and will therefore not provide us with useful information.

\subsubsection{Intermediate states}
\label{sec:towers-summing-intermediate}

Finally, let us try to account for those states in the tower that do not satisfy either of the previous approximations, \textit{i.e.} the modes for which $\alpha_n\approx 1$. In this regime, we can approximate ${\rm H}(\alpha)\approx c_1 - c_2\alpha$ (with $c_1\approx 0.46$, $c_2\approx 0.18$) and the equation that determines $\cut{strong}^{\rm grav}$ becomes  
\begin{align}
    q^2 &\approx \Mp^2 - \frac{q^2}{240\pi^2}\cdot\left\{N_L  + \sum_{n=n_L+1}^{n_H-1} d_n\cdot\left(c_1 - c_2\cdot f(n)^2\frac{4m^2}{q^2}\right) + \frac{q^2}{4m^2}\cdot\sigma_H \right\} \,.
\end{align}
Note that the first new term is simply a correction to the number of light states. If there are $N_I = \sum_{n_L+1}^{n_H-1} d_n$ intermediate states, then this correction can be interpreted as a correction to the number of light states as $N_L\to N_{L+I} = N_L + c_1\cdot N_I$, where each intermediate state counts as rougly half of a truly light state. The second new term involves the sum
\begin{equation}
    \vartheta_I = \sum_{n=n_L+1}^{n_H-1} d_n\cdot f(n)^2 \,,
\end{equation}
in terms of which the equation becomes
\begin{align}
    q^2 &\approx \Mp^2 - \frac{q^2}{240\pi^2}\cdot\left\{N_{L+I} - c_2\cdot \frac{4m^2}{q^2} \cdot \vartheta_I + \frac{q^2}{4m^2}\cdot\sigma_H \right\} \,.
\end{align} 
The solution can then be written as
\begin{align}
    \cut{strong}^{\rm grav} = \Mp \sqrt{\frac{240\pi^2}{N_{L+I} + 240\pi^2}}\sqrt{2\xi_H\left(-1 + \sqrt{1 + \frac{1+\xi_I}{\xi_H}} \right)} \,,
\end{align}
where we have defined, analogously to $\xi_H$,
\begin{equation}
    \xi_I = c_2\cdot\frac{\vartheta_I}{60\pi^2}\cdot\frac{m^2}{\Mp^2} \,.
\end{equation}
Once again we can see the effects of the intermediate states as a correction to the previous result. In contrast with the heavy-state correction, the intermediate states appear in two different ways: as a correction to the number of light-states, $N_{L+I}$, and as a correction to the contribution of the heavy states. This is not entirely surprising, since the intermediate states define the split between light and heavy states, which makes it reasonable to expect corrections to both contributions. 

The first contribution is simply taking into account the presence of these states with masses near $\cut{strong}^{\rm grav}$ in the counting of light states. We can understand it as getting as close as possible to this scale while still counting as light ``enough''. 

On the other hand, the more states we count as being close to $\cut{strong}^{\rm grav}$, the larger $\vartheta_I$ will be and therefore the larger $\xi_I$ will become. Note that in the presence of the intermediate states, the two limits we considered for $\xi_H$ are affected---for small $\xi_H$ (i.e. when we increase $n_H$), the square root contribution approaches $\sqrt{2}(1+\xi_I)^{1/4}\xi_H^{1/4}$ (c.f. \eqref{eq:strong-coupling-scale-vs-xi-small-xi}); for large $\xi_H$ (i.e. when we decrease $n_H$), this contribution approaches $\sqrt{1+\xi_I}$ (c.f. \eqref{eq:strong-coupling-scale-vs-xi-large-xi}).

While it is difficult to give a precise definition of which states count as intermediate, we can build an intuition for when there can be a large number of them. There can only be many states in a tower whose masses are near the scale $\cut{strong}^{\rm grav}$ if $m/\cut{strong}^{\rm grav}\ll 1$ or if the degeneracy of a few levels with masses near this scale is very large. 
The key point we want to make is that including these intermediate terms, to first approximation, only corrects the contributions from light and heavy states, without qualitatively changing the results. Ultimately, the best way to account for all modes is to solve the equation numerically, which avoids the arbitrariness in our definitions of $n_I$ and $n_H$ (and consequently $N_I$).

\subsection{Towers and EFTs}
\label{sec:towers-and-EFTs}

Let us close the analysis by focusing on the main goal of this work. We wanted to understand whether a bottom-up determination of the ``species scale'' for an infinite tower of states was compatible with the regime of validity of the EFT. If it was, we wanted to learn how much could be inferred from the bottom-up perspective alone, without any knowledge of the UV origin of the tower. 

As we already argued, the computation itself is compatible with the EFT framework as long as one keeps the infinite tower in the theory and keeps the graviton external momenta below the EFT cutoff. A key observation is that this cutoff should not be related to the mass scale of the tower, since the tower was never integrated out to generate operators suppressed by that scale. It is also not necessarily the strong coupling scale that we compute, which is a logically independent scale (cf. discussion in Section \ref{sc:EFT-background}). Additionally, the tower must be a convergent one, in the sense described in Section \ref{sec:towers-convergence}, which appears to require at least an EFT origin for the tower of states (in contrast, for example, with a tower of string oscillator modes). One might try to use a simple regularisation technique (e.g. $\zeta$-function regularisation) to evaluate the divergent sums, but one must be careful to ensure our method still applies. It would be interesting to extend our results to include these towers as well.

In fact, for a string oscillator tower, the scale obtained through a state counting argument \textit{will not} correspond to the top-down expectation that the effective description breaks down at the string scale $m_s$ (or from an estimate using black hole arguments) \cite{Blumenhagen:2023yws,Blumenhagen:2023tev}. Indeed, from the counting argument one obtains a scale \textit{above} $m_s$, enhanced by a logarithmic factor $\log\qty(\Mp/m_s)$ commonly interpreted as a correction \cite{Castellano:2024bna,Castellano:2022bvr}. Since this ``correction'' is what counts the large number of species responsible for lowering the strong coupling scale with respect to $\Mp$ (at least from the bottom-up perspective), it would be inconsistent to simultaneously disregard it in $\cut{strong}$ (or $\cut{sp}$). We would argue instead that the field theory computation should simply not be trusted for the case of a string tower, in accordance with our previous comments. While the top-down studies of the quantum gravity scale (e.g. as the scale suppressing higher-curvature corrections computed within a specific string theory setup) are in no way affected by this conclusion, one should not force a ``match'' between these results and an EFT estimate based on counting arguments for string towers.

The original species scale proposal \cite{Dvali:2007hz,Dvali:2007wp} gives what we called the ``light-states-only'' approximation \eqref{eq:cut-tower-light} that essentially neglects the modes above the strong coupling scale. Even though these heavy states contribute in a subleading way, a large number of them might give a significant contribution. Including them in our analysis, together with the light modes, slightly reduces the scale compared to the ``light-states-only'' estimate \eqref{eq:cut-tower-light-heavy}. However, for a large enough number of light states, this correction will be small and the strong coupling scale will still behave as $\Mp/\sqrt{N_L}$.
As an example of a convergent tower, we used a Kaluza-Klein tower that arises from the compactification of an extra dimension to illustrate these results. As expected, for $m_{\rm KK}\ll\Mp$, the result is very close to the light-state-only approximation. 
\newpage
\section{Conclusion}
\label{sec:conclusion}

The idea that quantum gravity effects lie far above the energy scales we might probe has for a long time been used as an argument against the phenomenological interest of proposals such as String Theory. Perhaps more importantly, it meant that we could not hope for experimental clues at low enough energies that could tell us where to look for such a fundamental theory. On the other hand, the huge Landscape of possible vacua turned a highly constrained fundamental theory into an unwieldy number of low-energy effective theories \cite{Susskind:2003kw,Taylor:2015xtz}, from which one could hardly extract specific testable predictions. Amidst this state of affairs, the Swampland programme \cite{Vafa:2005ui} took off as an approach that studied not specific elements of the Landscape, but rather universal patterns that could potentially hint at fundamental principles of quantum gravity \cite{Palti:2019pca,vanBeest:2021lhn,Brennan:2017rbf,Grana:2021zvf,Agmon:2022thq,VanRiet:2023pnx}. 

One such pattern is the appearance of towers of states that become exponentially lighter in the asymptotic limits of the moduli space of the effective theories \cite{SdC}. Together with the observation that the scale at which an effective description of gravity breaks down can be much smaller in the presence of a large number of fields \cite{Han_2005,Dvali:2007hz,Dvali:2007wp,Arkani-Hamed:2005zuc,Distler:2005hi,Dvali:2010vm}, the conjectures led to a wave of work on a key question: what is the \textit{real} scale of quantum gravity? 

However the interplay between EFT methods and infinite towers of massive states raised concerns of validity and consistency \cite{Burgess:2023pnk}. At face value, there appears to be a tension between the scale separation required for the validity of the EFT expansion and the very nature of these infinite towers that seems to prevent such a clean separation of scales. In particular, the perturbative derivation of the so-called ``species scale'' \cite{Donoghue_1994,Han_2005,Anber:2011ut,Aydemir:2012nz,Calmet:2017omb,Caron-Huot:2022ugt}, $\cut{sp}$, in the presence of towers, relies on the validity of this interplay and it is therefore important to understand whether it can be trusted.

In this paper we have outlined in full detail the perturbative argument for $\cut{sp}$, with particular emphasis on the assumptions and regimes of validity of the result. Having understood this in detail, we could then study in Section \ref{sec:towers} the full effect of an infinite tower of states on this scale. We argue that the seeming tension between the use of towers and the validity of the EFT arises only because the answer to the following question is not always clear:
\begin{itemize}
    \item What is the EFT one must consider in the computation of the ``species scale''?
    \item What is the ``species scale'' \textit{really} and how does it relate to an EFT cutoff?
\end{itemize}

With regards to cutting infinite towers and the lack of scale separation necessary to define the EFT, one often conflicts different effective theories when thinking about the towers. We are so used to integrating out infinite towers of states in order to define a valid low-energy EFT that we instinctively take the effective theory to be the one where the tower was integrated out. The whole framework of EFTs then tells us (correctly) that this theory should break down at energies above the mass of the first state to be integrated out \cite{Burgess:2020tbq}. Processes involving energies above this cutoff would therefore not find a valid answer within this EFT. On the other hand, nor could we perform the 1-loop computation required to run the perturbative argument for the ``species scale''---once the tower has been integrated out, there is no tower to speak of. 

Thus the EFT must be the one that includes the infinite tower, the one where the tower was not integrated out and so no infinite series of higher-order operators is controlled by the mass scale of the tower. While one may worry that an effective theory with infinitely many massive states is not well defined, there is no reason to believe that it should break down at the scale of the first state. With this in mind, we ensure the validity of the EFT by keeping the external momenta involved in the process below the very scale one is interested in finding, the scale at which the effective description of gravity breaks down \textit{due to the states within the EFT}. 

Furthermore, one must be careful with the physical interpretation of the ``species scale''. Due to the nature of gravitational interactions, the distinction between EFT cutoff and a strong coupling scale is murkier than it would have been in a theory like QED. From a bottom-up perspective and based solely on the perturbative argument outlined in this work, one can only take the ``species scale'' as an upper bound on the regime of validity of our EFT. As we showed in Section \ref{sec:1-loop-computation}, from this point-of-view the strong coupling scale and the EFT cutoff associated with higher-curvature corrections to GR are logically distinct, and the latter cannot be determined from this argument alone. We must therefore keep in mind the possibility that some other EFT cutoff appears at a different scale, possibly due to other states (unrelated to the tower) having been integrated out.

What propelled the study of this scale within the Swampland programme was again the observation of a pattern: even if the bottom-up argument does not guaranty a correspondence between strong coupling and the EFT cutoff, many examples were found where the ``species scale'' was indeed the scale controlling higher-curvature corrections. \textit{Observing} this connection requires a top-down perspective, through the computation of such corrections in a controlled string-theoretic setup \cite{Castellano:2023aum,vandeHeisteeg:2023dlw,Calderon-Infante:2025ldq}---it constitutes the \textit{gathering of evidence} within a specific example of quantum gravity. Identifying this scale from a bottom-up perspective (which typically leads one to make assumptions about fundamental properties of the UV theory) would not only strengthen our confidence on a quantum gravity scale much lower than $\Mp$, but also likely provide a hint towards the fundamental principles that underly the top-down observations \cite{Caron-Huot:2021rmr,Caron-Huot:2022ugt,Caron-Huot:2024lbf,Caron-Huot:2024tsk,Alberte:2020jsk}.

Ultimately there is no fundamental conflict between infinite towers and the EFT framework when it comes to the ``species scale''. For certain towers, such as a simple KK tower from a circle compactification, the perturbative argument can be used consistently to estimate the strong coupling scale; for others, like the string tower, the argument fails and the perturbative EFT estimate will not match the scale expected from top-down reasoning.  
Furthermore, the ``species scale'' is not guaranteed, by the perturbative argument alone, to be the scale suppressing higher-curvature corrections, but this is mostly of no consequence for all top-down studies of this quantum gravity scale---it rather becomes important if one tries to push the bottom-up argument too far.

\acknowledgments
We would like to thank
Christian Aoufia, Alberto Castellano, Sumer Jaitly, Joel Karlsson, Andriana Makridou, Miguel Montero, Fernando Quevedo, Ignacio Ruiz, Angel Uranga, Irene Valenzuela, and Max Wiesner for useful discussions. We are also grateful to Alberto Castellano, Miguel Montero,
and Angel Uranga for valuable feedback on the paper. 
B.B. acknowledges the support of an Atraccion del Talento Fellowship 2022--T1/TIC-23956 from Comunidad de Madrid, the grant Europa Excelencia EUR2024-153547, and the Spanish Research Agency (Agencia Estatal de Investigacion) through the grants IFT Centro de Excelencia Severo Ochoa CEX2020-001007-S and PID2021-123017NB-I00. J.M. acknowledges funding from the FWO projects G003523N, G094523N, and G0E2723N, as well as by the Odysseus grant G0F9516N from the FWO, and the KU Leuven grant C14/21/086.

\newpage
\appendix
{
\allowdisplaybreaks

\section{Details of 1-loop computation}
\label{ap:details-1loop-computation}

In this appendix we provide the explicit details of the 1-loop computation. As a consistency check, we will confirm the validity of the Ward identity at each stage of the computation. This verification was in fact crucial to ensure the absence of typos.

\subsection{Initial check}

We begin by testing if we had no typos in deriving an applying the Feynman rules by contracting the propagator with $q^\mu$ to verify the Ward identity. Analising the first factor in $A_{\mu\nu\rho\sigma}$ we find
\begin{align}
    &q^\mu(2k_\mu k_\nu-k_\mu q_\nu-q_\mu k_\nu-\eta_{\mu\nu}(k^2-k\vdot q+m^2)) \nonumber\\
    =&~2(k\vdot q)k_\nu-(k\vdot q)q_\nu-q^2 k_\nu-q_\nu(k^2-(k\vdot q)+m^2) \nonumber\\
    =&-k_\nu(q^2-2(k\vdot q))-q_\nu(k^2+m^2)+k_\nu(k^2+m^2)-k_\nu(k^2+m^2) \nonumber\\
    =&-k_\nu((k-q)^2+m^2)+(k_\nu-q_\nu)(k^2+m^2) \,,
\end{align}
where in the penultimate line we added and subtracted $k_\nu (k^2+m^2)$.

All in all,
\begin{align}
    q^\mu (d_{33})_{\mu\nu\rho\sigma} &= \frac{1}{\Mp^2}\int\frac{\dd[4]{k}}{(2\pi)^4}\frac{q^\mu A_{\mu\nu\rho\sigma}}{(k^2+m^2-i\epsilon)((k-q)^2+m^2-i\epsilon)} \nonumber \\
    &= \frac{1}{\Mp^2}\cdot 2\int\frac{\dd[4]{k}}{(2\pi)^4}\frac{-k_\nu\qty(2k_\rho k_\sigma-k_\rho q_\sigma-q_\rho k_\sigma-\eta_{\rho\sigma}(k^2-k\vdot q+m^2))}{k^2+m^2-i\epsilon} \,,
\end{align}
where we have used the invariance under $k\to q-k$ to double up the two terms we got in $q^\mu A_{\mu\nu\rho\sigma}$ and then cancel one of the factors in the denominator. Since the denominator is invariant under $k\to -k$, terms with odd powers of $k$ do not contribute and we obtain
\begin{align}
    q^\mu (d_{33})_{\mu\nu\rho\sigma} &= \frac{1}{\Mp^2}\cdot 2\int\frac{\dd[4]{k}}{(2\pi)^4}\frac{k_\nu k_\rho q_\sigma+k_\nu q_\rho k_\sigma-k_\nu \eta_{\rho\sigma}(k\vdot q)}{k^2+m^2-i\epsilon} \,.
\end{align}
Now we use rotational invariance to replace $k^\mu k^\nu\to\frac{1}{d}k^2 \eta^{\mu\nu}$, and obtain
\begin{align}
    q^\mu (d_{33})_{\mu\nu\rho\sigma} &= \frac{1}{\Mp^2}\cdot\frac{4}{d}\cdot\frac12(\eta_{\nu\rho}q_\sigma+\eta_{\nu\sigma}q_\rho-\eta_{\rho\sigma}q_\nu)
    \cdot\mu^{4-d}\int\frac{\dd[d]{k}}{(2\pi)^d}\frac{k^2}{k^2+m^2-i\epsilon} \nonumber\\
    &=\frac{1}{\Mp^2}\cdot q^\mu P_{\mu\nu\rho\sigma}\cdot \mu^{4-d}\int\frac{\dd[d]{k}}{(2\pi)^d}\frac{\frac{4}{d}k^2}{k^2+m^2-i\epsilon} \,,
\end{align}
Adding the contribution from the second diagram (with the quartic coupling)
\begin{equation}
    (d_4)_{\mu\nu\rho\sigma}=\frac{1}{\Mp^2}\mu^{4-d}\int\frac{\dd[d]{k}}{(2\pi)^d}\frac{\qty(\big(1-\frac{4}{d}\big)k^2 + m^2)P_{\mu\nu\rho\sigma}}{k^2+m^2-i\epsilon} \,,
\end{equation}
we find
\begin{equation}
    q^\mu(d_{33}+d_4)_{\mu\nu\rho\sigma}=\frac{1}{\Mp^2}q^\mu P_{\mu\nu\rho\sigma}\cdot \mu^{4-d}\int\frac{\dd[d]{k}}{(2\pi)^d}\frac{k^2+m^2}{k^2+m^2-i\epsilon}
    \stackrel{?}{=} 0 \,.
\end{equation}
Let us evaluate that last integral in dim-reg. We have (now in Euclidean signature)
\begin{align}
    \mu^{4-d}\int\frac{\dd[d]{k}}{(2\pi)^d}\frac{k^2+m^2}{k^2+m^2} 
    &=~\mu^{4-d}\frac{2\pi^{\frac{d}{2}}}{\Gamma\qty(\frac{d}{2})}\qty(\int_0^\infty\frac{\dd{k}}{(2\pi)^d}\frac{k^{d+1}}{k^2+m^2} + m^2\int_0^\infty\frac{\dd{k}}{(2\pi)^d}\frac{k^{d-1}}{k^2+m^2})  \nonumber\\
    &=\mu^{4-d}\frac{m^d}{(4\pi)^{\frac{d}{2}}}\qty(\frac{\Gamma\qty(-\frac{d}{2})\Gamma\qty(1+\frac{d}{2})}{\Gamma\qty(\frac{d}{2})}+\frac{\Gamma\qty(1-\frac{d}{2})\Gamma\qty(\frac{d}{2})}{\Gamma\qty(\frac{d}{2})}) \nonumber\\
    &=\mu^{4-d}\frac{m^d}{(4\pi)^{\frac{d}{2}}}\qty(\Gamma\qty(-\frac{d}{2})\cdot\frac{d}{2}-\frac{d}{2}\cdot\Gamma\qty(-\frac{d}{2}))=0 \,.
\end{align}
Note how we did not need to take the $d\to 4$ limit (or $q^2\to$ anything for that matter) to prove this identity. As long as the integral is computed correctly this identity should hold before any limits are taken.

\subsection{Feyman Parametrisation}

We use the following standard result,

\begin{equation}
    \frac{1}{AB}=\int_0^1\dd{x}\frac{1}{\qty(x A+(1-x)B)^2} \,,
\end{equation}
which holds for $A>0,~B>0$ (\textit{i.e.} in Euclidean signature), or $A=a-i\epsilon,~B=b-i\epsilon$ (\textit{i.e.} Lorentzian signature). We can therefore write

\begin{align}
    (d_{33})_{\mu\nu\rho\sigma} &= \frac{1}{\Mp^2}\int\frac{\dd[4]{k}}{(2\pi)^4}\frac{A_{\mu\nu\rho\sigma}}{(k^2+m^2-i\epsilon)((k-q)^2+m^2-i\epsilon)} \nonumber\\
    &= \frac{1}{\Mp^2}\int\frac{\dd[4]{k}}{(2\pi)^4}\int_0^1\dd{x}\frac{A_{\mu\nu\rho\sigma}}{\qty[x\qty((k-q)^2+m^2-i\epsilon)+(1-x)(k^2+m^2-i\epsilon)]^2} \nonumber\\
    &= \frac{1}{\Mp^2}\int\frac{\dd[4]{k}}{(2\pi)^4}\int_0^1\dd{x}\frac{A_{\mu\nu\rho\sigma}}{\qty[k^2+m^2-i\epsilon+x(-2k\vdot q+q^2)]^2} \,,
\end{align}
and change variables by defining
\begin{equation}
    \ell_\mu=k_\mu- xq_\mu\Longrightarrow k_\mu=\ell_\mu+x q_\mu \,,
\end{equation}
to obtain
\begin{align}
    (d_{33})_{\mu\nu\rho\sigma} &= \frac{1}{\Mp^2}\int\frac{\dd[4]{\ell}}{(2\pi)^4}\int_0^1\dd{x}\frac{A_{\mu\nu\rho\sigma}}{\qty[\ell^2+m^2+x(1-x)q^2-i\epsilon]^2} \,.
\end{align}
The numerator becomes
\begin{align}
    A_{\mu\nu\rho\sigma} &= \qty(2k_\mu k_\nu-k_\mu q_\nu-q_\mu k_\nu-\eta_{\mu\nu}(k^2-k\vdot q+m^2))\cdot \nonumber\\
    &\quad\cdot\qty(2k_\rho k_\sigma-k_\rho q_\sigma-q_\rho k_\sigma-\eta_{\rho\sigma}(k^2-k\vdot q+m^2)) \nonumber\\
    &= \Big(2\ell_\mu \ell_\nu-2x(1-x)q_\mu q_\nu-2(1-2x)\ell_{(\mu}q_{\nu)}-\nonumber\\
    &\quad -\eta_{\mu\nu}\qty(\ell^2-x(1-x)q^2-(1-2 x)(\ell\vdot q)+m^2)\Big)\cdot(\mu\nu\leftrightarrow\rho\sigma) \,.
\end{align}
With these changes we have lost the explicit symmetry of the integrand under $k_\mu\to q_\mu-k_\mu$,
\begin{align}
    A_{\mu\nu\rho\sigma} 
    &= \Big((2\ell_\mu \ell_\nu - \eta_{\mu\nu}\ell^2) - x(1-x)(2q_\mu q_\nu -\eta_{\mu\nu}q^2) - \eta_{\mu\nu}m^2 \nonumber\\
    &\quad\quad - (1-2x)(2\delta^\alpha_{(\mu} \delta^\beta_{\nu)} - \eta_{\mu\nu}\eta^{\alpha\beta})\ell_\alpha q_\beta  \Big)\cdot(\mu\nu\leftrightarrow\rho\sigma),
\end{align}
 which is now realised as a symmetry under the combined transformation
\begin{equation}
    \ell_\mu\to-\ell_\mu,\quad x\to1-x \,.
\end{equation}

\subsubsection{Checking Ward Identity}

To ensure we have not made any mistakes let us double check that the Ward identity still holds. Note how the quartic integral is unaffected by this trick. Once more focusing on the first factor of $A_{\mu\nu\rho\sigma}$ we find
\begin{align}
    &q^\mu\Big(2\ell_\mu \ell_\nu-2x(1-x)q_\mu q_\nu-2(1-2x)\ell_{(\mu}q_{\nu)} 
    -\eta_{\mu\nu}\qty(\ell^2-x(1-x)q^2-(1-2 x)(\ell\vdot q)+m^2)\Big) \nonumber \\
    &= \ell_\nu\qty(2(q\vdot\ell)-(1-2x)q^2) - q_\nu\qty(\ell^2 + m^2 + x(1-x)q^2) \,.
\end{align}
The first term is quadratic in $\ell$ so only terms which are even in $\ell$ coming from the other factor will contribute. The second term is odd in $\ell$ so only terms which are odd in $\ell$ in the second factor contribute. The remaining terms directly cancel a propagator so only terms even in $\ell$ will contribute. Let us consider these three integrals separately.

For reference, the terms even in $\ell$ in the second factor are
\begin{align}
    \qty(2\ell_\rho \ell_\sigma-2x(1-x)q_\rho q_\sigma-\eta_{\rho\sigma}\qty(\ell^2-x(1-x)q^2+m^2)) \,,
\end{align}
and the terms odd in $\ell$ are
\begin{align}
    -(1-2x)\qty(2\ell_{(\rho}q_{\sigma)}-\eta_{\rho\sigma}(\ell\vdot q)) \,.
\end{align}

Let us look at these integrals in turn (where $d$ is kept unfixed and in Euclidean signature). We will need the substitutions (valid under the integrals in $\ell$)
\begin{subequations}
\begin{align}
    \ell_\mu\ell_\nu&\to\frac{1}{d}\ell^2\eta_{\mu\nu} \,, \\
    \ell_\mu\ell_\nu\ell_\rho\ell_\sigma&\to\frac{1}{d(d+2)}\ell^4\qty(\eta_{\mu\nu}\eta_{\rho\sigma}+\eta_{\mu\rho}\eta_{\nu\sigma}+\eta_{\mu\sigma}\eta_{\rho\nu}) \,.
\end{align}
\end{subequations}

\noindent\textbf{First term}
\begin{align}
    I_1=\int\frac{\dd[d]{\ell}}{(2\pi)^d}\int_0^1\dd{x} \frac{ 2\ell_\nu(q\vdot\ell)\qty(2\ell_\rho \ell_\sigma-2x(1-x)q_\rho q_\sigma-\eta_{\rho\sigma}\qty(\ell^2-x(1-x)q^2+m^2))}{\qty[\ell^2+m^2+x(1-x)q^2]^2}
\end{align}
Let us split this even further by powers of $\ell$ (note that the $\eta_{\rho\sigma}$ term has the opposite sign in the $q^2$ so it does not cancel with the denominator),

\begin{align}
    I_1^{(4)} &= \int\frac{\dd[d]{\ell}}{(2\pi)^d}\int_0^1\dd{x}\frac{ 2\ell_\nu(q\vdot\ell)\qty(2\ell_\rho \ell_\sigma-\eta_{\rho\sigma}\ell^2)}{\qty[\ell^2+m^2+x(1-x)q^2]^2}\nonumber  \\
    &= 2q^\mu\qty(2\delta^\lambda_\rho \delta^\tau_\sigma-\eta_{\rho\sigma}\eta^{\lambda\tau})\int\frac{\dd[d]{\ell}}{(2\pi)^d}\int_0^1\dd{x}\frac{\ell_\mu\ell_\nu\ell_\lambda\ell_\tau}{\qty[\ell^2+m^2+x(1-x)q^2]^2} \nonumber\\
    &= 2q^\mu\qty(2\delta^\lambda_\rho \delta^\tau_\sigma-\eta_{\rho\sigma}\eta^{\lambda\tau})\frac{1}{d(d+2)}\qty(\eta_{\mu\nu}\eta_{\lambda\tau}+\eta_{\mu\lambda}\eta_{\nu\tau}+\eta_{\mu\tau}\eta_{\lambda\nu}) \nonumber\\
    &\quad \cdot\int\frac{\dd[d]{\ell}}{(2\pi)^d}\int_0^1\dd{x}\frac{\ell^4}{\qty[\ell^2+m^2+x(1-x)q^2]^2} \nonumber\\
    &= \frac{2}{d(d+2)}\qty(-d q_\nu\eta_{\rho\sigma}+2q_\rho\eta_{\nu\sigma}+2 q_\sigma \eta_{\rho\nu})\nonumber\\
    &\quad\cdot\int\frac{\dd[d]{\ell}}{(2\pi)^d}\int_0^1\dd{x}\frac{\ell^4}{\qty[\ell^2+m^2+x(1-x)q^2]^2} \,,
\end{align}

\begin{align}
    I_1^{(2)} &= \int\frac{\dd[d]{\ell}}{(2\pi)^d}\int_0^1\dd{x}\frac{2\ell_\nu (q\vdot\ell)\qty(-2x(1-x)q_\rho q_\sigma-\eta_{\rho\sigma}\qty(-x(1-x)q^2+m^2))}{\qty[\ell^2+m^2+x(1-x)q^2]^2} \nonumber\\
    &= \int_0^1\dd{x}\qty(-2x(1-x)q_\rho q_\sigma-\eta_{\rho\sigma}\qty(-x(1-x)q^2+m^2))2q^\mu\cdot\nonumber\\
    &\quad\cdot\int\frac{\dd[d]{\ell}}{(2\pi)^d}\frac{\ell_\mu\ell_\nu }{\qty[\ell^2+m^2+x(1-x)q^2]^2}\nonumber\\
    &=\int_0^1\dd{x}\frac{2}{d}q_\nu\qty(-2x(1-x)q_\rho q_\sigma-\eta_{\rho\sigma}\qty(-x(1-x)q^2+m^2))\cdot\nonumber\\
    &\quad\cdot\int\frac{\dd[d]{\ell}}{(2\pi)^d}\frac{\ell^2}{\qty[\ell^2+m^2+x(1-x)q^2]^2} \,.
\end{align}

\vskip 2em
\noindent\textbf{Second term}
\begin{align}
    I_2 &= \int\frac{\dd[d]{\ell}}{(2\pi)^d}\int_0^1\dd{x}\frac{\ell_\nu(1-2x)q^2(1-2x)\qty(2\ell_{(\rho}q_{\sigma)}-\eta_{\rho\sigma}(\ell\vdot q))}{\qty[\ell^2+m^2+x(1-x)q^2]^2} \nonumber\\
    &= \int_0^1\dd{x}(1-2x)^2q^2\qty(\delta^\mu_{\rho}q_{\sigma}+\delta^\mu_{\sigma}q_{\rho}-\eta_{\rho\sigma}q^\mu)\int\frac{\dd[d]{\ell}}{(2\pi)^d}\frac{\ell_\mu\ell_\nu}{\qty[\ell^2+m^2+x(1-x)q^2]^2}\nonumber\\
    &= \int_0^1\dd{x}(1-2x)^2\frac{q^2}{d}\qty(\eta_{\rho\nu}q_{\sigma}+\eta_{\sigma\nu}q_{\rho}-\eta_{\rho\sigma}q_\nu)\int\frac{\dd[d]{\ell}}{(2\pi)^d}\frac{\ell^2}{\qty[\ell^2+m^2+x(1-x)q^2]^2} \,.
\end{align}

\noindent\textbf{Third term}
\begin{equation}
    I_3=\int\frac{\dd[d]{\ell}}{(2\pi)^d}\int_0^1\dd{x}\frac{-q_\nu\qty(2\ell_\rho \ell_\sigma-2x(1-x)q_\rho q_\sigma-\eta_{\rho\sigma}\qty(\ell^2-x(1-x)q^2+m^2))}{\ell^2+m^2+x(1-x)q^2} \,.
\end{equation}
We can also split this into powers of $\ell$ (again noting that the last term does not cancel with the denominator due to the different sign in $q^2$).
\begin{align}
    I_3^{(2)}&=\int\frac{\dd[d]{\ell}}{(2\pi)^d}\int_0^1\dd{x}\frac{-q_\nu\qty(2\ell_\rho \ell_\sigma-\eta_{\rho\sigma}\ell^2)}{\ell^2+m^2+x(1-x)q^2} \nonumber\\
    &=\int_0^1\dd{x} q_\nu\qty(-2\delta^\lambda_\rho \delta^\tau_\sigma + \eta_{\rho\sigma}\eta^{\lambda\tau})\int\frac{\dd[d]{\ell}}{(2\pi)^d}\frac{\ell_\lambda\ell_\tau}{\ell^2+m^2+x(1-x)q^2} \nonumber\\
    &=\int_0^1\dd{x}\frac{d-2}{d}q_\nu\eta_{\rho\sigma}\int\frac{\dd[d]{\ell}}{(2\pi)^d}\frac{\ell^2}{\ell^2+m^2+x(1-x)q^2} \,, \\
    I_3^{(0)} &= \int\frac{\dd[d]{\ell}}{(2\pi)^d}\int_0^1\dd{x}\frac{-q_\nu\qty(-2x(1-x)q_\rho q_\sigma-\eta_{\rho\sigma}\qty(-x(1-x)q^2+m^2))}{\ell^2+m^2+x(1-x)q^2} \,.
\end{align}
Now we use the following results
\begin{subequations}
\begin{align}
    \int\frac{\dd[d]{\ell}}{(2\pi)^d}\frac{\ell^4}{\qty[\ell^2+M^2]^2} &= \frac{M^d}{(4\pi)^\frac{d}{2}}\qty(1+\frac{d}{2})\cdot\frac{d}{2}\cdot\Gamma\qty(-\frac{d}{2}) \,, \\
    \int\frac{\dd[d]{\ell}}{(2\pi)^d}\frac{\ell^2}{\qty[\ell^2+M^2]^2} &= \frac{M^{d-2}}{(4\pi)^\frac{d}{2}}\qty(-\frac{d}{2})\cdot\frac{d}{2}\cdot\Gamma\qty(-\frac{d}{2}) \,, \\
    \int\frac{\dd[d]{\ell}}{(2\pi)^d}\frac{\ell^2}{\ell^2+M^2} &= \frac{M^d}{(4\pi)^\frac{d}{2}}\cdot\frac{d}{2}\cdot\Gamma\qty(-\frac{d}{2}) \,, \\
    \int\frac{\dd[d]{\ell}}{(2\pi)^d}\frac{1}{\ell^2+M^2} &= \frac{M^{d-2}}{(4\pi)^\frac{d}{2}}\qty(-\frac{d}{2})\cdot\Gamma\qty(-\frac{d}{2}) \,,
\end{align}    
\end{subequations}
where $M^2=m^2+x(1-x)q^2$, to obtain
\begin{subequations}
\begin{align}
    I_1^{(4)} &= \frac{1}{2(4\pi)^\frac{d}{2}}\qty(-d q_\nu\eta_{\rho\sigma}+2q_\rho\eta_{\nu\sigma}+2 q_\sigma \eta_{\rho\nu})\Gamma\qty(-\frac{d}{2})\int_0^1\dd{x}M^d \,, \\
    I_1^{(2)} &= \frac{d}{2(4\pi)^\frac{d}{2}}\Gamma\qty(-\frac{d}{2})q_\nu\int_0^1\dd{x}\qty(2x(1-x)q_\rho q_\sigma+\eta_{\rho\sigma}\qty(-x(1-x)q^2+m^2))M^{d-2} \,, \\
    I_2 &= -\frac{q^2d}{4(4\pi)^\frac{d}{2}}\Gamma\qty(-\frac{d}{2})\qty(\eta_{\rho\nu}q_{\sigma}+\eta_{\sigma\nu}q_{\rho}-\eta_{\rho\sigma}q_\nu)\int_0^1\dd{x}(1-2x)^2M^{d-2} \,, \\
    I_3^{(2)} &= q_\nu\eta_{\rho\sigma}\frac{d-2}{2(4\pi)^\frac{d}{2}}\Gamma\qty(-\frac{d}{2})\int_0^1\dd{x}M^d \,, \\
    I_3^{(0)} &= -\frac{d}{2(4\pi)^\frac{d}{2}}\Gamma\qty(-\frac{d}{2})q_\nu \cdot\int_0^1\dd{x}\qty(2x(1-x)q_\rho q_\sigma+\eta_{\rho\sigma}\qty(-x(1-x)q^2+m^2))M^{d-2} \,.
\end{align}    
\end{subequations}
We can immediately see some cancellations. $I_1^{(2)}=-I_3^{(0)}$ and the first terms of $I_3^{(2)}$ and $I_1^{(4)}$ also cancel. In total we get
\begin{align}
    q^\mu(d_{33})_{\mu\nu\rho\sigma} &= \frac{1}{\Mp^2}\frac{2}{(4\pi)^\frac{d}{2}}q^\mu P_{\mu\nu\rho\sigma}\Gamma\qty(-\frac{d}{2})\int_0^1\dd{x}\qty(M^d-q^2\frac{d}{4}(1-2x)^2M^{d-2}) \,,
\end{align}
which is tantalisingly close to what we want. 
Using the fact that 
\begin{equation}
    \dv{M^d}{x}=\frac{d}{2}q^2(1-2x)M^{d-2} \,,
\end{equation}
we can integrate by parts to obtain
\begin{equation}
    \int_0^1 \dd{x} ~M^{d-2} (1-2x)^2 q^2 = - \frac{4}{d} m^d + \frac{4}{d}\int_0^1 \dd{x} M^d \,.
\end{equation}
Plugging that result in we find
\begin{align}
    q^\mu(d_{33})_{\mu\nu\rho\sigma} &= \frac{1}{\Mp^2}\frac{2}{(4\pi)^\frac{d}{2}}q^\mu P_{\mu\nu\rho\sigma}\Gamma\qty(-\frac{d}{2})\cdot m^d  \,, \\
    q^\mu(d_4)_{\mu\nu\rho\sigma} &= -\frac{1}{\Mp^2}\frac{2}{(4\pi)^{\frac{d}{2}}} q^\mu P_{\mu\nu\rho\sigma} \Gamma\qty(-\frac{d}{2}) \cdot m^d  \,,
\end{align}
which does cancel, thereby confirming the Ward identity is satisfied.
\subsection{Computing the integrals}

Our job is now to compute the integrals without first contracting with $q^\mu$. We start from
\begin{align}
    A_{\mu\nu\rho\sigma} 
    &= \Big((2\ell_\mu \ell_\nu - \eta_{\mu\nu}\ell^2) - x(1-x)(2q_\mu q_\nu -\eta_{\mu\nu}q^2) - \eta_{\mu\nu}m^2 \nonumber\\
    &\quad\quad - (1-2x)(2\delta^\alpha_{(\mu} \delta^\beta_{\nu)} - \eta_{\mu\nu}\eta^{\alpha\beta})\ell_\alpha q_\beta  \Big)\cdot(\mu\nu\leftrightarrow\rho\sigma) \,,
\end{align}
where we note how the first line is even in $\ell$ and the second line is odd in $\ell$. The only non-zero contributions will therefore be the product between the even terms or between the odd terms,
\begin{align}
    A_{\mu\nu\rho\sigma}^\text{even}
    &= \qty(\qty(2\ell_\mu \ell_\nu - \eta_{\mu\nu}\ell^2) - x\qty(1-x)\qty(2q_\mu q_\nu -\eta_{\mu\nu}q^2) - \eta_{\mu\nu}m^2)\cdot(\mu\nu\leftrightarrow\rho\sigma) \,, \\
    A_{\mu\nu\rho\sigma}^\text{odd} 
    &= \qty(- \qty(1-2x)\qty(2\delta^\alpha_{(\mu} \delta^\beta_{\nu)} - \eta_{\mu\nu}\eta^{\alpha\beta})\ell_\alpha q_\beta  )\cdot(\mu\nu\leftrightarrow\rho\sigma) \,.
\end{align}
After this we can also consider different powers of $\ell$ separately. Let us do these integrals in turn.

\vskip 2em
\noindent\textbf{Even-quartic}
\begin{align}
    I^{(\text{e},4)}&=\int\frac{\dd[d]{\ell}}{(2\pi)^d}\int_0^1\dd{x}\frac{\qty(2\ell_\mu \ell_\nu - \eta_{\mu\nu}\ell^2)\qty(2\ell_\rho \ell_\sigma - \eta_{\rho\sigma}\ell^2)}{\qty[\ell^2+m^2+x(1-x)q^2]^2} \nonumber\\
    &=\int_0^1\dd{x}\qty(4\delta^\alpha_\mu\delta^\beta_\nu\delta^\gamma_\rho\delta^\delta_\sigma-2\eta_{\mu\nu}\eta^{\alpha\beta}\delta^\gamma_\rho\delta^\delta_\sigma-2\eta_{\rho\sigma}\eta^{\gamma\delta}\delta^\alpha_\mu\delta^\beta_\nu+\eta_{\mu\nu}\eta_{\rho\sigma}\eta^{\alpha\beta}\eta^{\gamma\delta})\cdot\nonumber \\
    &\quad\cdot\int\frac{\dd[d]{\ell}}{(2\pi)^d}\frac{\ell_\alpha\ell_\beta\ell_\gamma\ell_\delta}{\qty[\ell^2+m^2+x(1-x)q^2]^2} \nonumber\\
    &=\int_0^1\dd{x}\qty(4\delta^\alpha_\mu\delta^\beta_\nu\delta^\gamma_\rho\delta^\delta_\sigma-2\eta_{\mu\nu}\eta^{\alpha\beta}\delta^\gamma_\rho\delta^\delta_\sigma-2\eta_{\rho\sigma}\eta^{\gamma\delta}\delta^\alpha_\mu\delta^\beta_\nu+\eta_{\mu\nu}\eta_{\rho\sigma}\eta^{\alpha\beta}\eta^{\gamma\delta})\cdot\nonumber \\
    &\quad\cdot\frac{1}{d(d+2)}\qty(\eta_{\alpha\beta}\eta_{\gamma\delta}+\eta_{\alpha\gamma}\eta_{\beta\delta}+\eta_{\alpha\delta}\eta_{\gamma\beta})\int\frac{\dd[d]{\ell}}{(2\pi)^d}\frac{\ell^4}{\qty[\ell^2+m^2+x(1-x)q^2]^2} \nonumber\\
    &=\frac{d\qty(d-2)\eta_{\mu\nu}\eta_{\rho\sigma}+8P_{\mu\nu\rho\sigma}}{d(d+2)}\int_0^1\dd{x}\int\frac{\dd[d]{\ell}}{(2\pi)^d}\frac{\ell^4}{\qty[\ell^2+m^2+x(1-x)q^2]^2} \nonumber\\
    &=\frac{d\qty(d-2)\eta_{\mu\nu}\eta_{\rho\sigma}+8P_{\mu\nu\rho\sigma}}{4(4\pi)^\frac{d}{2}}\cdot\Gamma\qty(-\frac{d}{2})\int_0^1\dd{x}M^d \,.
\end{align}

\vskip 1em
\noindent\textbf{Even-quadratic}
\begin{align}
    I^{(\text{e},2)}=&\int\frac{\dd[d]{\ell}}{(2\pi)^d}\int_0^1\dd{x}\frac{\qty(2\ell_\mu \ell_\nu - \eta_{\mu\nu}\ell^2)\qty(- x\qty(1-x)\qty(2q_\rho q_\sigma -\eta_{\rho\sigma}q^2) - \eta_{\rho\sigma}m^2)}{\qty[\ell^2+m^2+x(1-x)q^2]^2} \nonumber\\
    &+\int\frac{\dd[d]{\ell}}{(2\pi)^d}\int_0^1\dd{x}\frac{\qty(- x\qty(1-x)\qty(2q_\mu q_\nu -\eta_{\mu\nu}q^2) - \eta_{\mu\nu}m^2)\qty(2\ell_\rho \ell_\sigma - \eta_{\rho\sigma}\ell^2)}{\qty[\ell^2+m^2+x(1-x)q^2]^2} \nonumber\\
    =&\int_0^1\dd{x}\bigg(\qty(- x\qty(1-x)\qty(2q_\rho q_\sigma -\eta_{\rho\sigma}q^2) - \eta_{\rho\sigma}m^2)\qty(2\delta^\alpha_\mu\delta^\beta_\nu-\eta_{\mu\nu}\eta^{\alpha\beta}) \nonumber\\
    &+\qty(- x\qty(1-x)\qty(2q_\mu q_\nu -\eta_{\mu\nu}q^2) - \eta_{\mu\nu}m^2)\qty(2\delta^\alpha_\rho\delta^\beta_\sigma-\eta_{\rho\sigma}\eta^{\alpha\beta})\bigg)\cdot\nonumber\\
    &\cdot\int\frac{\dd[d]{\ell}}{(2\pi)^d}\frac{\ell_\alpha\ell_\beta}{\qty[\ell^2+m^2+x(1-x)q^2]^2} \nonumber\\
    =&\int_0^1\dd{x}\qty(\frac{2}{d}-1)\qty(-2 x\qty(1-x)\qty(q_\rho q_\sigma\eta_{\mu\nu}+q_\mu q_\nu\eta_{\rho\sigma} )+2x\qty(1-x)\eta_{\rho\sigma}\eta_{\mu\nu}q^2 -2 \eta_{\rho\sigma}m^2\eta_{\mu\nu})\cdot\nonumber\\
    &\cdot\int\frac{\dd[d]{\ell}}{(2\pi)^d}\frac{\ell^2}{\qty[\ell^2+m^2+x(1-x)q^2]^2} \nonumber\\
    =&\int_0^1\dd{x}M^{d-2}\qty(-2 x\qty(1-x)\qty(q_\rho q_\sigma\eta_{\mu\nu}+q_\mu q_\nu\eta_{\rho\sigma} )+2x\qty(1-x)\eta_{\mu\nu}\eta_{\rho\sigma}q^2 -2 \eta_{\mu\nu}\eta_{\rho\sigma}m^2)\cdot\nonumber\\
    &\cdot\frac{d(d-2)}{4(4\pi)^\frac{d}{2}}\cdot\Gamma\qty(-\frac{d}{2}) \,.
\end{align}

\vskip 1em
\noindent\textbf{Even-constant}
\begin{align}
    I^{(\text{e},0)}=&\int_0^1\dd{x}\qty(- x\qty(1-x)\qty(2q_\mu q_\nu -\eta_{\mu\nu}q^2) - \eta_{\mu\nu}m^2)\qty(- x\qty(1-x)\qty(2q_\rho q_\sigma -\eta_{\rho\sigma}q^2) - \eta_{\rho\sigma}m^2)\cdot\nonumber\\
    \cdot&\int\frac{\dd[d]{\ell}}{(2\pi)^d}\frac{1}{\qty[\ell^2+m^2+x(1-x)q^2]^2} \,.
\end{align}
The integral we need to compute is new. We need to use 
\begin{equation}
    \int\frac{\dd[d]{\ell}}{(2\pi)^d}\frac{1}{\qty[\ell^2+M^2]^2}=\frac{M^{d-4}}{(4\pi)^\frac{d}{2}}\qty(1-\frac{d}{2})\qty(-\frac{d}{2})\Gamma\qty(-\frac{d}{2}) \,,
\end{equation}
to obtain
\begin{align}
    I^{(\text{e},0)} &= \int_0^1\dd{x}\frac{M^{d-4}}{(4\pi)^\frac{d}{2}}\qty(1-\frac{d}{2})\qty(-\frac{d}{2})\Gamma\qty(-\frac{d}{2})\qty(- x\qty(1-x)\qty(2q_\mu q_\nu -\eta_{\mu\nu}q^2) - \eta_{\mu\nu}m^2)\cdot\nonumber\\
    &\quad\cdot\qty(- x\qty(1-x)\qty(2q_\rho q_\sigma -\eta_{\rho\sigma}q^2) - \eta_{\rho\sigma}m^2) \nonumber\\
    &=\Gamma\qty(-\frac{d}{2})\frac{d(d-2)}{4(4\pi)^\frac{d}{2}}\int_0^1\dd{x}M^{d-4}\cdot\nonumber\\
    &\quad\cdot\Bigg(x^2(1-x)^2\qty(4q_\mu q_\nu q_\rho q_\sigma-2 q_\mu q_\nu q^2\eta_{\rho\sigma}-2q_\rho q_\sigma q^2\eta_{\mu\nu}+\eta_{\mu\nu}\eta_{\rho\sigma}q^4) \nonumber\\
    &\quad + 2x(1-x)\qty(q_\mu q_\nu\eta_{\rho\sigma}m^2+q_\rho q_\sigma \eta_{\mu\nu} m^2-\eta_{\mu\nu}\eta_{\rho\sigma}q^2m^2)+\eta_{\mu\nu}\eta_{\rho\sigma}m^4\Bigg) \,.
\end{align}

\vskip 1em
\noindent\textbf{Odd}
\begin{align}
    I^{(\text{o},2)}=&\int\frac{\dd[d]{\ell}}{(2\pi)^2}\int_0^1\dd{x}\frac{\qty(- \qty(1-2x)\qty(2\delta^\alpha_{(\mu} \delta^\beta_{\nu)} - \eta_{\mu\nu}\eta^{\alpha\beta})\ell_\alpha q_\beta  )\cdot(\mu\nu\leftrightarrow\rho\sigma)}{\qty[\ell^2+m^2+x(1-x)q^2]^2} \nonumber\\
    =&\int_0^1\dd{x}\qty(1-2x)^2\qty(\delta^\alpha_\mu q_\nu+\delta^\alpha_\nu q_\mu - \eta_{\mu\nu}q^\alpha) \qty(\delta^\gamma_{\rho} q_\sigma+\delta^\gamma_\sigma q_\rho - \eta_{\rho\sigma}q^\gamma)\cdot\nonumber\\
    \cdot&\int\frac{\dd[d]{\ell}}{(2\pi)^d}\frac{\ell_\alpha\ell_\gamma}{\qty[\ell^2+m^2+x(1-x)q^2]^2} \nonumber\\   
    =&~\qty(\eta_{\mu\rho} q_\nu q_\sigma+\eta_{\mu\sigma} q_\nu q_\rho+\eta_{\nu\rho}q_\mu q_\sigma+\eta_{\nu\sigma}q_\mu q_\rho-2\eta_{\rho\sigma}q_\mu q_\nu-2 \eta_{\mu\nu}q_\rho q_\sigma+\eta_{\mu\nu}\eta_{\rho\sigma}q^2)\cdot\nonumber\\
    \cdot&\frac{1}{d}\int_0^1\dd{x}\qty(1-2x)^2\int\frac{\dd[d]{\ell}}{(2\pi)^d}\frac{\ell^2}{\qty[\ell^2+m^2+x(1-x)q^2]^2} \nonumber\\
    =&-\qty(\eta_{\mu\rho} q_\nu q_\sigma+\eta_{\mu\sigma} q_\nu q_\rho+\eta_{\nu\rho}q_\mu q_\sigma+\eta_{\nu\sigma}q_\mu q_\rho-2\eta_{\rho\sigma}q_\mu q_\nu-2 \eta_{\mu\nu}q_\rho q_\sigma+\eta_{\mu\nu}\eta_{\rho\sigma}q^2)\cdot\nonumber\\
    \cdot&~\frac{d}{4(4\pi)^\frac{d}{2}}\cdot\Gamma\qty(-\frac{d}{2})\int_0^1\dd{x}\qty(1-2x)^2M^{d-2} \,.
\end{align}
Now we need to pair them together. Note how $M$ depends on $q^2$ so we cannot just naively count powers of $q^2$. We instead take the following split into index structures
\begin{align}
    \Mp^2\cdot\Pi^\text{1-loop}_{\mu\nu\rho\sigma}(q) 
        =~
        & c_1\cdot (\eta_{\mu\sigma}\eta_{\nu\rho} + \eta_{\mu\rho}\eta_{\nu\sigma} ) 
        + c_2\cdot \eta_{\mu\nu}\eta_{\rho\sigma} \nonumber \\
        & + c_3\cdot \qty(\eta_{\mu\rho}\frac{q_\nu q_\sigma}{q^2} + \eta_{\mu\sigma}\frac{q_\nu q_\rho}{q^2} + \eta_{\nu\rho}\frac{q_\mu q_\sigma}{q^2} + \eta_{\nu\sigma}\frac{q_\mu q_\rho}{q^2})  \nonumber \\
        & + c_4\cdot \qty(\eta_{\mu\nu}\frac{q_\rho q_\sigma}{q^2} + \eta_{\rho\sigma}\frac{q_\mu q_\nu}{q^2}) \nonumber \\
        & + c_5\cdot \frac{q_\mu q_\nu q_\rho q_\sigma}{q^4} \,.
    \label{eq:index-structure}
\end{align}
We therefore write
\begin{subequations}
\begin{align}
    I^{(\text{e},4)}&=c_1^{(\text{e},4)}(\eta_{\mu\sigma}\eta_{\nu\rho} + \eta_{\mu\rho}\eta_{\nu\sigma} ) +c_2^{(\text{e},4)}\eta_{\mu\nu}\eta_{\rho\sigma} \,, \\
    I^{(\text{e},2)}&=c_2^{(\text{e},2)}\eta_{\mu\nu}\eta_{\rho\sigma}+c_4^{(\text{e},2)}\qty(\eta_{\mu\nu}\frac{q_\rho q_\sigma}{q^2} + \eta_{\rho\sigma}\frac{q_\mu q_\nu}{q^2}) \,, \\
    I^{(\text{e},0)}&=c_2^{(\text{e,0})}\eta_{\mu\nu}\eta_{\rho\sigma}+c_4^{(\text{e},0)}\qty(\eta_{\mu\nu}\frac{q_\rho q_\sigma}{q^2} + \eta_{\rho\sigma}\frac{q_\mu q_\nu}{q^2})+c_5^{(\text{e},0)} \frac{q_\mu q_\nu q_\rho q_\sigma}{q^4} \,, \\
    I^{(\text{o},2)}&=c_2^{(\text{o},2)}\eta_{\mu\nu}\eta_{\rho\sigma}+ c_3^{(\text{o},2)} \qty(\eta_{\mu\rho}\frac{q_\nu q_\sigma}{q^2} + \eta_{\mu\sigma}\frac{q_\nu q_\rho}{q^2} + \eta_{\nu\rho}\frac{q_\mu q_\sigma}{q^2} + \eta_{\nu\sigma}\frac{q_\mu q_\rho}{q^2}) \nonumber \\
    &\quad +c_4^{(\text{o},2)}\qty(\eta_{\mu\nu}\frac{q_\rho q_\sigma}{q^2} + \eta_{\rho\sigma}\frac{q_\mu q_\nu}{q^2}) \,,
\end{align}
\end{subequations}
where
\begin{subequations}
\begin{align}
    c_1^{(\text{e},4)}&=\frac{1}{(4\pi)^\frac{d}{2}}\Gamma\qty(-\frac{d}{2})\int_0^1\dd{x}M^d \,, \\
    c_2^{(\text{e},4)}&=\frac{d\qty(d-2)-4}{4(4\pi)^\frac{d}{2}}\Gamma\qty(-\frac{d}{2})\int_0^1\dd{x}M^d \,, \\
    c_2^{(\text{e},2)}&=-\frac{d(d-2)}{2(4\pi)^\frac{d}{2}}\Gamma\qty(-\frac{d}{2})\int_0^1\dd{x}M^{d-2}\qty(m^2-x\qty(1-x)q^2) \,, \\
    c_4^{(\text{e},2)}&=-\frac{d(d-2)}{2(4\pi)^\frac{d}{2}}\Gamma\qty(-\frac{d}{2})\int_0^1\dd{x}M^{d-2} x\qty(1-x)q^2 \,, \\
    c_2^{(\text{e},0)}&=\frac{d(d-2)}{4(4\pi)^\frac{d}{2}}\Gamma\qty(-\frac{d}{2})\int_0^1\dd{x}M^{d-4}\qty(m^2-x(1-x)q^2)^2 \,, \\
    c_4^{(\text{e},0)}&=\frac{d(d-2)}{2(4\pi)^\frac{d}{2}}\Gamma\qty(-\frac{d}{2})\int_0^1\dd{x}M^{d-4}x(1-x)q^2\qty(m^2-x(1-x)q^2) \,, \\
    c_5^{(\text{e},0)}&=\frac{d(d-2)}{(4\pi)^\frac{d}{2}}\Gamma\qty(-\frac{d}{2})\int_0^1\dd{x}M^{d-4}x^2(1-x)^2q^4 \,, \\
    c_2^{(\text{o},2)}&=-\frac{d}{4(4\pi)^\frac{d}{2}}\Gamma\qty(-\frac{d}{2})\int_0^1\dd{x}M^{d-2}\qty(1-2x)^2q^2 \,, \\
    c_3^{(\text{o},2)}&=-\frac{d}{4(4\pi)^\frac{d}{2}}\Gamma\qty(-\frac{d}{2})\int_0^1\dd{x}M^{d-2}\qty(1-2x)^2q^2 \,, \\
    c_4^{(\text{o},2)}&=\frac{d}{2(4\pi)^\frac{d}{2}}\cdot\Gamma\qty(-\frac{d}{2})\int_0^1\dd{x}M^{d-2}\qty(1-2x)^2q^2 \,. 
\end{align}    
\end{subequations}
Let us sum the individual contributions to each $c_i$. Since there is a common factor of $\frac{1}{(4\pi)^\frac{d}{2}}\Gamma\qty(-\frac{d}{2})$, let us not write it. 
\begin{subequations}
\begin{align}
    c_1^{(\text{e},4)}&= \int_0^1\dd{x}M^d \,, \\
    c_2^{(\text{e},4)}&=\qty(\frac{d\qty(d-2)}{4} - 1)\int_0^1\dd{x}M^d \,, \\
    c_2^{(\text{o},2)}&=-\frac{d}{4}\int_0^1\dd{x}M^{d-2}\qty(1-2x)^2q^2 \,, \\
    c_2^{(\text{e},2)}&=-\frac{d(d-2)}{2}\int_0^1\dd{x}M^{d-2}\qty(m^2-x\qty(1-x)q^2) \,, \\
    c_2^{(\text{e},0)}&=\frac{d(d-2)}{4}\int_0^1\dd{x}M^{d-4}\Big(m^4 - 2x(1-x)q^2\cdot m^2 + x^2(1-x)^2q^4 \Big) \,, \\
    c_3^{(\text{o},2)}&=-\frac{d}{4}\int_0^1\dd{x}M^{d-2}\qty(1-2x)^2q^2 \,, \\
    c_4^{(\text{e},2)}&=-\frac{d(d-2)}{2}\int_0^1\dd{x}M^{d-2} x\qty(1-x)q^2 \,, \\
    c_4^{(\text{o},2)}&=\frac{2d}{4}\int_0^1\dd{x}M^{d-2}\qty(1-2x)^2q^2 \,, \\
    c_4^{(\text{e},0)}&=\frac{d(d-2)}{2}\int_0^1\dd{x}M^{d-4}\Big(x(1-x)q^2\cdot m^2 - x^2(1-x)^2q^4 \Big) \,, \\
    c_5^{(\text{e},0)}&=d(d-2)\cdot\int_0^1\dd{x}M^{d-4}x^2(1-x)^2q^4 \,.
\end{align}    
\end{subequations}
Recall also the result for $d_4$,
\begin{equation}
    \Mp^2\cdot(d_4)_{\mu\nu\rho\sigma} = -(\eta_{\mu\sigma}\eta_{\nu\rho} + \eta_{\mu\rho}\eta_{\nu\sigma} - \eta_{\mu\nu}\eta_{\rho\sigma}) \cdot\frac{1}{(4\pi)^\frac{d}{2}}\Gamma\qty(-\frac{d}{2})\cdot m^d \,. 
    \label{eq:d4finitefinal}
\end{equation}
We once again use integration by parts to write
\begin{align}
    \int_0^1 \dd{x} ~M^{d-2} (1-2x)^2 q^2 = - \frac{4}{d} m^d + \frac{4}{d}\int_0^1 \dd{x} M^d \,.
\end{align}
Using the definition of $M$ we can trivially write (just put the last term to the LHS)
\begin{align}
    M^{d-2} x(1-x) q^2 = M^d - m^2 M^{d-2} \,.
    \label{eq:triviald-2}
\end{align}
Similarly for integrals with powers of $M^{d-4}$ we can write
\begin{align}
    M^{d-4}\qty(m^2+x(1-x)q^2)^2 &= M^d\nonumber\\
    M^{d-4}x^2(1-x)^2q^4 &= M^d-m^4M^{d-4}-2x(1-x)q^2m^2M^{d-4} \nonumber\\
    &= M^d-2m^2 M^{d-2}+m^4 M^{d-4} \,,
\end{align}
where in going to the last line we used (\ref{eq:triviald-2}).
We then find
\begin{subequations}
\begin{align}
    c_1^{(\text{e},4)}&= \int_0^1\dd{x}M^d \,, \\
    c_2^{(\text{e},4)}&=\qty(\frac{d\qty(d-2)}{4} - 1)\int_0^1\dd{x}M^d \,, \\
    c_2^{(\text{o},2)}&= m^d - \int_0^1 dx ~M^d \,, \\
    c_2^{(\text{e},2)}&=\frac{d(d-2)}{2}\qty(\int_0^1\dd{x} M^{d} - 2m^2 \int_0^1 \dd{x} M^{d-2}) \,, \\
    c_2^{(\text{e},0)}&=\frac{d(d-2)}{4}\qty(\int_0^1 \dd{x} M^d - 4m^2\int_0^1 \dd{x} M^{d-2} + 4m^4 \int_0^1 \dd{x} M^{d-4}) \,, \\
    c_3^{(\text{o},2)}&= m^d - \int_0^1 dx ~M^d \,, \\
    c_4^{(\text{e},2)}&=-\frac{d(d-2)}{2} \qty(\int_0^1 \dd{x} M^d - m^2 \int_0^1 \dd{x} M^{d-2}) \,, \\
    c_4^{(\text{o},2)}&= -2m^d + 2\int_0^1 dx ~M^d \,, \\
    c_4^{(\text{e},0)}&=\frac{d(d-2)}{2}\qty(-\int_0^1 \dd{x} M^{d} + 3m^2 \int_0^1 \dd{x} M^{d-2} - 2m^4  \int_0^1 \dd{x} M^{d-4}) \,,\\
    c_5^{(\text{e},0)}&=d(d-2)\cdot\qty(\int_0^1 \dd{x} M^d - 2m^2\int_0^1 \dd{x} M^{d-2} + m^4\int_0^1 \dd{x} M^{d-4}) \,,
\end{align}    
\end{subequations}
which (together with (\ref{eq:d4finitefinal})) leads to 
\begin{subequations}
\begin{align}
    c_1 &= - m^d + \int_0^1\dd{x} ~M^d \,, \\
    c_2 &= 2m^d + \qty(d(d-2) - 2)\int_0^1 \dd{x} M^d - 2\cdot d(d-2)\cdot m^2\int_0^1 \dd{x} M^{d-2} \\ 
    &\quad + d(d-2)\cdot m^4 \int_0^1 \dd{x} M^{d-4} \,,\\ 
    c_3 &= m^d - \int_0^1 \dd{x} ~M^d \,, \\
    c_4 &= - 2m^d - (d(d-2) - 2)\int_0^1 \dd{x} M^d + 2\cdot d(d-2) \cdot m^2\int_0^1 \dd{x} M^{d-2} \\
    &\quad - d(d-2)\cdot m^4\int_0^1 \dd{x} M^{d-4} \,, \\
    c_5 &= d(d-2)\int_0^1 \dd{x} M^d - 2\cdot d(d-2)\cdot m^2 \int_0^1 \dd{x} M^{d-2} \\ 
    &+ d(d-2)\cdot m^4 \int_0^1 \dd{x} M^{d-4} \,. 
\end{align}    
\end{subequations}

\subsubsection{Checking Ward identity}

Contracting \eqref{eq:index-structure} with $q^\mu$, we find that the coefficients must satisfy
\begin{subequations}
\begin{align}
    c_1 + c_3 &= 0 \,, \\
    c_2 + c_4 &= 0 \,, \\
    2c_3 + c_4 + c_5 &= 0 \,.
\end{align}    
\end{subequations}
Using the results above, we can confirm that this is indeed satisfied.

\subsection{The result}

Due to the relations that the coefficients satisfy, only 3 are independent and need to be computed, involving 3 different types of integrals,
\begin{align}
    \int_0^1 \dd{x} M^{d-a} \,, \quad\quad a=0,2,4 \,.
\end{align}

Expanding around $d=4$ (with $d=4-\epsilon$) before integrating, introducing the function
\begin{align}
    \text{F}(x) = \sqrt{1 + x}\cdot\text{arctanh}\qty(\frac{1}{\sqrt{1+x}}) \,,
\end{align}
and using the variable $\alpha=\frac{4m^2}{q^2}$, we find the result quoted in the main body of the manuscript.
\begin{subequations}
\begin{align}
    c_1 =&~ \frac{m^4}{30\pi^2} \qty(1 - \text{F}(\alpha)) + \frac{m^2q^2}{15\pi^2}\qty(\frac{5}{16\epsilon} + \frac{43}{96} - \frac14 \text{F}(\alpha) - \frac{5}{32}\gamma - \frac{5}{32}\log\qty(\frac{m^2}{4\pi\mu^2})) \nonumber \\
    &+ \frac{q^4}{15\pi^2}\qty(\frac{1}{32\epsilon} + \frac{23}{480} - \frac{1}{32}\text{F}(\alpha) - \frac{1}{64}\gamma - \frac{1}{64}\log\qty(\frac{m^2}{4\pi\mu^2})) \,, \\
    c_5 =&~ \frac{m^4}{10\pi^2}\qty(1 - \text{F}(\alpha)) + \frac{m^2q^2}{15\pi^2}\qty(-\frac38 + \frac12~\text{F}(\alpha)) \nonumber \\
    & +\frac{q^4}{15\pi^2}\qty(\frac{1}{4\epsilon} + \frac{47}{240} - \frac14 \text{F}(\alpha) - \frac18\gamma - \frac18\log\qty(\frac{m^2}{4\pi\mu^2})) \,.
\end{align}    
\end{subequations}
}

\printbibliography

@article{Burgess_2004,
   title={Quantum Gravity in Everyday Life: General Relativity as an Effective Field Theory},
   volume={7},
   ISSN={1433-8351},
   url={http://dx.doi.org/10.12942/lrr-2004-5},
   DOI={10.12942/lrr-2004-5},
   number={1},
   journal={Living Reviews in Relativity},
   publisher={Springer Science and Business Media LLC},
   author={Burgess, Cliff P.},
   year={2004},
   month=apr 
}

@misc{donoghue2017epfllecturesgeneralrelativity,
      title={EPFL Lectures on General Relativity as a Quantum Field Theory}, 
      author={John F. Donoghue and Mikhail M. Ivanov and Andrey Shkerin},
      year={2017},
      eprint={1702.00319},
      archivePrefix={arXiv},
      primaryClass={hep-th},
      url={https://arxiv.org/abs/1702.00319}, 
}

@inbook{Donoghue:2022eay,
    author = "Donoghue, John F.",
    title = "{Quantum General Relativity and Effective Field Theory}",
    eprint = "2211.09902",
    archivePrefix = "arXiv",
    primaryClass = "hep-th",
    doi = "10.1007/978-981-19-3079-9_1-1",
    year = "2023"
}

@article{Donoghue_1994,
   title={General relativity as an effective field theory: The leading quantum corrections},
   volume={50},
   ISSN={0556-2821},
   url={http://dx.doi.org/10.1103/PhysRevD.50.3874},
   DOI={10.1103/physrevd.50.3874},
   number={6},
   journal={Physical Review D},
   publisher={American Physical Society (APS)},
   author={Donoghue, John F.},
   year={1994},
   month=sep, pages={3874–3888} 
}

@article{Vafa:2005ui,
    author = "Vafa, Cumrun",
    title = "{The String landscape and the swampland}",
    eprint = "hep-th/0509212",
    archivePrefix = "arXiv",
    reportNumber = "HUTP-05-A043",
    month = "9",
    year = "2005"
}

@article{Palti:2019pca,
    author = "Palti, Eran",
    title = "{The Swampland: Introduction and Review}",
    eprint = "1903.06239",
    archivePrefix = "arXiv",
    primaryClass = "hep-th",
    reportNumber = "MPP-2019-53",
    doi = "10.1002/prop.201900037",
    journal = "Fortsch. Phys.",
    volume = "67",
    number = "6",
    pages = "1900037",
    year = "2019"
}

@article{vanBeest:2021lhn,
    author = "van Beest, Marieke and Calder\'on-Infante, Jos\'e and Mirfendereski, Delaram and Valenzuela, Irene",
    title = "{Lectures on the Swampland Program in String Compactifications}",
    eprint = "2102.01111",
    archivePrefix = "arXiv",
    primaryClass = "hep-th",
    doi = "10.1016/j.physrep.2022.09.002",
    journal = "Phys. Rept.",
    volume = "989",
    pages = "1--50",
    year = "2022"
}

@article{Brennan:2017rbf,
    author = "Brennan, T. Daniel and Carta, Federico and Vafa, Cumrun",
    title = "{The String Landscape, the Swampland, and the Missing Corner}",
    eprint = "1711.00864",
    archivePrefix = "arXiv",
    primaryClass = "hep-th",
    reportNumber = "IFT-UAM-CSIC-17-105",
    doi = "10.22323/1.305.0015",
    journal = "PoS",
    volume = "TASI2017",
    pages = "015",
    year = "2017"
}

@article{Grana:2021zvf,
    author = "Gra\~na, Mariana and Herr\'aez, Alvaro",
    title = "{The Swampland Conjectures: A Bridge from Quantum Gravity to Particle Physics}",
    eprint = "2107.00087",
    archivePrefix = "arXiv",
    primaryClass = "hep-th",
    doi = "10.3390/universe7080273",
    journal = "Universe",
    volume = "7",
    number = "8",
    pages = "273",
    year = "2021"
}

@article{Agmon:2022thq,
    author = "Agmon, Nathan Benjamin and Bedroya, Alek and Kang, Monica Jinwoo and Vafa, Cumrun",
    title = "{Lectures on the string landscape and the Swampland}",
    eprint = "2212.06187",
    archivePrefix = "arXiv",
    primaryClass = "hep-th",
    month = "12",
    year = "2022"
}

@article{VanRiet:2023pnx,
    author = "Van Riet, Thomas and Zoccarato, Gianluca",
    title = "{Beginners lectures on flux compactifications and related Swampland topics}",
    eprint = "2305.01722",
    archivePrefix = "arXiv",
    primaryClass = "hep-th",
    doi = "10.1016/j.physrep.2023.11.003",
    journal = "Phys. Rept.",
    volume = "1049",
    pages = "1--51",
    year = "2024"
}

@article{Susskind:2003kw,
    author = "Susskind, Leonard",
    editor = "Carr, Bernard J.",
    title = "{The Anthropic landscape of string theory}",
    eprint = "hep-th/0302219",
    archivePrefix = "arXiv",
    pages = "247--266",
    month = "2",
    year = "2003"
}

@article{Taylor:2015xtz,
    author = "Taylor, Washington and Wang, Yi-Nan",
    title = "{The F-theory geometry with most flux vacua}",
    eprint = "1511.03209",
    archivePrefix = "arXiv",
    primaryClass = "hep-th",
    reportNumber = "MIT-CTP-4732",
    doi = "10.1007/JHEP12(2015)164",
    journal = "JHEP",
    volume = "12",
    pages = "164",
    year = "2015"
}

@article{SdC,
    author = "Ooguri, Hirosi and Vafa, Cumrun",
    title = "{On the Geometry of the String Landscape and the Swampland}",
    eprint = "hep-th/0605264",
    archivePrefix = "arXiv",
    reportNumber = "CALT-68-2600, HUTP-06-A017",
    doi = "10.1016/j.nuclphysb.2006.10.033",
    journal = "Nucl. Phys. B",
    volume = "766",
    pages = "21--33",
    year = "2007"
}

@article{Harlow:2015lma,
    author = "Harlow, Daniel",
    title = "{Wormholes, Emergent Gauge Fields, and the Weak Gravity Conjecture}",
    eprint = "1510.07911",
    archivePrefix = "arXiv",
    primaryClass = "hep-th",
    doi = "10.1007/JHEP01(2016)122",
    journal = "JHEP",
    volume = "01",
    pages = "122",
    year = "2016"
}

@article{Grimm:2018ohb,
    author = "Grimm, Thomas W. and Palti, Eran and Valenzuela, Irene",
    title = "{Infinite Distances in Field Space and Massless Towers of States}",
    eprint = "1802.08264",
    archivePrefix = "arXiv",
    primaryClass = "hep-th",
    doi = "10.1007/JHEP08(2018)143",
    journal = "JHEP",
    volume = "08",
    pages = "143",
    year = "2018"
}

@article{Heidenreich:2018kpg,
    author = "Heidenreich, Ben and Reece, Matthew and Rudelius, Tom",
    title = "{Emergence of Weak Coupling at Large Distance in Quantum Gravity}",
    eprint = "1802.08698",
    archivePrefix = "arXiv",
    primaryClass = "hep-th",
    doi = "10.1103/PhysRevLett.121.051601",
    journal = "Phys. Rev. Lett.",
    volume = "121",
    number = "5",
    pages = "051601",
    year = "2018"
}

@article{Blumenhagen:2024ydy,
    author = "Blumenhagen, Ralph and Cribiori, Niccol\`o and Gligovic, Aleksandar and Paraskevopoulou, Antonia",
    title = "{Emergence of R$^{4}$-terms in M-theory}",
    eprint = "2404.01371",
    archivePrefix = "arXiv",
    primaryClass = "hep-th",
    reportNumber = "MPP-2024-72",
    doi = "10.1007/JHEP07(2024)018",
    journal = "JHEP",
    volume = "07",
    pages = "018",
    year = "2024"
}

@article{Blumenhagen:2023xmk,
    author = "Blumenhagen, Ralph and Cribiori, Niccol\`o and Gligovic, Aleksandar and Paraskevopoulou, Antonia",
    title = "{Emergent M-theory limit}",
    eprint = "2309.11554",
    archivePrefix = "arXiv",
    primaryClass = "hep-th",
    reportNumber = "MPP-2023-193",
    doi = "10.1103/PhysRevD.109.L021901",
    journal = "Phys. Rev. D",
    volume = "109",
    number = "2",
    pages = "L021901",
    year = "2024"
}

@article{Blumenhagen:2023tev,
    author = "Blumenhagen, Ralph and Cribiori, Niccol\`o and Gligovic, Aleksandar and Paraskevopoulou, Antonia",
    title = "{Demystifying the Emergence Proposal}",
    eprint = "2309.11551",
    archivePrefix = "arXiv",
    primaryClass = "hep-th",
    reportNumber = "MPP-2023-192",
    doi = "10.1007/JHEP04(2024)053",
    journal = "JHEP",
    volume = "04",
    pages = "053",
    year = "2024"
}

@article{Blumenhagen:2023yws,
    author = "Blumenhagen, Ralph and Gligovic, Aleksandar and Paraskevopoulou, Antonia",
    title = "{The emergence proposal and the emergent string}",
    eprint = "2305.10490",
    archivePrefix = "arXiv",
    primaryClass = "hep-th",
    reportNumber = "MPP-2023-96",
    doi = "10.1007/JHEP10(2023)145",
    journal = "JHEP",
    volume = "10",
    pages = "145",
    year = "2023"
}

@article{Veneziano:2001ah,
    author = "Veneziano, G.",
    title = "{Large N bounds on, and compositeness limit of, gauge and gravitational interactions}",
    eprint = "hep-th/0110129",
    archivePrefix = "arXiv",
    reportNumber = "CERN-TH-2001-278",
    doi = "10.1088/1126-6708/2002/06/051",
    journal = "JHEP",
    volume = "06",
    pages = "051",
    year = "2002"
}

@article{Han_2005,
   title={Scale of quantum gravity},
   volume={616},
   ISSN={0370-2693},
   url={http://dx.doi.org/10.1016/j.physletb.2005.04.040},
   DOI={10.1016/j.physletb.2005.04.040},
   number={3–4},
   journal={Physics Letters B},
   publisher={Elsevier BV},
   author={Han, Tao and Willenbrock, Scott},
   year={2005},
   month=jun, pages={215–220} 
}

@article{Dvali:2007hz,
    author = "Dvali, Gia",
    title = "{Black Holes and Large N Species Solution to the Hierarchy Problem}",
    eprint = "0706.2050",
    archivePrefix = "arXiv",
    primaryClass = "hep-th",
    doi = "10.1002/prop.201000009",
    journal = "Fortsch. Phys.",
    volume = "58",
    pages = "528--536",
    year = "2010"
}

@article{Dvali:2007wp,
    author = "Dvali, Gia and Redi, Michele",
    title = "{Black Hole Bound on the Number of Species and Quantum Gravity at LHC}",
    eprint = "0710.4344",
    archivePrefix = "arXiv",
    primaryClass = "hep-th",
    doi = "10.1103/PhysRevD.77.045027",
    journal = "Phys. Rev. D",
    volume = "77",
    pages = "045027",
    year = "2008"
}

@article{Dvali:2010vm,
    author = "Dvali, Gia and Gomez, Cesar",
    title = "{Species and Strings}",
    eprint = "1004.3744",
    archivePrefix = "arXiv",
    primaryClass = "hep-th",
    reportNumber = "CERN-PH-TH-2010-069, IFT-UAM-CSIC-10-25",
    month = "4",
    year = "2010"
}

@article{Arkani-Hamed:2005zuc,
    author = "Arkani-Hamed, Nima and Dimopoulos, Savas and Kachru, Shamit",
    title = "{Predictive landscapes and new physics at a TeV}",
    eprint = "hep-th/0501082",
    archivePrefix = "arXiv",
    reportNumber = "SLAC-PUB-10928, HUTP-05-A0001, SU-ITP-04-44",
    month = "1",
    year = "2005"
}

@phdthesis{Castellano:2024bna,
    author = "Castellano, Alberto",
    title = "{The Quantum Gravity Scale and the Swampland}",
    eprint = "2409.10003",
    archivePrefix = "arXiv",
    primaryClass = "hep-th",
    school = "U. Autonoma, Madrid (main)",
    year = "2024"
}

@article{Etheredge:2024tok,
    author = "Etheredge, Muldrow and Heidenreich, Ben and Rudelius, Tom and Ruiz, Ignacio and Valenzuela, Irene",
    title = "{Taxonomy of Infinite Distance Limits}",
    eprint = "2405.20332",
    archivePrefix = "arXiv",
    primaryClass = "hep-th",
    reportNumber = "ACFI-T24-04, CERN-TH-2024-067, IFT-UAM/CSIC-23-64",
    month = "5",
    year = "2024"
}

@article{Bedroya:2024uva,
    author = "Bedroya, Alek and Vafa, Cumrun and Wu, David H.",
    title = "{The Tale of Three Scales: the Planck, the Species, and the Black Hole Scales}",
    eprint = "2403.18005",
    archivePrefix = "arXiv",
    primaryClass = "hep-th",
    month = "3",
    year = "2024"
}

@article{Rudelius:2023spc,
    author = "Rudelius, Tom",
    title = "{Persistence of the pattern in the interior of 5d moduli spaces}",
    eprint = "2312.00120",
    archivePrefix = "arXiv",
    primaryClass = "hep-th",
    doi = "10.1016/j.physletb.2024.138640",
    journal = "Phys. Lett. B",
    volume = "853",
    pages = "138640",
    year = "2024"
}

@article{Castellano:2023jjt,
    author = "Castellano, Alberto and Ruiz, Ignacio and Valenzuela, Irene",
    title = "{Stringy evidence for a universal pattern at infinite distance}",
    eprint = "2311.01536",
    archivePrefix = "arXiv",
    primaryClass = "hep-th",
    doi = "10.1007/JHEP06(2024)037",
    journal = "JHEP",
    volume = "06",
    pages = "037",
    year = "2024"
}

@article{Castellano:2023stg,
    author = "Castellano, Alberto and Ruiz, Ignacio and Valenzuela, Irene",
    title = "{Universal Pattern in Quantum Gravity at Infinite Distance}",
    eprint = "2311.01501",
    archivePrefix = "arXiv",
    primaryClass = "hep-th",
    doi = "10.1103/PhysRevLett.132.181601",
    journal = "Phys. Rev. Lett.",
    volume = "132",
    number = "18",
    pages = "181601",
    year = "2024"
}

@article{Burgess:2023pnk,
    author = "Burgess, C. P. and Quevedo, F.",
    title = "{Perils of towers in the swamp: dark dimensions and the robustness of EFTs}",
    eprint = "2304.03902",
    archivePrefix = "arXiv",
    primaryClass = "hep-th",
    doi = "10.1007/JHEP09(2023)159",
    journal = "JHEP",
    volume = "09",
    pages = "159",
    year = "2023"
}

@article{vandeHeisteeg:2023ubh,
    author = "van de Heisteeg, Damian and Vafa, Cumrun and Wiesner, Max",
    title = "{Bounds on Species Scale and the Distance Conjecture}",
    eprint = "2303.13580",
    archivePrefix = "arXiv",
    primaryClass = "hep-th",
    doi = "10.1002/prop.202300143",
    journal = "Fortsch. Phys.",
    volume = "71",
    number = "10-11",
    pages = "2300143",
    year = "2023"
}

@article{Cribiori:2022nke,
    author = {Cribiori, Niccol\`o and L\"ust, Dieter and Staudt, Georgina},
    title = "{Black hole entropy and moduli-dependent species scale}",
    eprint = "2212.10286",
    archivePrefix = "arXiv",
    primaryClass = "hep-th",
    reportNumber = "LMU-ASC 56/22, MPP-2022-289",
    doi = "10.1016/j.physletb.2023.138113",
    journal = "Phys. Lett. B",
    volume = "844",
    pages = "138113",
    year = "2023"
}

@article{Fichet:2022ixi,
    author = "Fichet, Sylvain and Megias, Eugenio and Quiros, Mariano",
    title = "{Continuum effective field theories, gravity, and holography}",
    eprint = "2208.12273",
    archivePrefix = "arXiv",
    primaryClass = "hep-ph",
    doi = "10.1103/PhysRevD.107.096016",
    journal = "Phys. Rev. D",
    volume = "107",
    number = "9",
    pages = "096016",
    year = "2023"
}

@article{Castellano:2022bvr,
    author = "Castellano, Alberto and Herr\'aez, Alvaro and Ib\'a\~nez, Luis E.",
    title = "{The emergence proposal in quantum gravity and the species scale}",
    eprint = "2212.03908",
    archivePrefix = "arXiv",
    primaryClass = "hep-th",
    reportNumber = "IFT-UAM/CSIC-22-149",
    doi = "10.1007/JHEP06(2023)047",
    journal = "JHEP",
    volume = "06",
    pages = "047",
    year = "2023"
}

@article{vandeHeisteeg:2022btw,
    author = "van de Heisteeg, Damian and Vafa, Cumrun and Wiesner, Max and Wu, David H.",
    title = "{Moduli-dependent species scale}",
    eprint = "2212.06841",
    archivePrefix = "arXiv",
    primaryClass = "hep-th",
    doi = "10.4310/bpam.2024.v1.n1.a1",
    journal = "Beijing J. Pure Appl. Math.",
    volume = "1",
    number = "1",
    pages = "1--41",
    year = "2024"
}

@article{Andriot:2023isc,
    author = "Andriot, David",
    title = "{Bumping into the Species Scale with the Scalar Potential}",
    eprint = "2305.07480",
    archivePrefix = "arXiv",
    primaryClass = "hep-th",
    doi = "10.1002/prop.202300139",
    journal = "Fortsch. Phys.",
    volume = "71",
    number = "10-11",
    pages = "2300139",
    year = "2023"
}

@article{Cribiori:2023sch,
    author = {Cribiori, Niccol\`o and L\"ust, Dieter},
    title = "{A Note on Modular Invariant Species Scale and Potentials}",
    eprint = "2306.08673",
    archivePrefix = "arXiv",
    primaryClass = "hep-th",
    reportNumber = "LMU-ASC 20/23, MPP-2023-126",
    doi = "10.1002/prop.202300150",
    journal = "Fortsch. Phys.",
    volume = "71",
    number = "10-11",
    pages = "2300150",
    year = "2023"
}

@article{Calderon-Infante:2023uhz,
    author = "Calder\'on-Infante, Jos\'e and Delgado, Matilda and Uranga, Angel M.",
    title = "{Emergence of species scale black hole horizons}",
    eprint = "2310.04488",
    archivePrefix = "arXiv",
    primaryClass = "hep-th",
    reportNumber = "IFT-UAM/CSIC-23-123, CERN-TH-2023-173",
    doi = "10.1007/JHEP01(2024)003",
    journal = "JHEP",
    volume = "01",
    pages = "003",
    year = "2024"
}

@article{vandeHeisteeg:2023dlw,
    author = "van de Heisteeg, Damian and Vafa, Cumrun and Wiesner, Max and Wu, David H.",
    title = "{Species scale in diverse dimensions}",
    eprint = "2310.07213",
    archivePrefix = "arXiv",
    primaryClass = "hep-th",
    doi = "10.1007/JHEP05(2024)112",
    journal = "JHEP",
    volume = "05",
    pages = "112",
    year = "2024"
}

@article{Castellano:2023aum,
    author = "Castellano, Alberto and Herr\'aez, Alvaro and Ib\'a\~nez, Luis E.",
    title = "{On the Species Scale, Modular Invariance and the Gravitational EFT expansion}",
    eprint = "2310.07708",
    archivePrefix = "arXiv",
    primaryClass = "hep-th",
    month = "10",
    year = "2023"
}

@article{Scalisi:2024jhq,
    author = "Scalisi, Marco",
    title = "{Species Scale and Primordial Gravitational Waves}",
    eprint = "2401.09533",
    archivePrefix = "arXiv",
    primaryClass = "hep-th",
    reportNumber = "MPP-2024-7",
    doi = "10.1002/prop.202400033",
    journal = "Fortsch. Phys.",
    volume = "72",
    number = "6",
    pages = "2400033",
    year = "2024"
}

@article{Martucci:2024trp,
    author = "Martucci, Luca and Risso, Nicol\`o and Valenti, Alessandro and Vecchi, Luca",
    title = "{Wormholes in the axiverse, and the species scale}",
    eprint = "2404.14489",
    archivePrefix = "arXiv",
    primaryClass = "hep-th",
    doi = "10.1007/JHEP07(2024)240",
    journal = "JHEP",
    volume = "07",
    pages = "240",
    year = "2024"
}

@article{Casas:2024jbw,
    author = "Casas, Gonzalo F. and Ib\'a\~nez, Luis E.",
    title = "{Modular Invariant Starobinsky Inflation and the Species Scale}",
    eprint = "2407.12081",
    archivePrefix = "arXiv",
    primaryClass = "hep-th",
    reportNumber = "IFT-UAM/CSIC-24-105",
    month = "7",
    year = "2024"
}

@article{Seo:2024zzs,
    author = "Seo, Min-Seok",
    title = "{Axion species scale and axion weak gravity conjecture-like bound}",
    eprint = "2407.16156",
    archivePrefix = "arXiv",
    primaryClass = "hep-th",
    doi = "10.1007/JHEP11(2024)082",
    journal = "JHEP",
    volume = "11",
    pages = "082",
    year = "2024"
}

@article{Aoki:2024ixq,
    author = "Aoki, Shuntaro and Otsuka, Hajime",
    title = "{Inflationary constraints on the moduli-dependent species scale in modular invariant theories}",
    eprint = "2411.08467",
    archivePrefix = "arXiv",
    primaryClass = "hep-th",
    reportNumber = "CTPU-PTC-24-34, KYUSHU-HET-302",
    month = "11",
    year = "2024"
}

@article{Distler:2005hi,
    author = "Distler, Jacques and Varadarajan, Uday",
    title = "{Random polynomials and the friendly landscape}",
    eprint = "hep-th/0507090",
    archivePrefix = "arXiv",
    reportNumber = "UTTG-07-05",
    month = "7",
    year = "2005"
}

@article{Basile:2024dqq,
    author = "Basile, Ivano and Cribiori, Niccol\`o and Lust, Dieter and Montella, Carmine",
    title = "{Minimal black holes and species thermodynamics}",
    eprint = "2401.06851",
    archivePrefix = "arXiv",
    primaryClass = "hep-th",
    doi = "10.1007/JHEP06(2024)127",
    journal = "JHEP",
    volume = "06",
    pages = "127",
    year = "2024"
}

@article{Herraez:2024kux,
    author = {Herr\'aez, Alvaro and L\"ust, Dieter and Masias, Joaquin and Scalisi, Marco},
    title = "{On the Origin of Species Thermodynamics and the Black Hole - Tower Correspondence}",
    eprint = "2406.17851",
    archivePrefix = "arXiv",
    primaryClass = "hep-th",
    reportNumber = "MPP-2024-123, LMU-ASC 08/24",
    month = "6",
    year = "2024"
}

@article{Cribiori:2023ffn,
    author = "Cribiori, Niccol\`o and Lust, Dieter and Montella, Carmine",
    title = "{Species entropy and thermodynamics}",
    eprint = "2305.10489",
    archivePrefix = "arXiv",
    primaryClass = "hep-th",
    reportNumber = "LMU-ASC 18/23, MPP-2023-97",
    doi = "10.1007/JHEP10(2023)059",
    journal = "JHEP",
    volume = "10",
    pages = "059",
    year = "2023"
}

@article{Alberte:2020jsk,
    author = "Alberte, Lasma and de Rham, Claudia and Jaitly, Sumer and Tolley, Andrew J.",
    title = "{Positivity Bounds and the Massless Spin-2 Pole}",
    eprint = "2007.12667",
    archivePrefix = "arXiv",
    primaryClass = "hep-th",
    reportNumber = "Imperial/TP/2020/LA/02",
    doi = "10.1103/PhysRevD.102.125023",
    journal = "Phys. Rev. D",
    volume = "102",
    number = "12",
    pages = "125023",
    year = "2020"
}

@article{Caron-Huot:2022ugt,
    author = "Caron-Huot, Simon and Li, Yue-Zhou and Parra-Martinez, Julio and Simmons-Duffin, David",
    title = "{Causality constraints on corrections to Einstein gravity}",
    eprint = "2201.06602",
    archivePrefix = "arXiv",
    primaryClass = "hep-th",
    reportNumber = "CALT-TH 2021-003",
    doi = "10.1007/JHEP05(2023)122",
    journal = "JHEP",
    volume = "05",
    pages = "122",
    year = "2023"
}

@article{Caron-Huot:2024lbf,
    author = "Caron-Huot, Simon and Li, Yue-Zhou",
    title = "{Gravity and a universal cutoff for field theory}",
    eprint = "2408.06440",
    archivePrefix = "arXiv",
    primaryClass = "hep-th",
    month = "8",
    year = "2024"
}

@article{Caron-Huot:2024tsk,
    author = "Caron-Huot, Simon and Tokuda, Junsei",
    title = "{String loops and gravitational positivity bounds: imprint of light particles at high energies}",
    eprint = "2406.07606",
    archivePrefix = "arXiv",
    primaryClass = "hep-th",
    reportNumber = "CTPU-PTC-24-17",
    doi = "10.1007/JHEP11(2024)055",
    journal = "JHEP",
    volume = "11",
    pages = "055",
    year = "2024"
}

@article{Caron-Huot:2021rmr,
    author = "Caron-Huot, Simon and Mazac, Dalimil and Rastelli, Leonardo and Simmons-Duffin, David",
    title = "{Sharp boundaries for the swampland}",
    eprint = "2102.08951",
    archivePrefix = "arXiv",
    primaryClass = "hep-th",
    doi = "10.1007/JHEP07(2021)110",
    journal = "JHEP",
    volume = "07",
    pages = "110",
    year = "2021"
}

@article{Anber:2011ut,
    author = "Anber, Mohamed M. and Donoghue, John F.",
    title = "{On the running of the gravitational constant}",
    eprint = "1111.2875",
    archivePrefix = "arXiv",
    primaryClass = "hep-th",
    doi = "10.1103/PhysRevD.85.104016",
    journal = "Phys. Rev. D",
    volume = "85",
    pages = "104016",
    year = "2012"
}

@article{Aydemir:2012nz,
    author = "Aydemir, Ufuk and Anber, Mohamed M. and Donoghue, John F.",
    title = "{Self-healing of unitarity in effective field theories and the onset of new physics}",
    eprint = "1203.5153",
    archivePrefix = "arXiv",
    primaryClass = "hep-ph",
    doi = "10.1103/PhysRevD.86.014025",
    journal = "Phys. Rev. D",
    volume = "86",
    pages = "014025",
    year = "2012"
}

@article{Calmet:2017omb,
    author = "Calmet, X. and Casadio, R. and Kamenshchik, A. Yu. and Teryaev, O. V.",
    title = "{Graviton propagator, renormalization scale and black-hole like states}",
    eprint = "1708.01485",
    archivePrefix = "arXiv",
    primaryClass = "hep-th",
    doi = "10.1016/j.physletb.2017.09.080",
    journal = "Phys. Lett. B",
    volume = "774",
    pages = "332--337",
    year = "2017"
}

@article{Branchina:2023ogv,
    author = "Branchina, Carlo and Branchina, Vincenzo and Contino, Filippo and Pernace, Arcangelo",
    title = "{Does the Cosmological Constant really indicate the existence of a Dark Dimension?}",
    eprint = "2308.16548",
    archivePrefix = "arXiv",
    primaryClass = "hep-th",
    month = "8",
    year = "2023"
}

@article{Aldazabal:2000cn,
    author = "Aldazabal, G. and Franco, S. and Ibanez, Luis E. and Rabadan, R. and Uranga, A. M.",
    title = "{Intersecting brane worlds}",
    eprint = "hep-ph/0011132",
    archivePrefix = "arXiv",
    reportNumber = "CAB-IB-2919500, CERN-TH-2000-320, MIT-CTP-3042, FTUAM-00-23, IFT-UAM-CSIC-00-37",
    doi = "10.1088/1126-6708/2001/02/047",
    journal = "JHEP",
    volume = "02",
    pages = "047",
    year = "2001"
}

@article{Casas:2024ttx,
    author = "Casas, Gonzalo F. and Ib\'a\~nez, Luis E. and Marchesano, Fernando",
    title = "{Yukawa Couplings at Infinite Distance and Swampland Towers in Chiral Theories}",
    eprint = "2403.09775",
    archivePrefix = "arXiv",
    primaryClass = "hep-th",
    reportNumber = "IFT-UAM/CSIC-24-40",
    month = "3",
    year = "2024"
}

@book{Burgess:2020tbq,
    author = "Burgess, C. P.",
    title = "{Introduction to Effective Field Theory}",
    doi = "10.1017/9781139048040",
    isbn = "978-1-139-04804-0, 978-0-521-19547-8",
    publisher = "Cambridge University Press",
    month = "12",
    year = "2020"
}

@book{Green:2012oqa,
    author = "Green, Michael B. and Schwarz, John H. and Witten, Edward",
    title = "{Superstring Theory Vol. 1}: {25th Anniversary Edition}",
    doi = "10.1017/CBO9781139248563",
    isbn = "978-1-139-53477-2, 978-1-107-02911-8",
    publisher = "Cambridge University Press",
    series = "Cambridge Monographs on Mathematical Physics",
    month = "11",
    year = "2012"
}

@article{navas2024review,
  title={Review of Particle Physics},
  author={Navas, S},
  journal={Phys. Rev. D},
  volume={110},
  pages={030001},
  year={2024}
}

@article{Fermi:1934hr,
    author = "Fermi, E.",
    title = "{An attempt of a theory of beta radiation. 1.}",
    reportNumber = "UCRL-TRANS-726",
    doi = "10.1007/BF01351864",
    journal = "Z. Phys.",
    volume = "88",
    pages = "161--177",
    year = "1934"
}

@article{Niedermaier:2006wt,
    author = "Niedermaier, Max and Reuter, Martin",
    title = "{The Asymptotic Safety Scenario in Quantum Gravity}",
    doi = "10.12942/lrr-2006-5",
    journal = "Living Rev. Rel.",
    volume = "9",
    pages = "5--173",
    year = "2006"
}

@inbook{Weinberg:1980gg,
    author = "Weinberg, Steven",
    title = "{ULTRAVIOLET DIVERGENCES IN QUANTUM THEORIES OF GRAVITATION}",
    booktitle = "{General Relativity}: {An Einstein Centenary Survey}",
    pages = "790--831",
    year = "1980"
}

@article{Gomis:1995jp,
    author = "Gomis, Joaquim and Weinberg, Steven",
    title = "{Are nonrenormalizable gauge theories renormalizable?}",
    eprint = "hep-th/9510087",
    archivePrefix = "arXiv",
    reportNumber = "RIMS-1036, UTTG-18-95",
    doi = "10.1016/0550-3213(96)00132-0",
    journal = "Nucl. Phys. B",
    volume = "469",
    pages = "473--487",
    year = "1996"
}

@article{Branchina:2023rgi,
    author = "Branchina, Carlo and Branchina, Vincenzo and Contino, Filippo",
    title = "{Naturalness and UV sensitivity in Kaluza-Klein theories}",
    eprint = "2304.08040",
    archivePrefix = "arXiv",
    primaryClass = "hep-th",
    doi = "10.1103/PhysRevD.108.045007",
    journal = "Phys. Rev. D",
    volume = "108",
    number = "4",
    pages = "045007",
    year = "2023"
}

@article{Branchina:2024ljd,
    author = "Branchina, Carlo and Branchina, Vincenzo and Contino, Filippo and Pernace, Arcangelo",
    title = "{Dark Dimension and the Effective Field Theory limit}",
    eprint = "2404.10068",
    archivePrefix = "arXiv",
    primaryClass = "hep-th",
    doi = "10.1142/S0219887824503031",
    month = "4",
    year = "2024"
}

@article{Basile:2023blg,
    author = {Basile, Ivano and L\"ust, Dieter and Montella, Carmine},
    title = "{Shedding black hole light on the emergent string conjecture}",
    eprint = "2311.12113",
    archivePrefix = "arXiv",
    primaryClass = "hep-th",
    reportNumber = "LMU-ASC 35/23, MPP-2023-262",
    doi = "10.1007/JHEP07(2024)208",
    journal = "JHEP",
    volume = "07",
    pages = "208",
    year = "2024"
}

@article{Ashoorioon:2011aa,
    author = "Ashoorioon, A. and Danielsson, U. and Sheikh-Jabbari, M. M.",
    title = "{1/N resolution to inflationary \ensuremath{\eta}-problem}",
    eprint = "1112.2272",
    archivePrefix = "arXiv",
    primaryClass = "hep-th",
    reportNumber = "UUITP-26-11, IPM-P-2011-052",
    doi = "10.1016/j.physletb.2012.06.034",
    journal = "Phys. Lett. B",
    volume = "713",
    pages = "353--357",
    year = "2012"
}

@article{DeLuca:2024fbc,
    author = "De Luca, G. Bruno and De Ponti, Nicol\`o and Mondino, Andrea and Tomasiello, Alessandro",
    title = "{Can you hear the Planck mass?}",
    eprint = "2406.00095",
    archivePrefix = "arXiv",
    primaryClass = "hep-th",
    doi = "10.1007/JHEP08(2024)123",
    journal = "JHEP",
    volume = "08",
    pages = "123",
    year = "2024"
}

@article{Calderon-Infante:2025ldq,
    author = "Calder\'on-Infante, Jos\'e and Castellano, Alberto and Herr\'aez, Alvaro",
    title = "{The Double EFT Expansion in Quantum Gravity}",
    eprint = "2501.14880",
    archivePrefix = "arXiv",
    primaryClass = "hep-th",
    month = "1",
    year = "2025"
}

\end{document}